# Extreme Precision Radial Velocity Working Group

# Final Report

July 2021

Edited by:
Jonathan Crass, Department of Physics, University of Notre Dame
Scott Gaudi, Department of Astronomy, The Ohio State University
Stephanie Leifer, Jet Propulsion Laboratory, California Institute of Technology

Crass, J.[1], Gaudi, B.S.[2], Leifer, S.[3], Beichman, C.[4], Bender, C.[5], Blackwood, G.[6], Burt, J.[6], Callas, J.L.[6], Cegla, H.[7], Diddams, S.[8], Dumusque, X.[9], Eastman, J.[10], Ford, E.[11], Fulton, B.J.[4], Gibson, R.[12], Halverson, S.[13], Haywood, R.[14], Hearty, F.[11], Howard, A.[15], Latham, D.[10], Loehner-Boettcher, J.[16], Mamajek, E.[6], Mortier, A.[17], Newman, P.[18], Plavchan, P.[18], Quirrenbach, A.[19], Reiners, A.[20], Robertson, P.[21], Roy, A.[22], Schwab, C.[23], Seifahrt, A.[24], Szentgyorgyi, A.[10], Terrien, R.[25], Teske, J.[26], Thompson, S.[27], Vasisht, G.[3]

1 Department of Physics, University of Notre Dame,
2 Department of Astronomy, The Ohio State University
3 NASA Jet Propulsion Laboratory, California Institute of Technology
4 NASA Exoplanet Science Institute, California Institute of Technology
5 University of Arizona
6 NASA Exoplanet Exploration Program, Jet Propulsion Laboratory
7 University of Geneva / University of Warwick
8 National Institute of Standards and Technology
9 Université de Genève
10 Harvard-Smithsonian Center for Astrophysics
11 Pennsylvania State University
12 Columbia University
13 Massachusetts Institute of Technology
14 Harvard-Smithsonian Center for Astrophysics / University of Exeter
15 California Institute of Technology
16 University Corporation for Atmospheric Research
17 Kavli Institute for Cosmology, University of Cambridge
18 George Mason University
19 Landessternwarte, University of Heidelberg
20 University of Göttingen
21 University of California, Irvine
22 Space Telescope Science Institute
23 Macquarie University
24 University of Chicago
25 Carleton College
26 Carnegie Observatories / Department of Terrestrial Magnetistm (now Earth and Planets Laboratory)
27 University of Cambridge

# Working Group Members, Participants, and Contributors

| Name/Role | Affiliation |
|---|---|
| **Steering Group** | |
| Gary Blackwood, *Co-Chair* | NASA Exoplanet Exploration Program, Jet Propulsion Laboratory |
| Scott Gaudi, *Co-Chair* | Ohio State University |
| Chas Beichman | NASA Exoplanet Science Institute, California Institute of Technology |
| Jennifer Burt, *Observation Strategies Co-Chair* | NASA Exoplanet Exploration Program, Jet Propulsion Laboratory |
| Heather Cegla, *Stellar Variability Group Co-Chair* | University of Geneva / University of Warwick |
| Eric Ford, *Data Processing/Analysis Co-Chair* | Pennsylvania State University |
| Andrew Howard, *Science Mission Drivers Co-Chair* | California Institute of Technology |
| David Latham | Harvard-Smithsonian Center for Astrophysics |
| Eric Mamajek | NASA Exoplanet Exploration Program, Jet Propulsion Laboratory |
| Peter Plavchan | George Mason University |
| Andreas Quirrenbach, *Resource Evaluation Co-Chair* | Landessternwarte, University of Heidelberg |
| **General Members** | |
| Chad Bender, *Science Mission Drivers Co-Chair & Tellurics Chair* | University of Arizona |
| John Callas | NASA Exoplanet Exploration Program, Jet Propulsion Laboratory |
| Jonathan Crass | University of Notre Dame |
| Scott Diddams, *Resource Evaluation Co-Chair* | National Institute of Standards and Technology |
| Xavier Dumusque | Université de Genève |
| Jason Eastman | Harvard-Smithsonian Center for Astrophysics |
| B.J. Fulton | NASA Exoplanet Science Institute, California Institute of Technology |
| Rose Gibson | Columbia University |
| Sam Halverson, *Performance & Error Budget Chair* | Massachusetts Institute of Technology |
| Raphaëlle Haywood, *Stellar Variability Co-Chair* | Harvard-Smithsonian Center for Astrophysics / University of Exeter |
| Fred Hearty | Pennsylvania State University |
| Stephanie Leifer, *Instrumentation Co-Chair* | NASA Jet Propulsion Laboratory, California Institute of Technology |
| Johannes Loehner-Boettcher | University Corporation for Atmospheric Research |
| Annelies Mortier | Kavli Institute for Cosmology, University of Cambridge |
| Patrick Newman | George Mason University |
| Ansgar Reiners | University of Göttingen |
| Paul Robertson | University of California, Irvine |
| Arpita Roy, *Data Processing/Analysis Co-Chair* | Space Telescope Science Institute |
| Christian Schwab | Macquarie University |
| Andreas Seifahrt | University of Chicago |
| Andrew Szentgyorgyi, *Instrumentation Co-Chair* | Harvard-Smithsonian Center for Astrophysics |
| Ryan Terrien | Carleton College |
| Johanna Teske, *Observing Strategies Co-Chair* | Carnegie Observatories / Department of Terrestrial Magnetistm (now Earth and Planets Laboratory) |
| Samantha Thompson | University of Cambridge |
| Gautam Vasisht | NASA Jet Propulsion Laboratory, California Institute of Technology |

| Name/Role | Affiliation |
|---|---|
| **Participants** | |
| Suzanne Aigrain | Oxford University |
| Megan Bedell | Flatiron Institute |
| Rebecca Bernstein | Carnegie Observatories |
| Ryan Blackman | Yale University |
| Cullen Blake | University of Pennsylvania |
| Lars Buchhave | Technical University of Denmark |
| Bill Chaplin | University of Birmingham |
| David Ciardi | NASA Exoplanet Science Institute, California Institute of Technology |
| Jessi Cisewski-Kehe | Yale University |
| Andrew Collier Cameron | University of St. Andrews |
| Matthew Cornachione | University of Utah |
| Debra Fischer | Yale University |
| Nadege Meunier | University of Grenoble |
| Joe Ninan | Pennsylvania State University |
| John O'Meara | W.M. Keck Observatory |
| Joel Ong | Yale University |
| Sharon Wang | Carnegie Observatories |
| Sven Wedemeyer | University of Oslo |
| Lily Zhao | Yale University |
| **Review Committee** | |
| Jacob Bean | University of Chicago |
| Steve Howell | NASA Ames Research Center |
| Michael McElwain | NASA Goddard Space Flight Center |
| Josh Winn | Princeton University |

# Acknowledgements

The Extreme Precision Radial Velocity (EPRV) Working Group gratefully acknowledges the National Aeronautics and Space Administration (NASA) and National Science Foundation (NSF) for establishing this body and supporting its effort to develop a roadmap for enabling EPRV detection of temperate, terrestrial exoplanets. We thank the Exoplanet Exploration Program Technical Advisory Committee for its oversight and review of the Working Group's findings. We also thank the Flatiron Institute and Carnegie Institution for Science for hosting in-person Working Group meetings.

We acknowledge the excellent discussions and scientific input by members of the International Team number 453, "Towards Earth-like Alien Worlds: Know thy star, know thy planet," supported by the International Space Science Institute (ISSI, Bern; Principal Investigator: Chris A. Watson). We thank Ryan Cloutier, Allen Davis, Colin Folsom, Claire Moutou, and Belinda Nicholson for their expert contributions to discussions on intrinsic magnetic variability. The Instrumentation Analysis Group thanks subject matter experts, specifically Francesco Pepe, Nem Jovanovic, and Rich Dekany, who presented material to us for use in the study. The Working Group thanks Laura Kreidberg for her input on atmospheric characterization.

Raphaëlle Haywood performed this work under contract with the California Institute of Technology (Caltech) / Jet Propulsion Laboratory (JPL) funded by NASA through the Sagan Fellowship Program executed by the NASA Exoplanet Science Institute, and with funding from the College of Engineering, Mathematics and Physical Sciences at the University of Exeter, UK.

H. M. Cegla acknowledges financial support from the National Centre for Competence in Research (NCCR) PlanetS, supported by the Swiss National Science Foundation (SNSF) through a CHaracterising ExOPlanets Satellite (CHEOPS) Fellowship, as well as funding from a UK Research and Innovation (UKRI) Future Leaders Fellowship, grant number MR/S035214/1.

Work by Scott Gaudi was partially supported by the Thomas Jefferson Chair for Space Exploration endowment at The Ohio State University.

Part of this work was carried out by the Jet Propulsion Laboratory, California Institute of Technology under contract with the National Aeronautics and Space Administration.

# Table of Contents

# Appendices

# Executive Summary

Precise mass measurements of exoplanets discovered by the direct imaging or transit technique are required to determine planet bulk properties and potential habitability. It is generally acknowledged that for the foreseeable future, the Extreme Precision Radial Velocity (EPRV) measurement technique, with precision at the level of <10 cm/s, is the only method potentially capable of detecting and measuring the masses of Earth analogs orbiting solar-type stars from the ground (see the discussions in the Astronomy and Astrophysics 2010 Decadal Survey report "*New Worlds, New Horizons*" and the National Academy of Sciences (NAS) Exoplanet Science Strategy (ESS) report, and references therein). Relative astrometry at the microarcsecond level also offers the potential of measuring the masses of Earth analogs orbiting solar-type stars (Unwin et al. 2008), but this requires going to space, and thus would be significantly more expensive than even the extensive ground-based EPRV initiative we recommend here.

The reflex radial velocity of a solar-type star being orbited by an Earth has an amplitude of 9 cm/s and a period of 1 year. Thus, EPRV measurements with a precision of better than approximately 10 cm/s (with a few cm/s stability over many years) are required to determine the masses and orbits of habitable-zone Earths orbiting nearby F, G, and K spectral-type stars.[1] Unfortunately, for nearly a decade, precision radial velocity (PRV) instruments and surveys have been unable to routinely reach radial velocity (RV) accuracies of less than roughly 1 m/s (Fischer et al. 2016; Dumusque et al. 2017). However, the next generation of EPRV instruments, which have been designed to have instrumental accuracies approaching 10 cm/s, have recently been developed (ESPRESSO, Pepe et al. 2010; EXPRES, Jurgenson et al. 2016; MAROON-X, Seifahrt et al. 2018; and NEID, Schwab et al. 2016). Maximizing the knowledge gained from these instruments and optimizing their output can be used as a stepping stone in forging a plausible pathway to detecting Earth analogs orbiting nearby solar-type stars and beginning a program that develops the tools needed to achieve this common goal.

Making EPRV science and technology development a critical component of both NASA and NSF program plans is crucial for reaching the goal of characterizing potentially habitable Earth-like planets and supporting future exoplanet direct imaging missions such as the Habitable Exoplanet Observatory (HabEx; Gaudi et al. 2020) or Large Ultraviolet Optical Infrared Surveyor (LUVOIR; The LUVOIR Team 2019). Not only may such efforts enable the mass measurement of spectroscopically characterized Earth analogs, they may also enable prioritization, discovery, and efficient scheduling of potential targets for study. Given the ground-based nature of EPRV programs, coordination and cooperation between NASA and NSF utilizes the strengths of each agency in attempting to achieve this ambitious goal.

In recognition of these facts, the 2018 National Academy of Sciences (NAS) Exoplanet Science Strategy (ESS) report recommended the development of EPRV measurements as a critical step toward the detection and characterization of habitable, Earth-analog planets:

> *NASA and NSF should establish a strategic initiative in extremely precise radial velocities to develop methods and facilities for measuring the masses of temperate terrestrial planets orbiting Sun-like stars.*

[1] Strictly speaking, if the noise in the RV signal is truly independent from observation to observation, then one could use a large number of observations to measure an amplitude less than the single measurement precision. However, it is not yet known if it will be possible to make measurements with independent measurement errors at the 10 cm/s level, particularly once accounting for stellar variability.

This is not a new recognition; the radial velocity detection of exoplanets was strongly endorsed by the Astronomy and Astrophysics 2010 Decadal Survey report "*New Worlds, New Horizons.*" Achieving the requisite RV precision and accuracy requires addressing four broad categories of RV signal uncertainty: stellar variability, telluric contamination, instrumental uncertainty, and pure statistical uncertainty.

In response to the NAS recommendation, NASA and NSF commissioned the 'Extreme Precision Radial Velocity Working Group' to recommend a ground-based program architecture and implementation plan to achieve the goal intended by the NAS. This report documents the findings of the EPRV Working Group, which was tasked to "deliver to the NASA Astrophysics Division (APD) and the NSF Division of Astronomical Sciences (AST) a […] recommendation of the best ground-based program architecture and implementation to achieve the goal of measuring the masses of temperate terrestrial planets orbiting Sun-like stars." We note that, while probe-class space-based EPRV and astrometry are promising candidates for detecting and measuring the masses of Earth analogs, a consideration of these alternatives is outside of the scope of this study.

The main findings and recommendations of the study are:

1. **There exist multiple plausible system architectures in terms of telescope size, longitude and latitude distribution, and dedication that could, if stellar variability mitigation, telluric mitigation, and instrumental precision goals are met, successfully acquire a set of measurements with the statistical precision required to detect Earth analogs.** These architectures could leverage existing telescopes, utilize new telescopes built for this purpose, or a mixture of both.
2. **The most significant obstacle to achieving EPRV capabilities is the intrinsic variability of planet host stars.** To address this, we must radically advance our understanding of the underlying stellar physics and its impact on RV measurements. **We recommend immediately implementing a *long-term, large-scale, interdisciplinary* research and analysis program in this area**. Such a program will include observational (solar and stellar), numerical, and theoretical efforts, and address key science questions by order of priority and importance. This research program is essential to determine the optimal pathway for stellar variability mitigation in a future ground-based EPRV survey.
3. **The extent of research required exceeds the capacity of the present EPRV community.** The scale of effort needed to address the stellar variability problem requires not only increasing the size of the EPRV community, but also increased cooperation within the community. Growing the number of experts prepared to advance the state-of-the-art will require training new graduate students, and ensuring that there are attractive postdoctoral opportunities for them to build on their EPRV expertise. However, there is a more immediate issue, in that **there are not presently enough personnel in the U.S. alone with expertise in high-precision RV, and solar and stellar physics to conduct the necessary analyses in the next few years.** Given both the scale of facilities and the breadth of expertise required, detecting Earth analogs will necessarily be an international endeavor, and the establishment of both domestic and international collaborations is crucial to increase the likelihood of success.
4. **Given the timescale required to develop and carry out a strategic EPRV initiative, knowledge retention in the field of EPRV science and technology is a key issue.** To date, such retention has been challenging, as a great deal of the work thus far has been conducted by graduate students and postdoctoral research fellows, many of whom have gone on to other fields following their academic programs. **The current cycle of short-term appointments is not conducive to the success of this extended program. A deliberate and urgent effort** **must be made to retain early career experts in high-precision RV science (including solar and stellar physics) and technology. We need to provide viable and attractive career opportunities, including (but not limited to) long-term, EPRV-dedicated postdoctoral positions of 4–5 years.**

5. **Establishment of an EPRV Research Coordination Network and Standing Advisory Committee is advised to help with collaboration and coordination of efforts (e.g., appropriate overlap of stellar target observations).** While each existing and upcoming EPRV instrument will have its own science program, by coordinating, the community will improve its ability to make meaningful comparisons across different instruments. Long-term oversight by a Research Coordination Network of current and future EPRV instruments globally would enable EPRV teams to establish and maintain strong, long-lasting synergies across institutions/countries and disciplines (solar, stellar, exoplanetary). Such guidance will enable the community to have shared, coordinated target lists and observing strategies/cadences amongst the instruments. This level of coordination will maximize the quality of the solar and stellar science that is needed to break the stellar variability barrier, and improve methodology and algorithms for analysis of spectroscopic time-series to detect and characterize low-mass planets.
6. **An EPRV program should curate and leverage datasets and knowledge gained from existing state-of-the-art instruments, as well as the next-generation instruments currently under development, construction, and commissioning:**
   a. Where feasible, leverage knowledge and experience gained from current EPRV instruments including HARPS/HARPS-N/HARPS3 (Mayor et al. 2003; Cosentino et al. 2012; Thompson et al. 2016), ESPRESSO, EXPRES, MAROON-X, and NEID.
   b. Perform high-cadence observations with solar feeds mounted on several spectrographs that demonstrate <50 cm/s stability over a day. The minimum specifications should be daily cadence, spectral resolution of greater than 100,000 and signal-to-noise ratio (SNR) per resolution element of greater than 300. Continuous solar monitoring (achieved using multiple spectrographs at overlapping longitudes) is crucial to track subtle instrumental systematics and validate stellar variability mitigation techniques over at least one solar cycle.
   c. To the extent possible, coordinate observations with major instruments on a small set of bright standard stars on a variety of timescales in order to create a comprehensive dataset that can be used to disentangle instrumental systematics from other sources of systematic uncertainties (with the same minimum specifications listed in 6b).
   d. Understand the potential performance and extensibility to the visible band of single-mode fiber-fed, diffraction-limited instruments with adaptive optics systems like the PAlomar Radial Velocity Instrument (PARVI; Gibson et al. 2020; Vasisht et al., in prep) and iLocater (Crepp et al. 2016) as a path to lower cost, essentially telescope-aperture independent spectrograph architectures for EPRV.
   e. Coordinate the data availability, data analysis techniques, statistical tools, and results from these projects.
7. Establishing confidence in detections of low-mass planets will be increasingly challenging. Therefore, **we view it as critical for the EPRV community to carry out a set of EPRV data challenges designed to evaluate the effectiveness and reliability of the advanced analysis methods that are being developed.** In order to reach this point, we recommend that the EPRV community design a comprehensive roadmap for a series of data challenges that would lead to an improved ability to extract science from EPRV observations and a better understanding of the capabilities and limitations of EPRV data analysis as a function of key properties of instruments, target stars, and survey strategy. The cadence of these data challenges should be sufficient to address the relevant open questions within the desired timeline. We anticipate one or two data challenges per year will be required. The topics of these individual challenges should be staggered such that they each engage different subsets of the EPRV community, in order to avoid undue burdens on its members.

8. Gaps in RV data pipelines and analysis tools should be addressed:
   a. **There should be a designated centralized repository of RV datasets** in a curated form (standardized formats) where researchers can apply their tools to publicly released datasets from multiple instruments. Examples of current archives that could be adapted for this purpose are the NASA Exoplanet Science Institute (NExScI) Exoplanet Archive and the Data Analysis Centre for Exoplanets hosted by the University of Geneva.
   b. **There should be a publicly available software pipeline package of data reduction and analysis algorithms.** This package would include a collection of pipelines, analytic tools, and modules that researchers could mix and match with their own algorithms. These should be seamlessly coordinated with the central RV data repository discussed in (a). Ideally, this package would contain a uniform data analysis pipeline that could be used to reduce and process data across instruments, and would be modular, customizable, and open-source for continued community-led development.
9. **EPRV technology testbeds, including full end-to-end systems coupled to solar feeds, should be established.** These testbeds would enable technology maturation ranging from key components to a full instrumental system architecture, analogous to those developed for exoplanet direct imaging programs. Such testbeds will provide a critical component of a technology maturation program that is needed to achieve improved instrumental precision over the current generation of instruments.
10. To address telluric line contamination in an EPRV survey, **extensive laboratory spectroscopic study is needed in the near-infrared and visible spectral regions using very long gas absorption cells, high intensity light sources, and cavity ring-down spectroscopy (CRDS) systems** to improve line lists and depths. Further, input line lists from the target stars in the EPRV survey will be needed for spectrum fitting.
11. **The potential cost of a comprehensive EPRV program to detect and characterize Earth analogs is comparable to that of a Medium-class Explorer (MIDEX) flight mission.**

Given these findings and recommendations, the EPRV Working Group has developed a notional 15-year implementation Roadmap for a funded development program. This Roadmap consists broadly of three stages of activities and is motivated by the Working Group's analysis that provided an 'existence proof' that a plausible set of architectures could, if all sources of systematic uncertainties can be modeled or controlled at or below the 10 cm/s level, survey and detect Earth analogs around the likely set of target stars in a reasonable amount of time.

The first stage is a coordinated set of activities over approximately 5 years that are needed to determine the feasibility of mitigating major sources of systematic errors (i.e., stellar variability, telluric contamination, and instrumental error) at the required level to detect Earth analogs. If it is determined that these systematic error sources can be reasonably mitigated, a second stage would be warranted and would have a high chance of success. This second stage would be a 5–10-year precursor survey of specific target stars, based on a strategically selected list of direct imaging mission targets. This list would be curated and maintained at a designated EPRV initiative science center or centers. The precursor program would allow a detailed characterization of host star variability and the development of mitigation strategies while also providing prioritization data for future observations. A full EPRV survey, which would utilize newly constructed advanced EPRV instruments to detect and measure the masses of Earth analogs around the set of target stars, would be the third and final stage spanning a further 5+ year period. Importantly, this schedule includes gateway reviews (notionally every two to five years) that would evaluate the progress towards the primary goals of each stage of the EPRV initiative. Gateway reviews will enable the refocusing of investment in key areas if required to advance progress, allow the adjustment of program goals, and would also include potential ‘off ramps,’ which could be invoked should it be deemed that a pathway towards a relevant stage goal is unlikely to be optimal or successful. The schedule also includes regular community meetings and workshops to maximize widespread engagement.

The Working Group identified the following activities that should be carried out during the three stages of the EPRV Roadmap. These include determining whether sources of systematic uncertainties can be modeled or controlled, while also addressing remaining unknowns regarding the optimal method for achieving the highest statistical precision for measuring planet masses. Additionally, they address programmatic considerations involved in setting up and maintaining the EPRV initiative.

### Stage 1: Roadmap to Determine the Feasibility and Likelihood of a Successful EPRV Survey

*Stellar Variability:*

- Support the adoption of solar feeds for spectrographs that demonstrate >100k spectral resolution, SNR > 300 and <50 cm/s stability over a day.
- Determine how well variability mitigation strategies built from Sun-as-a-star knowledge translate to other spectral types present in the proposed list of target stars.
- Determine the level of RV precision and accuracy enabled by stellar variability mitigation strategies and modeling, and the corresponding implications for the planet mass determinations.

*Instrumentation:*

- Maximize knowledge gained from current instruments still under development, construction, and commissioning, and that have recently begun operations.
- Establish or use existing testbeds for full instrument system development and component testing and characterization, and verification of software-based instrument simulators and select pipeline modules.
- Determine whether it will be possible to secure sufficient charge-coupled devices (CCDs) for a full EPRV survey in a timely manner.
- Determine whether CMOS detectors are an acceptable replacement for CCD detectors.
- Investigate the continued availability of gratings required for high-resolution spectrographs.
- Determine if there are alternative, cheaper, and more robust methods of fabricating gratings.
- Investigate the feasibility of securing robust, long-lived, high-stability calibration sources.

*Architecture:*

- Determine whether extreme adaptive optics (AO) in the visible combined with fiber injection of single-mode fiber-fed diffraction limited spectrographs present a viable and more desirable option than traditional seeing-limited RV instruments.

- Determine whether it would be practical to retrofit existing telescopes for dedicated, robotic operation for EPRV observations, or whether new telescopes are required. Determine which of these options presents the least risk and requires the fewest resources.
- Identify suppliers for multiple, large-aperture telescope systems and conduct site selection surveys, should the building of new telescopes be required for the full EPRV survey.

*Telluric Line Contamination:*

- Determine whether telluric contamination lines in spectra can be adequately mitigated, over a sufficiently broad wavelength range.

*Extensive and Detailed Theoretical Analysis:*

- Determine what combination of spectral bandwidth, resolution, SNR, and cadence is sufficient for the detection of Earth-analog systems.
- Explore the advantages of spectropolarimetry in EPRV observations (e.g., as implemented in SPIRou).
- Specify, based on lessons learned from observational data, the required quality of the spectra (including resolution and SNR) to detect Earth analogs, which then constrains the required effective apertures, observing time, and cadence.

*Software:*

- Support the development of a well-designed, well-engineered, and actively maintained open-source pipeline with the demonstrated ability to retrieve state-of-the-art results on EPRV data from multiple instruments.

*Programmatics:*

- Ensure the needed staffing of personnel with expertise in PRV, heliophysics, and stellar variability to conduct the necessary analysis as well as conduct an EPRV survey. This necessarily involves establishing formal collaborations with non-U.S. entities, and creating attractive employment paths for early career (graduate students, postdoctoral fellows, and non-tenure track researchers) experts in PRV science and technology.
- Establish a Research Coordination Network and Standing Advisory Committee.

### Stage 2: An EPRV Precursor Survey

Stage 1 will provide key knowledge needed to assess the viability of mitigating key error sources, which limit current EPRV capabilities. If it is determined that error sources are likely to be reasonably mitigated, Stage 2 of the program will proceed. This would be a 5–10-year precursor survey of specific target stars informed by the NASA Astrophysics Decadal Mission Concept Studies for LUVOIR-A, LUVOIR-B, HabEx, and Starshade Rendezvous target lists with additional criteria relevant for measuring precise radial velocities. The precursor program will use existing EPRV instruments that match the resolution, bandwidth, and precision requirements defined during Stage 1 of the program. The precursor survey should start as soon as feasible once instruments with appropriate performance and resources are available. This can occur prior to the completion of Stage 1 to maximize observational baselines and will allow a detailed characterization of host star variability and development of stellar variability mitigation strategies. The precursor survey will provide important prioritization criteria for a future Earth-analog survey.

### Stage 3: An EPRV Survey for Earth Analogs

Should all of the activities in Stages 1 and 2 of the Roadmap be completed in a successful and satisfactory manner, and the Advisory Committee deem that surveying the list of target stars identified for a direct imaging mission will result in the detection or exclusion of Earth analogs for the majority of the targets, then the initiative will move onto the third stage – a full EPRV survey. The full EPRV survey will require developing new instruments and may require building new telescope facilities. Construction may need to commence during the latter parts of Stage 2 to enable a timeline that is compatible with the proposed mission concepts that would perform direct imaging of Earth analogs around the nearest solar-type stars (≥2035 launch dates). Upon commissioning, the program will require a minimum observational period of 5 years (nominally 10 years) to reach its goal. The Working Group estimated that such a survey will cost roughly the equivalent of a NASA MIDEX mission.

# 1 INTRODUCTION AND MOTIVATION

## 1.1 Exoplanetary Mass Determination

The mass of an exoplanet is one of its most defining characteristics, and provides clues to its formation history, as well as information about its composition and atmosphere. Similarly, the exoplanet mass function, particularly when considered as a function of orbital semimajor axis, is an observable that provides a fundamental empirical test of the validity of ab initio theories of planet formation (e.g., Suzuki et al. 2018). For planets detected via transits, the mass of the planet provides the bulk density, and thus defines the basic nature of the planet (rocky, ice giant, or gas giant). When combined with an estimate of the radius,[2] mass is critical for understanding exoplanetary atmospheres and can be key to interpreting absorption or emission spectra (von Paris et al. 2013; Nayak et al. 2017). This characterization is particularly important for potentially habitable planets (Valencia, O'Connell & Sasselov 2007; Papuc & Davies 2008; Valencia & O'Connell 2009; van Heck & Tackley 2011; Dorn et al. 2018).

There are two primary methods for measuring the mass of a nearby extrasolar planet[3]: radial velocity (or Doppler shift) and astrometry. Each method presents its own set of opportunities and challenges. Radial velocity (RV) measurements only provide the minimum mass of an exoplanet. However, with an estimate of the host star mass and a constraint on its inclination (via transits, direct imaging, or astrometry), this can be turned into an absolute mass measurement. Astrometry, on the other hand, can be used to directly estimate the planet mass, with an estimate of the host star mass.

When considering nearby Earth analogs (e.g., Earth mass planets orbiting in the habitable zones of solar-type stars at distances of ~10 pc), both methods are challenged by the very small signal that must be detected. Assuming a circular orbit, the Doppler reflex amplitude, *K,* of an Earth analog orbiting a solar analog is:

$$K = (2\pi G/P)^{1/3} M_*^{-2/3} M_p sin(i) = 9\ cm/s (P/year)^{-1/3} \left(M_*/M_\odot\right)^{-2/3} \left(M_p/M_\oplus\right) sin(i),$$

where $P$ is the orbital period, $M_p$ and $M_*$ are the mass of the planet and host star, respectively, and we have assumed that $M_p \ll M_*$. Finally, sin(*i*) is the sine of the inclination of the planetary orbit normal with respect to the line-of-sight (such that sin(*i*)=90 is an edge-on orbit). The astrometric signal, $\alpha$, of an identical system located at a distance of $d$=10 pc from the sun is:

$$\alpha \ = \ (M_p/M_*)(a/d) \ = \ 0.3\, \mu as\ (M_p/M_\oplus)(a/AU)(d/10\, pc)^{-1},$$

where $a$ is the semimajor axis of the orbit. The amplitudes of both signals are very small, and are orders of magnitude less than the smallest relative RV or narrow angle astrometric signals that have been detected to date.

A crucial difference between RV and astrometry techniques is that detecting an astrometric signal with an amplitude of <10 $\mu as$ can *only* be accomplished in space due to the aberrations induced by the Earth's atmosphere. The radial velocity amplitude of an Earth analog orbiting a solar-type star is 9 cm/s with a 1-year period. Thus, RV measurements with a precision of better than ~10 cm/s (with a few cm/s stability over many years) are required to determine the masses and orbits of habitable-zone (HZ) Earths orbiting nearby F, G, and K spectral-type stars. Achieving such precisions (and more importantly, accuracies), are firmly in the regime of Extreme Precision RV (EPRV) measurements.

[2] The radius of a directly imaged potentially habitable planet can be constrained (to a factor of 0.5–2x) via its visible light spectrum (Nayak et al. 2017; Feng et al. 2018), or directly inferred from its thermal emission spectrum (e.g., Quanz et al. 2019).

[3] We note that transit timing variations (TTVs) can be used to measure the masses of nearby exoplanets, but this method requires a fairly specific architecture, e.g., a multiplanet system in which the planet is in 'dynamical contact' with the other planets in the system, and at least one of the planets transits. Thus, the TTV method of measuring the mass of an exoplanet is not applicable to most of the targets for a direct imaging mission.

## 1.2 The Definitions of RV Precision and Accuracy

The terms “RV precision” and “RV accuracy” are often used casually and interchangeably, as well as used in different contexts with different meanings. Thus, it is important to define what is meant by these terms as they are used in this document.

As described in a study report to the Exoplanet Exploration Program Analysis Group (ExoPAG; Plavchan et al. 2015), RV precision refers to the single measurement statistical uncertainty, generally assuming Gaussian statistics and no correlations between different observations. Defined in this way, the RV measurement precision of a given observation on a given star that is achievable is a function of characteristics including information content of the spectrum, which is itself a function of the signal-to-noise ratio (SNR) per resolution element, spectral grasp and wavelength range, spectral resolution, spectral type, projected stellar rotation speed, and RV extraction technique. This metric reflects the minimum irreducible measurement uncertainty that is achievable for any given observation.

In contrast, the accuracy quantifies the reliability of a RV measurement. A measurement may be precise (e.g., have small statistical uncertainties), but not accurate (e.g., the “true” value of the RV may be different from the estimated value by an amount that is large compared to the quoted precision). This may be due to unaccounted-for signals (e.g., intrinsic stellar variability, telluric contamination). These are often lumped together under the umbrella term `systematics.’ In the case of RV measurements, some authors casually refer to any unmodeled signal or excess noise as `RV jitter.’ However, jitter is best reserved to describe signals and/or noise that are modeled as a source of uncorrelated noise (i.e., high frequency compared to timescale of observations). It is important to specify not only the magnitude of any systematics, but also their timescale, e.g., measurements may be correlated over a few minutes, but uncorrelated over a few months. If the single measurement RV precisions are larger (i.e., less precise) than the magnitude of the RV signal (as is currently the case for Earth analogs), a robust detection of the RV signal would require taking a (weighted) average of many observations. In this case, the measurement precision for the velocity amplitude of a planet can be significantly smaller than the single measurement precision. In the case of an Earth analog with $K$~10 cm/s and orbital period of ~1 year, if the single measurement precision is ~1 m/s, then achieving a SNR~10 detection requires averaging over ~$10^4$ data points (in the case of uncorrelated Gaussian noise), and thus an accuracy of at least ~1 cm/s on the timescale of roughly a few months. In practice, correlated noise (e.g., residual noise after attempting to correct for intrinsic stellar signals) is likely to result in even more observations being required to push beyond the single measurement precision.

We note that RV jitter can have contributions from unmodeled instrumental effects, stellar variability, and telluric spectral lines. However, if one can successfully model these effects, then these systematics become signals that can be removed, resulting in measurements that are both precise and accurate. Alternatively, if these effects result in jitter that is uncorrelated over the relevant timescales, then one can obtain a measurement of the velocity amplitude that is significantly more precise than any single measurement by combining many observations. However, this generally requires significantly more measurements and thus resources. A potential, and perhaps likely, scenario is that one can successfully model a significant fraction of the stellar variability signal, but some residual signal remains as a source of correlated errors. The magnitude, timescale, and strength of correlations will be critical in determining the number of observations needed to reach science goals.

## 1.3 Sources of RV Systematic Errors

There are three broad categories of uncertainties in ground-based RV measurements (see Figure 1-1). The first is simply the photon noise uncertainty, $\sigma_{photon}$. This depends on many quantities, including the telescope aperture, instrument throughput, and exposure time. It also depends on the spectral grasp of the instrument, the spectral resolution, and the intrinsic RV information of the star (e.g., the number, depths, and widths of the photospheric absorption lines – see Bouchy, Pepe & Queloz 2001 and Beatty & Gaudi 2015 for more details). The photon noise uncertainty represents the minimum achievable uncertainty for any given measurement. The second source of uncertainty we broadly label as the facility uncertainty, $\sigma_{facility}$. This is the uncertainty due to the instrument, data reduction pipeline, and contamination from telluric absorption lines. Facility uncertainty is often detailed in instrument error budgets and these have been produced for most of the next-generation EPRV instruments that have recently, or will soon, come online. There is an additional uncertainty due to stellar variability, $\sigma_{star}$, arising from brightness and velocity variations on the surfaces of Sun-like stars. These phenomena include magneto-convection (granulation and supergranulation), pulsations, faculae/plage, starspots, and flares.

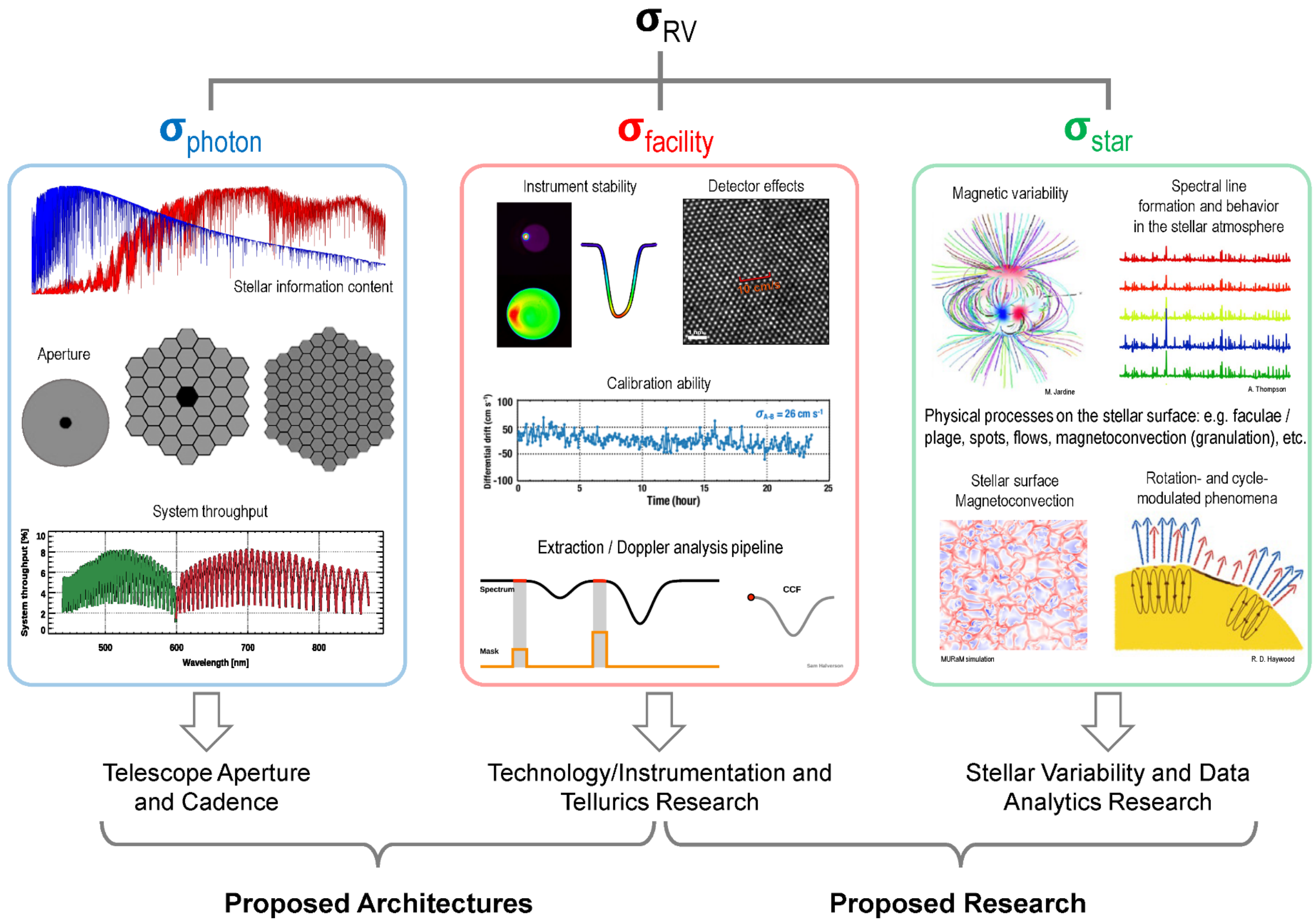

**FIGURE 1-1.** Various sources of RV uncertainty, grouped into three broad categories: photon noise, facility, and stellar variability. The first two sources depend on the telescope, survey strategy, wavelength range and resolution, stellar RV information content, instrument (including stability, detector effects, and calibration), and data reduction pipeline. The last category depends on the various sources of stellar variability (some of which are shown in the figure), as well as the timescale over which those sources of variability manifest themselves. Figure by Sam Halverson.

The latter two sources of uncertainty, if they are not mitigated, represent systematic errors. It is important to specify not only their magnitude but also the timescale over which these effects are correlated. If the uncertainties due to stellar variability and facility uncertainty are not mitigated at the required level, then they provide the minimum irreducible uncertainty. Conversely, if the net photon noise uncertainty (e.g., after averaging many individual data points) is significantly larger than the uncertainties due to the facility or stellar variability, then there is little point in attempting to mitigate these other sources of uncertainty. Thus, the approach adopted by the Working Group (WG) has been to first determine if there exists one or more system architectures (consisting of a set of telescopes and instruments), that could achieve the photon noise uncertainty and cadence required to detect Earth analogs orbiting nearby solar-type stars (Section 4). If such an architecture can be identified, then it provides motivation to perform the research and analysis to improve the precision of RV instruments, and mitigate RV systematic uncertainties caused by stellar variability, instrumental effects, and tellurics.

## 1.4 The Challenge of Stellar Variability

Understanding and modeling intrinsic stellar variability is critical to achieving EPRV capabilities. Our limited knowledge of Sun-like stars poses the most significant threat to confirming and characterizing Earth-like worlds. Sun-like stars have intrinsic variability, primarily driven by convection, magnetic fields and their interplay, that induces correlated RV signals with 0.1–100 m/s amplitudes spanning timescales from seconds to decades (e.g., see reviews by Fischer et al. 2016; Dumusque et al. 2017; Collier Cameron 2018; Cegla 2019; Hatzes 2019, and references therein). **Intrinsic stellar variability currently precludes the confirmation and characterization of Earth-analogs** (NAS 2018, pp. 92–93).

When searching for planets with a stabilized spectrograph, traditionally, the stellar spectra are cross-correlated with a template and the radial velocity is determined from the centroid of the cross-correlation function (e.g., Baranne et al. 1996; Pepe et al. 2002). Consequently, any changes in the stellar line shapes can be mistaken as Doppler shifts. Time-varying inhomogeneities on stellar surfaces can lead to spectral line profile variation causing spurious RV measurements that mask or mimic planetary signals (e.g., CoRoT-7 “d”: Hatzes et al. 2010; Lanza et al. 2010; Pont et al. 2011; Haywood et al. 2014). The underlying physics and corresponding RV impacts are further detailed in Appendix A, Section A.2. **Understanding and modeling intrinsic stellar variability is critical to achieving EPRV.**

## 1.5 Achieving Extreme Precision Radial Velocity Measurements

The radial velocity amplitude of an Earth analog orbiting a solar-type star is 9 cm/s with a 1-year period. Thus, EPRV measurements with a precision of better than approximately 10 cm/s (with a few cm/s stability over many years) are required to determine the masses and orbits of habitable-zone Earths orbiting nearby F, G, and K spectral-type stars. Achieving such precisions (and more importantly, accuracies), is firmly in the regime of EPRV. The primary barriers to achieving such precisions and accuracies are four fold. First, the change in the shape of the stellar spectral lines as a function of time due to stellar variability must be removed on a timescale significantly shorter than the orbital period of an Earth analog (e.g., ~1 year). Second, the contamination of the spectrum due to telluric absorption features (both macro and micro telluric features) must be measured and eliminated (for the portion of the spectrum used to measure velocities), also at a timescale shorter than the orbital period of an Earth analog. Third, the instrumental uncertainty, specifically our ability to assign a given spectral feature to an absolute (or at least fixed) wavelength, must be controlled to much better than the systematic error requirement. Finally, the pure statistical power of the dataset must be high enough to enable sufficient information to not only detect the RV signal, but also be able to measure (and robustly disentangle) the above systematic uncertainties.

To date, astrophysical, telluric, and instrumental effects have limited reliable detections to amplitudes of approximately 1 m/s. Is there a path to separate stellar and instrumental variability from true Doppler shifts to lower this floor by roughly ***two orders of magnitude*** from the ground? In principle, the answer is yes: true Doppler shifts will shift all of the stellar spectral lines by the same value of $\Delta\lambda/\lambda$, regardless of the wavelength, the shape of the spectral lines, and the physical location(s) in the star's photosphere where the lines are being formed. Stellar, telluric, and instrumental effects will not behave in the manner. However, whether or not the Doppler signal can be cleanly separated from the other effects in practice has yet to be determined.

If systematic errors can be controlled or mitigated, RV monitoring of stars from the ground will allow for the detection and characterization of planetary orbits (typically excluding the longitude of ascending node and the inclination) of Earth analogs around solar-type stars. This will enable, at the very least, measurement of the minimum mass of such planets detected by proposed future space exoplanet imaging missions such as HabEx or LUVOIR. If these planets are detected before the launch of these missions, then knowing which stars have such planets and the position of the planets in their orbits can be used to prioritize potential targets for study, and may significantly increase the efficiency and/or yield of these missions as shown in Figure 1-2.

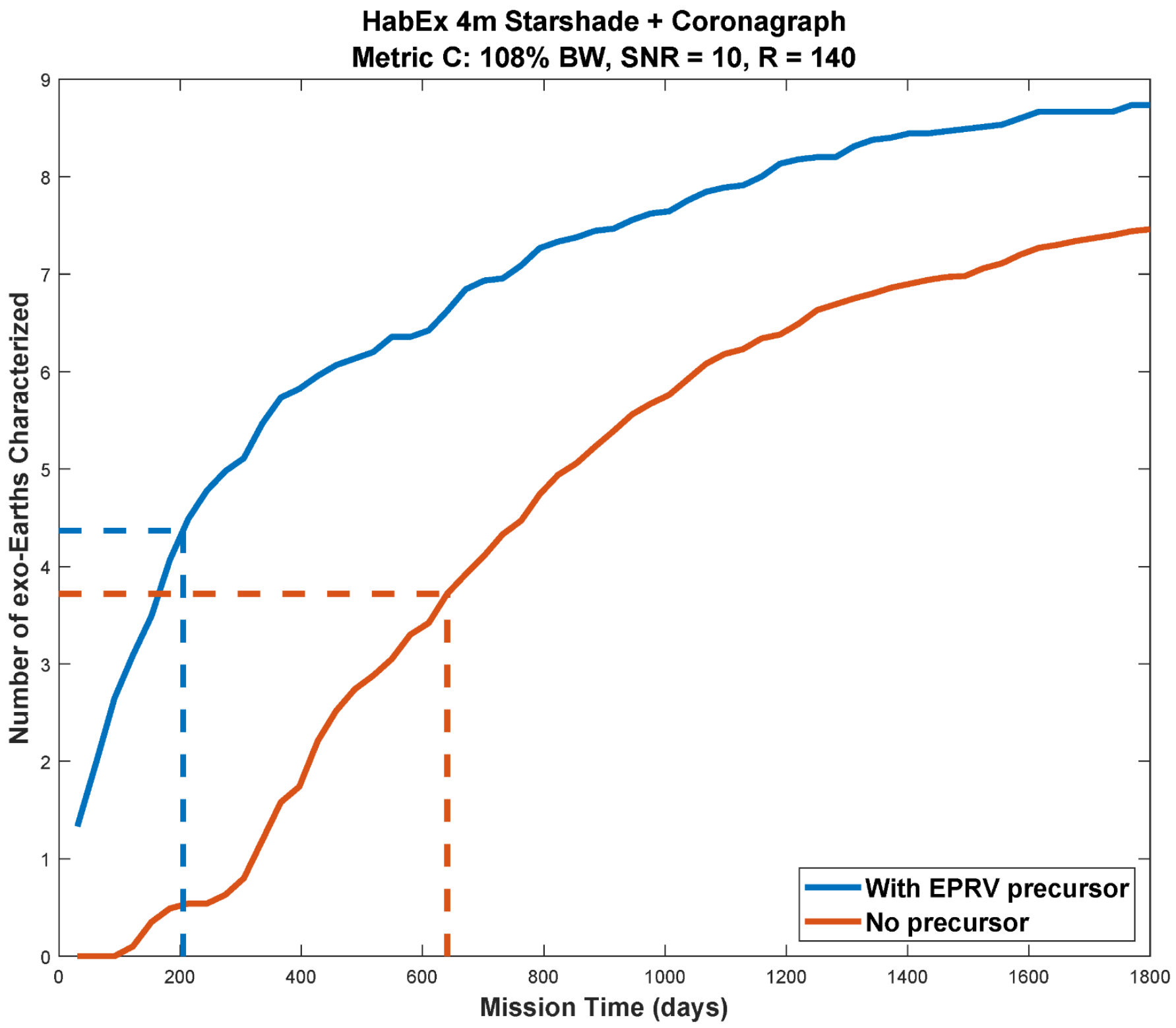

**FIGURE 1-2.** The simulated number of characterized exo-Earths for the HabEx mission (4 m starshade and chronograph architecture) with and without an EPRV precursor survey. EPRV precursor observations reduce the mission time to achieve 50% of the yield of characterized planets by a factor of 3, allowing high impact science to occur earlier in the mission, providing more time for follow up characterization, and mitigating the risk of early mission failure (adapted from Morgan et al. 2021 (submitted)).

Correspondingly, the precise measurement of the Doppler shifts of starlight resulting from the radial velocity of planet-hosting stars has been deemed a critical component of NASA's technology program, and was identified in the National Academy of Sciences (NAS) Exoplanet Science Strategy (ESS) report as a critical step toward the detection and characterization of habitable, Earth analog planets. The ESS report issued these findings:

- "Mass is the most fundamental property of a planet, and knowledge of a planet's mass (along with a knowledge of its radius) is essential to understand its bulk composition and to interpret spectroscopic features in its atmosphere. If scientists seek to study Earth-like planets orbiting Sun-like stars, they need to push mass measurements to the sensitivity required for such worlds."
- "The radial velocity method will continue to provide essential mass, orbit, and census information to support both transiting and directly imaged exoplanet science for the foreseeable future."
- "Radial velocity measurements are currently limited by variations in the stellar photosphere, instrumental stability and calibration, and spectral contamination from telluric lines. Progress will require new instruments installed on large telescopes, substantial allocations of observing time, advanced statistical methods for data analysis informed by theoretical modeling, and collaboration between observers, instrument builders, stellar astrophysicists, heliophysicists, and statisticians."

The report went on to issue the following recommendation:

> *NASA and NSF should establish a strategic initiative in extremely precise radial velocities to develop methods and facilities for measuring the masses of temperate terrestrial planets orbiting Sun-like stars.*

# 2 EPRV Working Group Structure, Scope, and Deliverables

In response to the NAS directive, NASA and the NSF commissioned the EPRV Working Group to recommend a ground-based program architecture and implementation plan to achieve the goal intended by the NAS. The task statement for the EPRV-WG was laid out in the Terms of Reference (ToR). The ToR defines the background, deliverables, the structure of participation, steering, and oversight of the Working Group, workflow, and schedule. Here we simply reiterate the deliverables of the EPRV-WG as stated in the ToR:

> *The EPRV Working Group is established to deliver to the NASA Astrophysics Division (APD) and the NSF Division of Astronomical Sciences (AST) a report, as a first step, that includes a recommendation of the best ground-based program architecture and implementation to achieve the goal of measuring the masses of temperate terrestrial planets orbiting Sun-like stars. In the Working Group deliverable recommendation and at decision points of the subsequent Initiative (if implemented) possible ground-based solutions will be identified. If ground-based solutions are not found then another study may be considered for a space-based solution. The recommendation will define a roadmap that NASA/APD and NSF/AST can carry out jointly or separately to achieve the necessary breakthrough in extreme precision radial velocity measurement.*
>
> *[Upon delivery of the report, the EPRV Working Group will subsequently be disbanded]. The report will include scope, schedule, and planning-level funding requirements. The report may include both directed and competed scope. No selection criteria for specific competed scope will be developed, and all products and deliberations of the EPRV Working Group will be conducted and documented in an open forum.*
>
> *NASA/APD and NSF/AST will discuss the report's findings within the context of the existing NASA-NSF Exoplanet Observational Research (NN-EXPLORE)[4] partnership agreement. NASA and NSF will consider the recommendations for implementation through their own processes.*

The NASA-NSF EPRV Working Group brought together many of the world's experts in EPRV to outline a program that can:

1. Determine whether or not achieving a few cm/s systematic precision is achievable, and if so,
2. Determine what fiducial survey architecture(s) (if any) may be able to survey the nearest few hundred solar-type stars and determine whether or not they have Earth analogs.

This goal, following the Exoplanet Science Strategy Recommendation and the Charter of the Working Group, was captured as the following decision statement, which was agreed by all members of the Working Group:

> **Decision Statement:** Recommend the best ground-based program architecture and implementation (a.k.a. Roadmap) to achieve the goal of measuring the masses of temperate terrestrial planets orbiting Sun-like stars.

The Working Group comprised members who represent the breadth of science, technology, engineering, and programmatic (schedule, cost) expertise necessary to deliver the program recommendations. A series of subgroups (termed Analysis Groups) were formed from the Working Group membership to define, analyze, and develop key program requirements or architectures (Table 2-1). Membership of subgroups was overseen by a steering group to ensure suitable expertise in each area and cross-membership between Analysis Groups where appropriate. Each Analysis Group had Chairs/Co-Chairs who facilitated coordination of each Analysis Group and were responsible for reporting back to the full Working Group.

[4] https://exoplanets.nasa.gov/exep/NNExplore/

The Working Group sessions commenced on June 6, 2019, and met via teleconferences typically every other week with analysis groups meeting at the same frequency. In addition, three multi-day, face-to-face Working Group meetings took place in St. Louis, MO, New York, NY, and Washington, D.C., with the last of these concluding on January 31, 2020.

The rest of this report outlines the EPRV-WG's deliverables, approach, findings, and recommendations. This report documents the methodology used by the EPRV Working Group to develop a roadmap for achieving the goal of measuring the masses of temperate terrestrial planets orbiting Sun-like stars, and details specific recommendations to that end.

**TABLE 2-1.** EPRV Analysis Groups.

| | Analysis Groups | | | | | | | |
|---|---|---|---|---|---|---|---|---|
| **Group:** | **A** | **B** | **C** | **D** | **E** | **F** | **G** | **H** |
| **Name:** | Science Mission Drivers | Performance & Error Budgets | Instrumentation | Stellar Variability | Observing Strategies | Data Processing/ Analysis | Resource Evaluation | Tellurics |
| **Chair or Co-Chairs:** | Andrew Howard<br>Chad Bender | Sam Halverson | Stephanie Leifer<br>Andrew Szentgyorgyi | Heather Cegla<br>Raphaëlle Haywood | Jennifer Burt<br>Johanna Teske | Eric Ford<br>Arpita Roy | Andres Quirrenbach<br>Scott Diddams | Chad Bender |
| **Goal or Tasks:** | Define nominal science requirements for a future EPRV program including spectral types, magnitude, wavelength range, etc. | Provide a nominal estimate of the required RV precision needed to enable the program defined by Group A. | Identify key technical needs, development strategies, possible architectures and hardware risks in delivering the program defined by Group A at the precision defined by Group B. | Define nominal observational requirements (resolution, SNR, cadence, etc.) needed to sufficiently mitigate stellar variability in RV measurements. | Assess the viability of different survey strategies and any proposed architectures to enable the program defined by Group A. | Identify critical needs to develop optimal data pipelines for future EPRV programs. | Assess cost, schedule and risk for proposed program architectures. | Determine observational and theoretical needs to achieve required telluric mitigation in the program defined by Group A. |
| **Group Report:** | - | - | Appendix A, Section A.1, p. A-1 | Appendix A, Section A.2, p. A-16 | Section 4, p. 4-1<br><br>Appendix A.3, p. A-28 | Appendix A.4, p. A-30 | Section 5.3.1, p. 5-2<br><br>Appendix A, Section A.1, p. A-1 | Appendix A.5, p. A-37 |

# 3 STUDY PROCESS AND METHODOLOGY

Determining the viability of any EPRV program required clearly defined goals and metrics for assessment. With the charge of recommending the optimal ground-based program architecture and implementation (a.k.a. roadmap) to measure the masses of temperate terrestrial planets orbiting Sun-like stars, the full Working Group defined a set of success criteria and requirements for achieving this goal (Section 3.1). The criteria were used to develop proposed instrument architectures (Section 3.2), which were assessed against their likelihood to meet the defined objectives (Section 4 and Section 5).

## 3.1 Success Criteria Formulation and Definition

The Working Group developed a list of requirements and desired attributes of an EPRV program to define the success criteria for fulfilling the ESS recommendation (shown in Tables 3-1 and 3-2). Many of the success criteria were designed around developing an instrument and program architecture with the ability to detect an Earth analog around a defined set of survey stars (Section 3.1.1). The desired attributes were grouped into categories of science, schedule, difficulty, cost, and other factors.

Six requirements (Table 3-1) were documented:

1. *Determine*[5] the feasibility by 2025 to detect earth-mass planets in the habitable zone of solar-type stars.
2. *Demonstrate*[6] the feasibility (i.e., provide validation by a combination of analysis and test) of detecting Earth-mass planets in the HZ of solar-type stars prior to the 2030 Decadal Survey.
3. Conduct a set of precursor surveys to characterize stellar variability.
4. *Demonstrate* the feasibility to survey (~100) stars on the 'green' target star list (See Section 3.1.1) at the level able to detect earth-mass planets in the HZ of solar-type stars.
5. By 2025 *demonstrate* on-sky capabilities to detect planets with $K = 30$ cm/sec and periods of ~100 days.
6. Capture knowledge from current and near-term instruments.

Sixteen weighted "Wants" (desires, or goals) were documented (Table 3-2). Four "Wants" emerged as key and driving:

1. Survey as many stars as possible on the "yellow" target star list (~100 additional stars; see Section 3.1.1).
2. Follow up temperature terrestrial transit discoveries to inform the mass-radius relation.
3. Provide the greatest relative probability of success to meet stellar variability mitigation requirements (Appendix A.2).
4. Minimize estimated cost of the program.

[5] In the context of the defined requirements, *determine* was defined as: To provide evidence through simulation, analysis, or actual measurement that the systems under consideration can, with further development, meet the specified requirements. This evaluation serves as a milestone prior to proceeding to the development phase.

[6] In the context of the defined requirements, *demonstrate* was defined as: To show through simulation, analysis, or actual measurement, bottom up or top down, that the systems under consideration are highly likely to meet the specified requirements. This evaluation serves as a milestone prior to proceeding to the science surveys (i.e., validation).

**TABLE 3-1.** EPRV success criteria – requirements.

| | Success Criteria | Technical Requirements and *Comments* |
|---|---|---|
| **M0a** | ***Determine* the feasibility** by 2025 to detect (with a well-characterized and sufficiently small false discovery rate) and measure the mass (msin(*i*) with <=10% fractional precision) of <=1 Earth mass planets that orbit a 1 $M_{Sun}$ main sequence star and receive insolation within 10% of that of the Earth ($Insolation_{Earth}$). | False discovery rate of <= 1/(alpha N_target_stars) for each star being surveyed based on EPRV data alone (i.e., not including additional evidence from transits, direct imaging, astrometry, etc.), where N_target_stars is the number of stars to be included in EPRV surveys (including all targets with significant observations, not just those receiving the most intensive EPRV observations) and alpha is a constant to fall in a range of 1–10 that should be set at a later date based on how well we can mitigate stellar variability; (2) a fractional precision of <=10% on $m_p \sin i_p$ (for RV in isolation). Validate methods of stellar variability mitigation, telluric mitigation, and statistical validation, key for the EPRV method, including using follow-up of transiting planets.<br><br>*Comment: Latitude (hemispheric) diversity in telescope; sufficient longitude diversity in telescope* |
| **M0b** | ***Demonstrate* the feasibility** to detect (with a well-characterized and sufficiently small false discovery rate) and measure the mass (msin(*i*) with <=10% fractional precision) of <=1 Earth mass planets that orbit a 1 $M_{Sun}$ main sequence star and receive insolation within 10% $Insolation_{Earth}$ prior to 2030 Decadal Survey. | Demonstrate = Validate, by a combination of analysis and test; Analysis Group defines nomenclature for terms<br><br>*Comment: Terrestrial implied by mass and insolation* |
| **M1a** | Design and execute a set of **precursor surveys and analysis activities** on the 'green' and 'yellow' stars on evolving target star list and on the Sun. | In order to characterize the stellar variability of the target stars. Evaluate the resources required to mitigate stellar variability to the required levels.<br><br>*Comment: The target list is those objects for which a HZ Earth analog has predicted spectroscopic exposure times < 60 days as calculated by a NASA mission concept study. The target list is provided by the Exoplanet Exploration Program (ExEP) Science Office and is informed by the NASA Astrophysics Decadal Mission Concept Studies for LUVOIR-A, LUVOIR-B, HabEx, and Starshade Rendezvous, with additional criteria relevant for measuring precise radial velocities. Targets are classified as **required (green)** or **desired (yellow)**. Required targets appear on the HabEx deep list, or two or more of the above noted study target lists, are restricted to spectral types F7–K9, and have literature rotation velocities of vsini < 5 km/s. Desired targets are not included in the required target sample, appear on at least one study list, expand the allowed spectral type range to include M-dwarfs, and have vsini < 10 km/s. The required list currently has ~100 targets; the desired list currently has ~125 targets.* |
| **M1b** | ***Demonstrate* the feasibility** to survey each of the 'green' stars on the evolving target list at the level of M0b. | Review progress early decade and triennially. Facilities and analysis required to do so.<br><br>*Comment: Actual commit-to star list would be after precursor surveys. Consequence is both hemispheres. Risk: Too little telescope time with current generation of instruments to learn lessons, inform next-generation instruments.* |
| **M2** | Meet Intermediate Milestone: **By 2025, demonstrate on-sky feasibility** with capabilities in-hand to detect *K* down to 30 cm/s for periods out to few hundred days using a statistical method that has been validated using simulated and/or observed spectra time-series | Demonstrate = Validate, by a combination of analysis and test. Analysis Group A (Science Mission Drivers) defines *K* |
| **M4** | **Capture knowledge** from current and near-future generation of instruments, surveys, analysis, and coordination activities to help inform development of future EPRV instruments. | *Comment: Implies more than static; also continue usage of products from operations as possible. Come back to solar and stellar activities* |

**TABLE 3-2.** Desired EPRV program attributes.

| | Success Criteria | Technical Requirements and *Comments* |
|---|---|---|
| | **Relative Science** | |
| **W1** | Survey as many 'yellow' stars as possible on the evolving target list | "Reflected Must M1b" |
| **W2** | Measure masses of temperate terrestrial planets orbiting M stars, not on yellow list | |
| **W3** | Use follow-up of transiting temperate terrestrial planets to inform the mass-radius relation from key transit discoveries | |
| **W4** | Validate methods of stellar variability mitigation, telluric mitigation, and statistical validation, key for the EPRV method, including using follow-up of transiting planets | *Need for current and near-future transit missions* |
| | **Relative Schedule** | |
| **W5** | Schedule: Start the precursor M1a surveys as soon as possible, so as to maximize impact at Preliminary Design Review (PDR) on design of direct imaging missions (e.g., HabEx, LUVOIR) | Impacts survey/operations. Launch readiness date (LRD) HabEx 2035. LRD LUVOIR 2039 before LRD of direct imaging missions<br><br>*Begin the survey at the performance level referenced in M0b as early as possible* |
| **W6** | Schedule: Start the survey as soon as possible, so as to maximize impact at PDR on design of direct imaging missions (e.g., HabEx, LUVOIR) | Impacts design of missions. HabEx PDR – Feb 2029. LUVOIR PDR (LRD − 5 = 2034 at time of writing).<br><br>*But still science value in exoplanet detection via EPRV independent of whether DI mission selected by Astro2020* |
| | **Relative Difficulty** | |
| **W7** | Prefer the architecture with the greatest relative probability of success to meet stellar variability requirement | Implies: greatest probability of success, and community confidence in the results |
| **W8** | Relative difficulty to secure required telescopes/instruments, fraction of time, and observing cadence and coordination between telescopes | |
| **W9** | Prefer the architecture with the greatest probability of success of achieving the survey referenced in M1b | Including, but not exclusive of, technical and schedule risk. Prefer the architecture with the lowest relative risk of successfully achieving the survey referenced in M1b |
| | **Relative Cost** | |
| **W10** | Least estimated cost | *Estimated costs should be plausible (consensus by the group)* |
| | **Other Factors** | |
| **W11** | Take advantage of opportunities for international collaboration and draw from as broad of a pool of relevant expertise and observing facilities as possible | |
| **W12** | Maximize use of, and knowledge and understanding of, existing facilities (observatories), infrastructure, and hardware (including detectors) | *All else being equal, use existing infrastructure rather than build new* |
| **W13** | Maximize broader impacts in society | Including, but not limited to, increasing underrepresented groups in the field, outreach, scientific credibility |
| **W14** | Encourage free exchange of ideas, including data and source codes | |
| **W15** | Implement as a coordinated and distributed program | |
| **W16** | Encourage collaboration between the subdisciplines in stellar astrophysics, heliophysics, instrumentation, statistics, and earth sciences (mitigating tellurics) | *ESS 2018: "Such an initiative should also strategically encourage the free exchange of ideas between the above-mentioned subdisciplines in stellar astrophysics, heliophysics, instrumentation, and statistics for overcoming the effects of stellar variability."* |

### 3.1.1 Target Star Selection

To identify a set of stars that would constitute a reasonable set of targets for a future EPRV survey, it was assumed that any EPRV program would provide precursor observations for future direct imaging missions. To this end, we have adopted and cross-matched the target lists from the HabEx, LUVOIR-A, LUVOIR-B, and Starshade Rendezvous target lists (Gaudi et al. 2020; The LUVOIR Team 2019; Seager 2018). The Working Group gathered archival information on stellar properties relevant for the estimation of RV information content including spectral type, effective temperature, apparent magnitude, rotational velocity, metallicity and surface gravity. Consensus was then reached on the following selection criteria for prioritization of these direct imaging targets to identify stars that are both most important to future direct imaging efforts, and most amenable to RV survey monitoring, and grouped them into three categories – dubbed 'green,' 'yellow,' and 'red' – by the following definitions:

- Green targets
  - Spectral types F7–K9,
  - vsin(*i*) < 5 km/s, and
  - On the HabEx 'deep survey' or '50 highest priority stars' lists (Gaudi et al. 2020), or on at least 2 other mission concept target lists (including LUVOIR-A, LUVOIR-B, HabEx 'master list,' Starshade Rendezvous).
- Yellow targets
  - Spectral types F7–M,
  - vsin(*i*) between 5 and 10 km/s, and/or
  - Appears on only one mission concept list.
- Red targets
  - Spectral type hotter than F7, and/or
  - vsin(*i*) > 10 km/s.

The green targets were designated by the EPRV Working Group as "Musts" for an EPRV survey, with the yellow targets designated as "Wants," and the red targets excluded from consideration due to inadequate RV information content. The Working Group discussed including a stellar activity criteria as part of the target star selection metrics (nominally based on Log R'HK values). It was agreed that while such a requirement may ultimately be needed, the current uncertainty regarding decade-long stability of activity indicators would make adopting any criteria at this time effectively arbitrary. Additionally, it is unclear which form(s) of stellar variability are likely to be dominant after future research in stellar variability mitigation, which would impact the ultimate criterion adopted. Therefore, the Working Group chose not to adopt a stellar variability criteria for this study, however, this should be reassessed during the subsequent EPRV program as further knowledge becomes available.

During simulations of proposed architectures, five stars were noted to be driving observing efficiencies and these were removed from target lists to improve program efficiency (Section 4.1). The resulting 'green prime' star list (101 stars) and associated stellar properties are listed in Appendix B, and shown distributed on the sky in Figure 3-1. These stars constitute the sample used as input for the EPRV survey simulations (Section 4). This list should not be considered exhaustive and other stars (i.e., those not included in the simulations) should be considered in any future stellar characterization efforts.

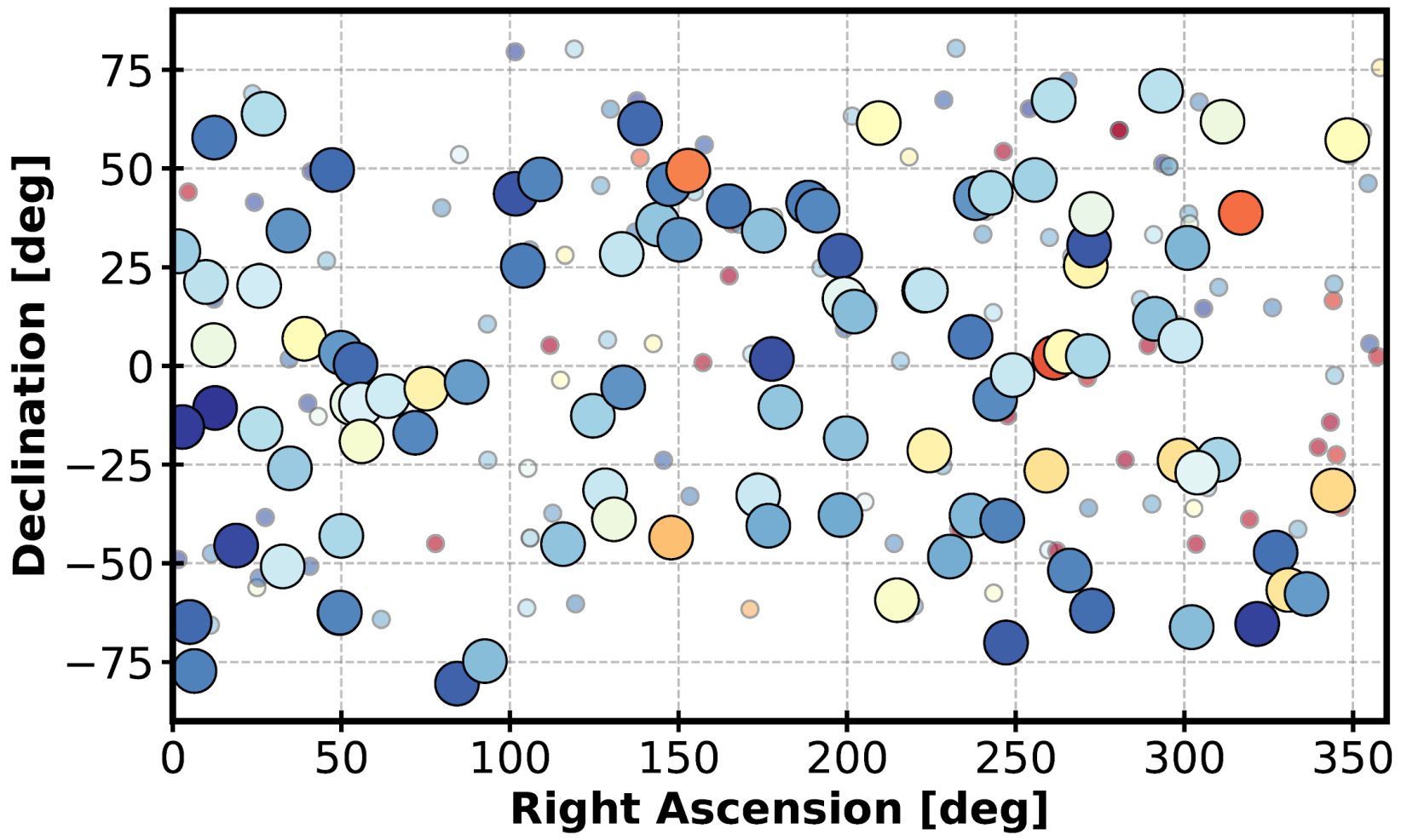

**FIGURE 3-1.** The 101 'green prime' stars are depicted as solid, large colored circles distributed on the sky. These stars constitute the sample used as input for the EPRV survey simulations (Section 4). The 92 additional "yellow" stars are shown as smaller circles with grey outlines. The color of the circles corresponds to the effective temperatures of the stars. Figure by Jennifer Burt.

## 3.2 Architecture Concepts

Notional ground-based telescope and instrument architectures were developed to meet the defined requirements. Each architecture concept was 'championed' by a specific Working Group member who, in an iterative process, worked with the Analysis Groups to define its attributes such as aperture size, spectrograph resolution, spectral grasp, telescope number and locations, observing cadence, and instrument noise floor. Where appropriate, specific Analysis Groups defined the requirements necessary to meet the program requirements. For example, the Stellar Variability Analysis Group defined minimum (and optimal) observing cadence, resolution and SNR requirements, which would be needed to provide a potential pathway to mitigating stellar variability.

Twelve notional architecture concepts were initially considered which were down-selected to seven for further detailed analysis (Table 3-3 and Sections 3.2.2–3.2.10). Architecture specifics evolved during the period of the Working Group study to improve their overall performance and to bring them into alignment with the program defined requirements and desired attributes. Changes were driven both by scientific performance considerations (e.g., target observing cadence) and programmatics (e.g., site availability, cost).

### *3.2.1 Architecture Commonalities*

Architectures consisted of a varying number of telescopes and apertures. Most telescopes were assumed to be dedicated 100% time EPRV programs except where noted in the architecture description (III, IV, VIIIa, and VIIIb). Telescopes were assumed to be positioned in a global network with six notional sites providing longitudinal and declination coverage, which was used by most architectures (Figure 3-2). Using numerous sites aids with meeting the cadence requirement for stellar variability mitigation defined by the Stellar Variability Analysis Group and guards against false positive detections caused by instrument systematics or observational aliases. The notional site locations were Mauna Kea in Hawaii, Kitt Peak in Arizona, and Calar Alto in Spain in the northern hemisphere, along with La Silla in Chile, Sutherland in South Africa, and Siding Springs in Australia in the southern hemisphere.

**TABLE 3-3.** Table of architecture properties. Architectures not listed - 0a, 0b, III, IV, and VII are described in the text.

| Architecture | I | IIa | IIb | V | VI | VIIIa | VIIIb |
|---|---|---|---|---|---|---|---|
| Champion | Jennifer Burt | Andrew Howard | Andrew Howard | Chas Beichman | Peter Plavchan | Benjamin Fulton | Benjamin Fulton |
| Telescopes | 6×2.4 m | 2×6 m + 4×4 m | 6×4m | 6×3 m | 6×1 m arrays | 2×10 m + 4×3.5 m | 2×10 m + 6×2.4 m |
| Input | Seeing-limited | Seeing-limited | Seeing-limited | Diffraction-limited | Seeing-limited | Seeing-limited | Seeing-limited |
| New or Existing Facilities | New | Existing | Existing | Existing | New | Existing | Existing (10 m), New (2.4 m) |
| Time Allocation | 100% | 100% | 100% | 100% | 100% | 25%, 100% | 25%, 100% |
| Spectral Grasp | 380–930 nm | 380–930 nm | 380–930 nm | 500–1700 nm | 500–800 nm | 380–930 nm | 380–930 nm |
| Resolution | 180k | 180k | 180k | 180k | 150k | 180k | 180k |
| Total System Efficiency | 6% | 6% | 6% | 7% | 6% | 6% | 6% |
| Instrument Noise Floor | 10 cm/s | 5 cm/s | 5 cm/s | 10 cm/s | 10 cm/s | 5 cm/s | 5 cm/s |
| Required Peak SNR/pix | 300 | 300 | 300 | 300 | 300 | 1000 for 10 m; 300 for 3.5 m | 1000 for 10 m; 300 for 2.4 m |
| Photon RV Precision per Epoch | 10 cm/s | 10 cm/s | 10 cm/s | 10 cm/s | 10 cm/s | 5 cm/s for 10 m; 15 cm/s for 3.5 m | 5 cm/s for the 10 m; 15 cm/s for 2.4 m |
| Goal Observation Cadence per Target | 1/night | 3/night | 3/night | 2/night | 1/night | 1/wk on 10 m; 1/night on 3.5 m | 1/wk on 10 m; 1/night on 2.4 m |

Each architecture was assumed to use spectrometers calibrated using laser frequency combs (LFCs) and/or etalons (which are themselves stabilized by an LFC) as wavelength calibration sources. Spectrographs for all of the seeing-limited architectures were assumed to be the same instrument design – a high resolution (R > 150,000), optical bandpass, EPRV spectrograph based on an improved version of the NN-EXPLORE Exoplanet Investigations with Doppler Spectroscopy (NEID) cross-dispersed echelle spectrograph (see Figure 3-3) located on the 3.4 m WIYN telescope (Robertson et al. 2016; Schwab et al. 2016). Spectrographs on the largest telescopes will require pupil slicing to achieve this high resolution. Architecture V differed from the other architectures as this utilized an adaptive optics (AO) fed diffraction-limited design with two distinct spectrometers.

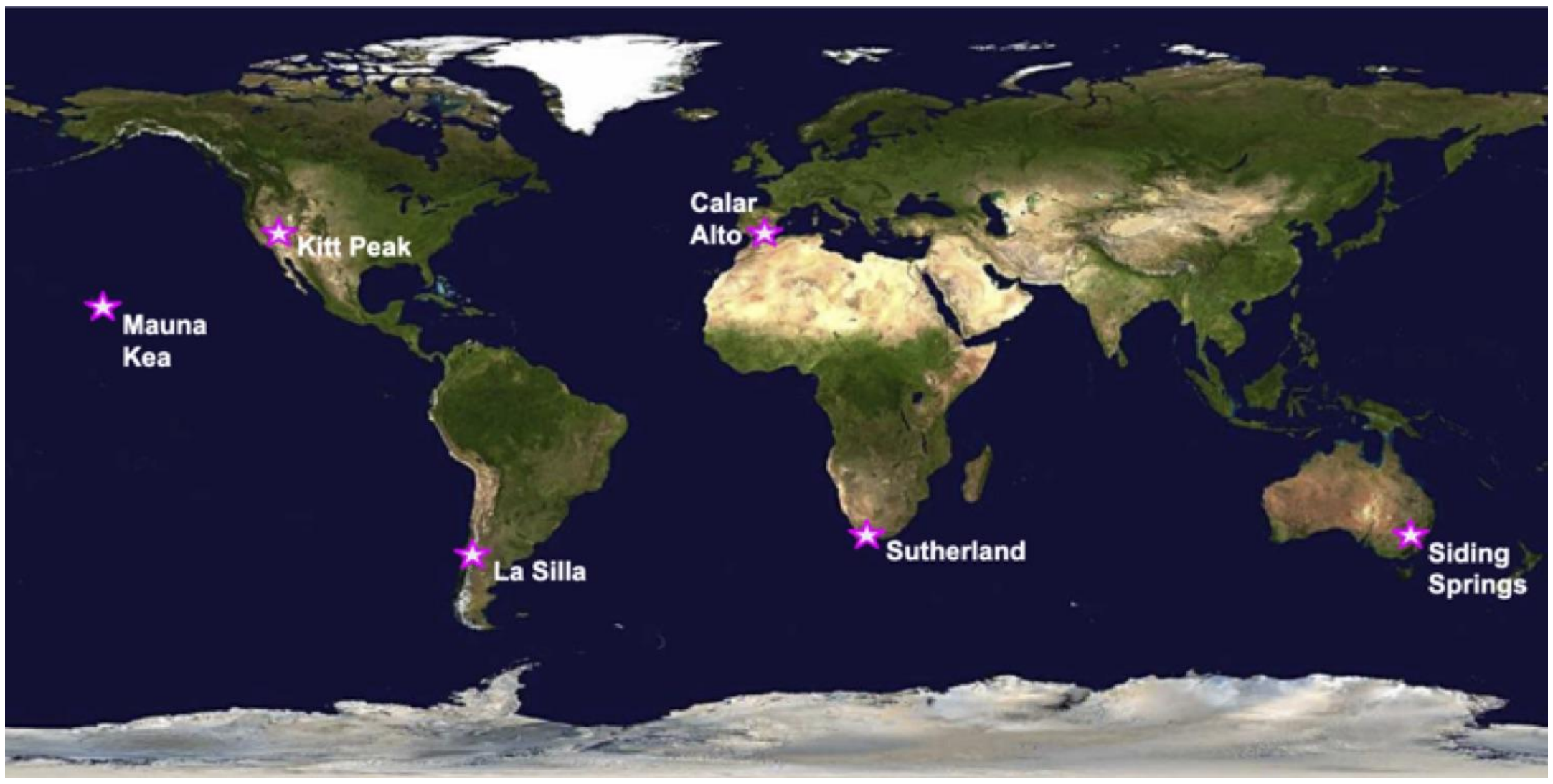

**FIGURE 3-2.** Locations of EPRV telescope facilities assumed in the study/survey simulation.

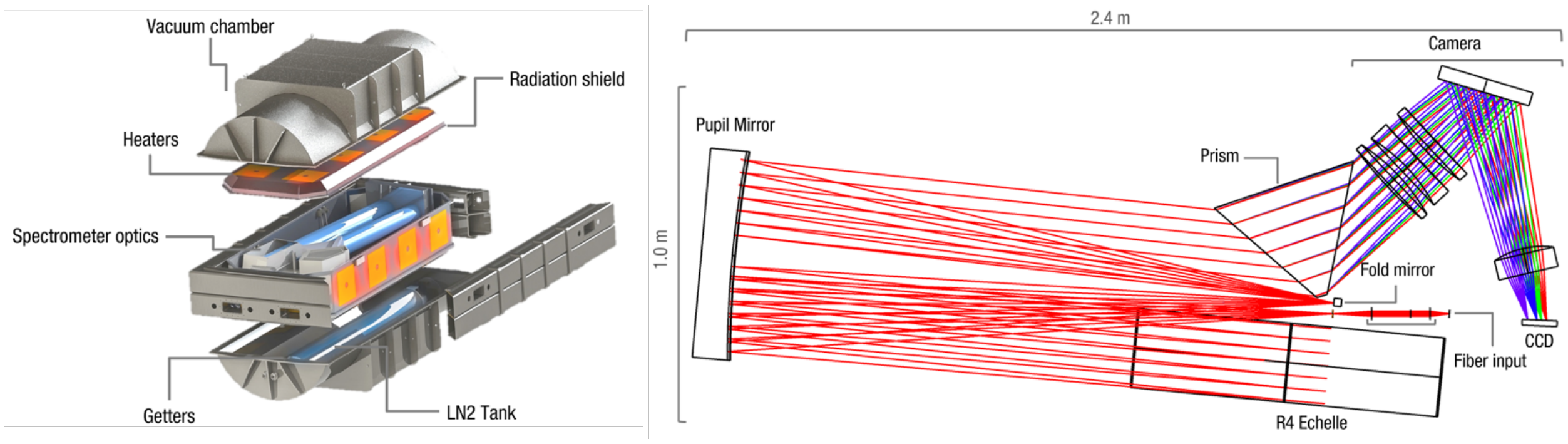

**FIGURE 3-3.** The Extreme Precision Doppler Spectrometer 'NEID'. *Left:* Rendering of the NEID instrument, showing the vacuum vessel and optomechanical components (Robertson et al. 2019). *Right:* Schematic of the NEID optical train (Schwab et al. 2016).

The standardized EPRV spectrograph adopted in the finalized architectures has the following properties listed in Table 3-4 (in addition to those listed in Table 3-3).

**TABLE 3-4.** EPRV spectrograph properties.

| Parameter | Value |
|---|---|
| Pixels per Resolution Element | 5 |
| Read Noise | 4.5 e- |
| Dark Current | 3 e-/hour |
| Well Depth | 90000 e- |
| Gain | 0.704225 ADU/e- |
| Detector Readout Time | 30 sec |

### 3.2.2 Architectures 0a/0b – 'Do Nothing'

One option was to "do nothing," and denoted as Architecture 0a. It represents the current path with no new initiatives, focused activities, or support beyond the current standard competed proposal calls, which are open to all disciplines in the astronomy and astrophysics communities. Architecture 0a extrapolates the RV precision that could be anticipated from the execution of existing programs. Architecture 0b was a modified version of this concept, which used existing state-of-the PRV facilities, but with the distinction of there being an additional investment in solar data collection, data analysis, and coordination to further inform the community. **All other architectures assume the implementation of Architecture 0b as a baseline, and then consider activities that would build upon it further.**

### 3.2.3 Architecture I – 6×2.4 m

This architecture uses a longitudinally dispersed network of six identical, robotic, 2.4 m telescopes distributed across the northern and southern hemispheres. The notional telescope concept is based on the Automated Planet Finder (APF) telescope (Vogt et al. 2014), which is a fully automated facility that has been executing precision RV surveys since 2014 and is the largest aperture robotic RV telescope currently in operation. Each facility is presumed to be equipped with a 10 cm solar feed to carry out high cadence, high SNR, daytime observations of the Sun, and is paired with an EPRV spectrograph with R = 180,000 that achieves 10 cm/s instrument stability. The perceived advantages of this architecture were that it might be (in principle) a relatively low-cost approach due to the use of identical hardware at each site, with spectrograph size comparable to existing systems. The telescopes would be dedicated facilities, spending 100% of nightly observations on an EPRV survey.

### 3.2.4 Architectures IIa and IIb – 2×6 m + 4×4 m, 6×4 m

Architecture II is similar in concept to Architecture I, except for the use of larger telescopes, with the goal of higher precision for a given exposure time through higher photon flux, and thereby SNR. The telescope and instrument configurations are defined to be identical with existing telescopes in this size class (4–6 m) at Kitt Peak, Calar Alto, Sutherland, and Siding Spring. Each facility would be equipped with an EPRV spectrograph with a noise floor of 5 cm/s and resolution of R = 180,000. Architecture IIa includes two 6 m telescopes with four 4 m telescopes, while 4 m apertures are exclusively used in Architecture IIb. Given the relatively large number of existing 4 m class telescopes, this architecture uses an approach that requires repurposing existing observatories.

### 3.2.5 Architecture III – 1×10 m + 1×8 m

Architecture III uses a pair of the current generation large telescopes (8–10 m) with one each being located in each hemisphere at Mauna Kea, Hawaii and La Silla or Cerro Tololo Inter-American Observatory. Each facility is equipped with an EPRV spectrograph with a 7 cm/s noise floor. Fifty percent of each telescope's time would be dedicated to an EPRV survey.

During the study, it became clear that facilities with broad latitude and longitude coverage would be necessary to provide sufficient cadence of observations to meet the program requirements. The limited number of telescopes in this architecture and the limited telescope time precluded it from meeting this need and therefore only a limited study was completed.

### 3.2.6 Architecture IV – Architecture VIII + 2×25 m

Architecture IV uses the configuration of Architecture VIII (2×10 m plus smaller apertures) supplemented by observations using Extremely Large Telescopes (ELTs). The ELTs are used for targeted follow up exclusively, and not for a primary RV survey. The telescopes that constitute the ELT network – the Giant Magellan Telescope (GMT; Johns et al. 2012) and Thirty Meter Telescope (TMT; Skidmore et al. 2015) have coverage in both hemispheres and are equipped with first-light instruments capable of completing EPRV studies.

The GMT provides visible coverage via the GMT-Consortium Large Earth Finder (G-CLEF; Szentgyorgyi et al. 2012) instrument, which is an optical band echelle spectrograph (350–900 nm) with a maximum resolution of R~110,000. Its specification for single-measurement radial velocity precision is 30 cm/sec. The TMT provides near-infrared coverage using Multi-Object Diffraction-limited High-Resolution Infrared Spectrograph (MODHIS; Mawet et al. 2019). MODHIS operates in the near-infrared (NIR) band (0.95–2.5 µm) and is fed using the TMT adaptive optics system. The working resolution is 100,000 and the design single-measurement radial velocity precision is 30 cm/sec. The telescope time dedicated to EPRV programs is initially assumed to be significant (90%) due to the timing of the instrument coming online at the facilities. This would decrease as the number of available instruments at the ELTs grows over time.

### 3.2.7 Architecture V – 6×3m (Adaptive Optics)

The distinguishing feature of Architecture V is its use of diffraction-limited spectrographs. Adaptive optics is used to feed single-mode fibers, which are then used to illuminate a pair of spectrographs, one at visible wavelengths and one in the near-infrared (NIR). The NIR system provides broad wavelength coverage to potentially help with mitigation of stellar variability. To minimize the impact of tellurics, only deep red or NIR wavelengths for which the atmospheric transmission is greater than 95% was assumed.

Architecture V is similar to Architecture I in that it uses six medium-sized (3 m) telescopes distributed around the globe. Each facility is equipped with a pair of EPRV spectrographs with a combined bandpass of 500–1700 nm with R = 180,000, split between four bands (two per spectrometer). A 10 cm/s noise floor is assumed for all spectrographs. A separate 430 nm spectrometer is proposed to monitor Ca H&K lines for stellar variability.

For both spectrographs, overall instrument efficiency was assumed to be the product of the internal spectrometer efficiency (20% in both cases), and the optical coupling efficiency. For single-mode fiber-fed instruments, the Strehl ratio can be used as an estimator for fiber coupling performance. A Strehl ratio that has been demonstrated on sky in 1" seeing was used to estimate the stated total system efficiency of 7%, after assuming 20% of the starlight is used to derive the corrections for a 40×40 element AO system. At R magnitude = 7.5, the optical coupling was 30% in the visible, and 80% in the NIR assuming improved coupling using phase-induced amplitude apodization (PIAA) foreoptics (Jovanovic et al. 2017). The instrument RV precision is a function of the AO performance as shown in Figure 3-4.

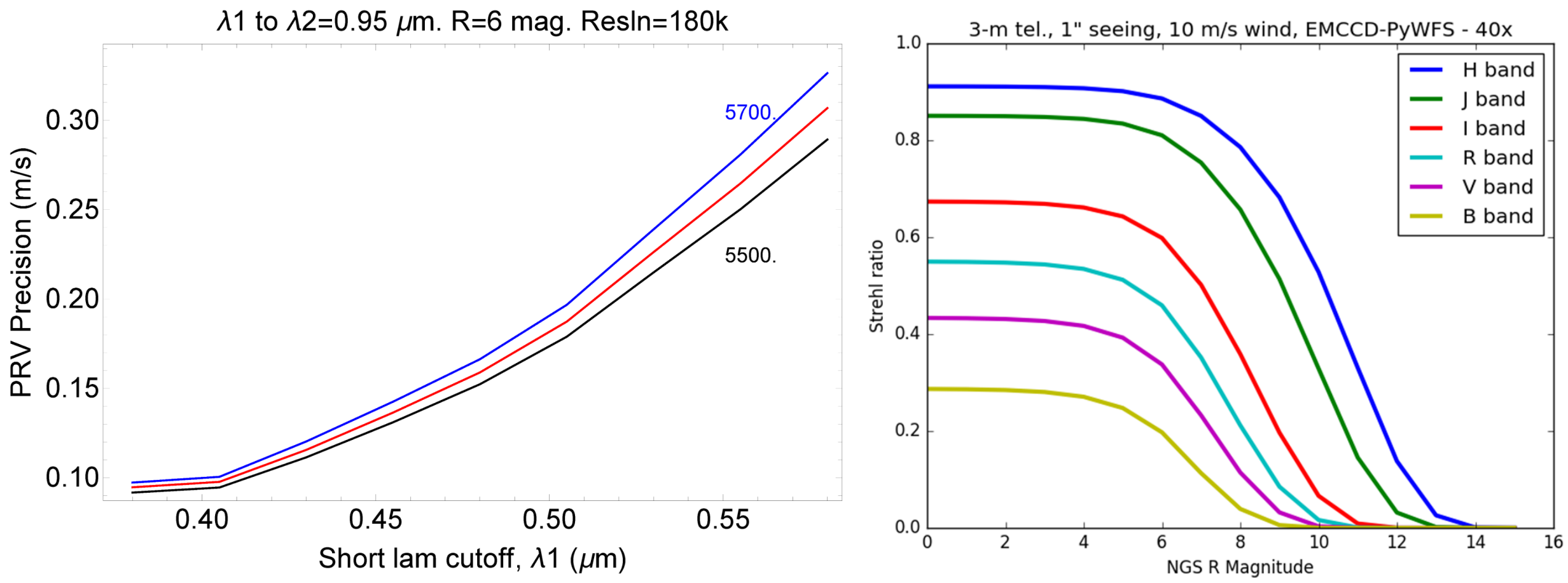

**FIGURE 3-4.** *Left:* Impact of shortest wavelength due to AO performance on achievable EPRV precision for a range of stellar effective temperatures. *Right:* A model of achievable Strehl ratio as a function of wavelength on a 3 m telescope with an advanced AO system utilizing a Pyramid wavefront sensor (Mawet, private communication).

### 3.2.8 Architecture VI – 6×1 m Arrays

This architecture is a Minerva-like (Swift et al. 2015) configuration, with an array of small (1 m) telescopes at each telescope site to achieve the light gathering power equivalent to a larger aperture at potentially lower cost. For the purposes of this study, each array is considered to be the same as a single 4 m class telescope (equivalent to Architecture IIb). The spectrograph has a reduced bandpass compared to other architectures due to the multiple fibers for each telescope in the array, and drives the difference in the assumed instrument parameters. Each facility would be equipped with an EPRV spectrograph with a noise floor of 10 cm/s and resolution of R = 150,000.

### 3.2.9 Architecture VII – Novel Technologies

The team included an architecture with advanced instrument concepts including novel spectrograph designs (e.g., externally dispersed, and advanced photonic designs). However, the schedule uncertainty and high risk associated with the development of these technology concepts rendered them unusable for the purpose of accurately assessing Architecture VII against other architectures in the study.

### 3.2.10 Architectures VIIIa and VIIIb – 2×10 m + 4×3.5 m / 6×2.4 m

Architecture VIIIa is a hybrid architecture that attempts to combine two methods of stellar variability mitigation – continuous, high cadence observations using small aperture telescopes, and extremely high SNR with high-resolution spectra using large apertures – into a single architecture. It consists of a worldwide network of four dedicated, robotic, 3.5 m telescopes at distinct locations, plus 25% of the time (at least one night per week or 25% of each night, with the latter preferred in order to protect against weather losses) on two 8–10 m class telescopes with one in each hemisphere. The smaller aperture telescopes operate with lower SNR and photon precision compared to the larger apertures.

Each facility would be equipped with an EPRV spectrograph with a noise floor of 5 cm/s and resolution of R = 180,000 with the spectrographs on the large apertures requiring slicing to achieve this resolution. The architecture delivers higher SNR on the large apertures but with reduced cadence. Architecture VIIIb is a slight modification to Architecture VIIIa where the 4×3.5 m telescopes are replaced with 6×2.4 m telescopes. The Keck and Gemini telescopes are possible choices for use with the 8–10 m apertures in this architecture.

# 4 Architecture Survey Simulation

To aid in assessing the likelihood of the proposed instrument architectures meeting the defined objectives (Tables 3-1 and 3-2), simulations of architecture performance were completed by the Observing Strategies Analysis Group. Simulations assumed a 10-year RV survey under the idealized conditions of complete mitigation of stellar variability and telluric line contamination, and single-measurement RV instrument precision capability of 5–15 cm/s depending on the specific architecture (Table 3-3).

Simulations used the 'green' target stars list discussed in Section 3.1.1. The goal was to determine if an architecture could detect an Earth-analog in orbit around each target star and the confidence in such a detection. Five stars were removed from the target list during the simulations to generate a 'green prime' list of 101 stars. The removed stars (HIP 100017, HIP 57939, HIP 32439, HIP 38908, and HIP 3583) required more than an hour per observation to reach a 10 cm/s precision goal on a 4 m telescope primarily due to a combination of spectral type, rotational velocity, and V mag (i.e., these were corner cases). The Working Group decided it was better and more realistic to omit this small number of stars in the simulations in order to obtain many more observations of other, less time intensive, targets. The resulting 'green prime' star list (101 stars) and associated stellar properties are listed in Appendix B.

For several architectures, the survey simulation was an iterative process. Deficiencies in architecture performance were highlighted in early simulations and were addressed by modest changes in architecture design to allow for an optimization of architecture parameters (e.g., a small increase in telescope aperture). The results presented are for the final architecture configurations. Additional details on the construction and execution of these observing simulations can be found in Newman et al. (in prep).

## 4.1 Survey Simulation Method

The green prime list of targets was broken into a northern hemisphere subset (≥ -5° declination, 51 stars), and a southern hemisphere subset (≤ +5° declination, 58 stars) for observing with telescopes in their respective hemispheres, with 9 stars in both lists near the celestial equator for cross-comparison. We did not quantitatively optimize the target lists for each individual telescope site to account for their specific location, seasons, and weather patterns. Instead, the particular declination cuts were chosen by consensus by the EPRV Working Group to have some non-zero overlap between the target lists of both hemispheres.

Parameters from each architecture concept including telescope size, available survey time (in fractions of a year), and several instrument parameters: wavelength range, spectrograph resolution, overall efficiency, target RV precision photon noise, and target spectroscopic SNR per resolution element were used in the simulations. From these parameters, estimated exposure times for all stars were calculated for each telescope/site/instrument combination as an intermediate step. These exposure times were then used for dispatch scheduler survey simulations. Two sets of exposure times were calculated – one that accounted for only the desired photon noise RV precision, and one that met both the desired photon noise RV precision and the desired spectroscopic SNR per resolution element (whichever was longer), to determine which requirement was driving the estimated exposure times and resultant survey cadence.

Six sites were assumed for all simulated architectures (Section 3.2.1). Those architectures with heterogeneous telescope compositions had the largest aperture telescopes located at Mauna Kea and Las Campanas. The use of a multiple site network provides redundancy against weather losses, the possibility of multiple observations per star on a given day, daily observations of equatorial targets for cross-calibration, and a carefully choreographed observing plan updated daily.

We made use of historical weather records for all observatory sites with the simplification that each night is assumed to be either entirely clear or entirely unusable, so there are no partial nights. Whether or not a night is lost due to weather is determined randomly at the start of the night, with the probabilities drawn from the historical monthly average. Monthly historical weather averages were not available for the Siding Spring and Sutherland locations and here yearly and semester averages were used respectively. For simplification, we assume that nights lost due to weather do not exhibit night-to-night correlations, even though in reality it is typical to lose multiple nights in a row from weather patterns.

Telescopes were assumed to operate by a dispatch scheduler. The scheduler is derived from the one used by the MINERVA (MINiature Exoplanet Radial Velocity Array) facility (Swift et al. 2015) and selects the target star in real time. To minimize simulation complexity, telescope sites were assumed to operate independently with no knowledge about when targets were last observed at other sites. Targets were prioritized via a combination of hour angle (highest weight at zenith), and time since last observation (a minimum of 2 hours, with linearly increasing weight thereafter). For observations, we enforce a SNR requirement of at least 300 per pixel at 550 nm and a RV photon noise requirement on all EPRV target observations as defined by each architecture (Table 3-3). Because many of the stars are inherently bright due to their F–K spectral types and proximity to Earth, satisfying both the SNR and RV uncertainty requirements generally requires a number of sequential "exposures" that are then binned into a single "observation" so as not to saturate the instrument's detector. Each observation is also required to span at least 5 minutes of on-sky time so as to average over the stellar pulsation modes (p-modes) that are expected in Sun-like stars (Chaplin et al. 2019).

For Architectures VIIIa/b, which have mixed aperture sizes, the simulated observing strategy involved observing all of the visible green list stars using 8–10 m class telescopes to SNR = 1000 and photon-limited uncertainty of 5 cm/s at least once per week. The smaller, dedicated, (3.5 m/2.4 m) telescopes observe the green list stars at least once per night when visible at a much lower SNR (SNR~200) and precision (15 cm/s).

## 4.2 Survey Simulation Assumptions and Their Effects

Single-measurement RV uncertainty in simulations has been assessed as the quadrature sum of an assumed instrumental noise floor, $\sigma_{inst}$ (between 5 and 10 cm/s per measurement, depending on the specific architecture) and the statistical photon noise uncertainty, $\sigma_{photon}$ (see Table 3-3). The distributions of both noise contributions are assumed to be Gaussian, and we assume that the individual measurements are uncorrelated. This implies that the single measurement uncertainty can never be smaller than $\sigma_{inst}$. However, it also implies the optimistic (and likely unrealistic) assumption that the instrumental systematics can also be reduced by averaging over many measurements, regardless of the timescale over which those measurements are taken.

A number of simplifying assumptions were made when completing the survey simulations. These range from optimistic to pessimistic with some assumptions having neutral or ambiguous effects. No attempt was made to quantify the overall impact of the assumptions. We emphasize that our assumption that the stellar variability can be completely accurately characterized and completely mitigated presents the largest uncertainty and highest overall risk in our study, and as such is the primary focus for the first stage of our recommended EPRV strategy. Given this, there is very little justification for consideration of more realistic choices for our other optimistic assumptions (edge-on orbits, no confusion to due multiple planets, no aliasing, dark nights, etc.).

Should it be demonstrated that stellar variability can be adequately characterized and mitigated, higher-fidelity simulations would be indicated.

### Optimistic Assumptions:

- Theoretical instrument concepts are adopted (including spectrographs, fiber feed, calibration sources, and AO systems) and these are assumed to be possible to build, and will exist to provide the required single-measurement instrument precision.
- Theoretical instruments are assumed to achieve their goals in terms of efficiency and noise characteristics.
- Stellar variability is assumed to be adequately characterized and mitigated during the data extraction of RV signals such that the remaining signal due to stellar variability is uncorrelated in time. For an exploration of the significance and implications of this assumption, see Luhn et al. 2021, in prep.
- Telluric line contamination is assumed to be limited and correctable so that large regions of the instrument's spectral grasp are not lost to telluric contamination.
- We assume that each host star harbors a single Earth analog, i.e., 1 $M_{Earth}$, with an orbital period chosen to be in the habitable zone based on host star type and in a circular orbit.
- We assume that orbital planes are edge-on as viewed from Earth and uncorrelated with the $v\sin(i)$ of their host star.
- Telescopes have negligible downtime.
- Observations are possible at minimal air mass. The atmospheric model adopted in the exposure time calculator is likely more transparent than at actual sites.
- Weather holds to historical rates. This is optimistic as climate change effects appear to be increasing clouds/haze at most sites.
- Weather conditions are uncorrelated.
- Time allocation is uniform (in reality, moonlight affects when stars can be observed well and introduces additional structure in the observing window function).
- We do not account for errors due to RV contamination from additional planets in the same planetary system. For an exploration of how additional planets are likely to impact the number of observations needed to measure the mass of a rocky planet, see He, Ford & Ragozzine 2021.
- We assume that the orbital period of each planet can be accurately determined (i.e., we do not consider whether the inferred orbital period might be an alias of the true period).

### Pessimistic Assumptions:

- Observatories cannot coordinate, thereby decreasing the survey efficiency if a given star is observed more times than needed (across multiple facilities) throughout a given night.
- Telescope sizes, particularly 4-, 6-, and 8-meter class observatories are at the lower limit of what each size class actually means for collecting area.
- Pointing limits (many telescopes can go below the 2 air mass/30 degree above the horizon limit that was adopted).
- Site selection does not include any options in eastern Europe or Asia.
- Only existing observatory sites can be used.

### Neutral Assumptions:

- The throughput of existing instruments has been adopted and the median throughput value across the instrument's bandpass has been used instead of the peak value.
- A constant spectrograph resolution is assumed across the full bandpass.
- An improved version of the NN-Explore's NEID cross-dispersed echelle spectrograph is used as the standard/baseline for future instruments.
- Site selection – possible alternatives may influence weather and declination effects, but these can vary in multiple directions.

## 4.3 Survey Simulation Results

The outcomes of the survey simulation are shown in Figure 4-1. Histograms of the SNR of detecting an Earth analog around each host star using the simulated 10-year survey are shown for each architecture. A minimum SNR threshold of 10 was adopted as the baseline requirement for a successful detection.

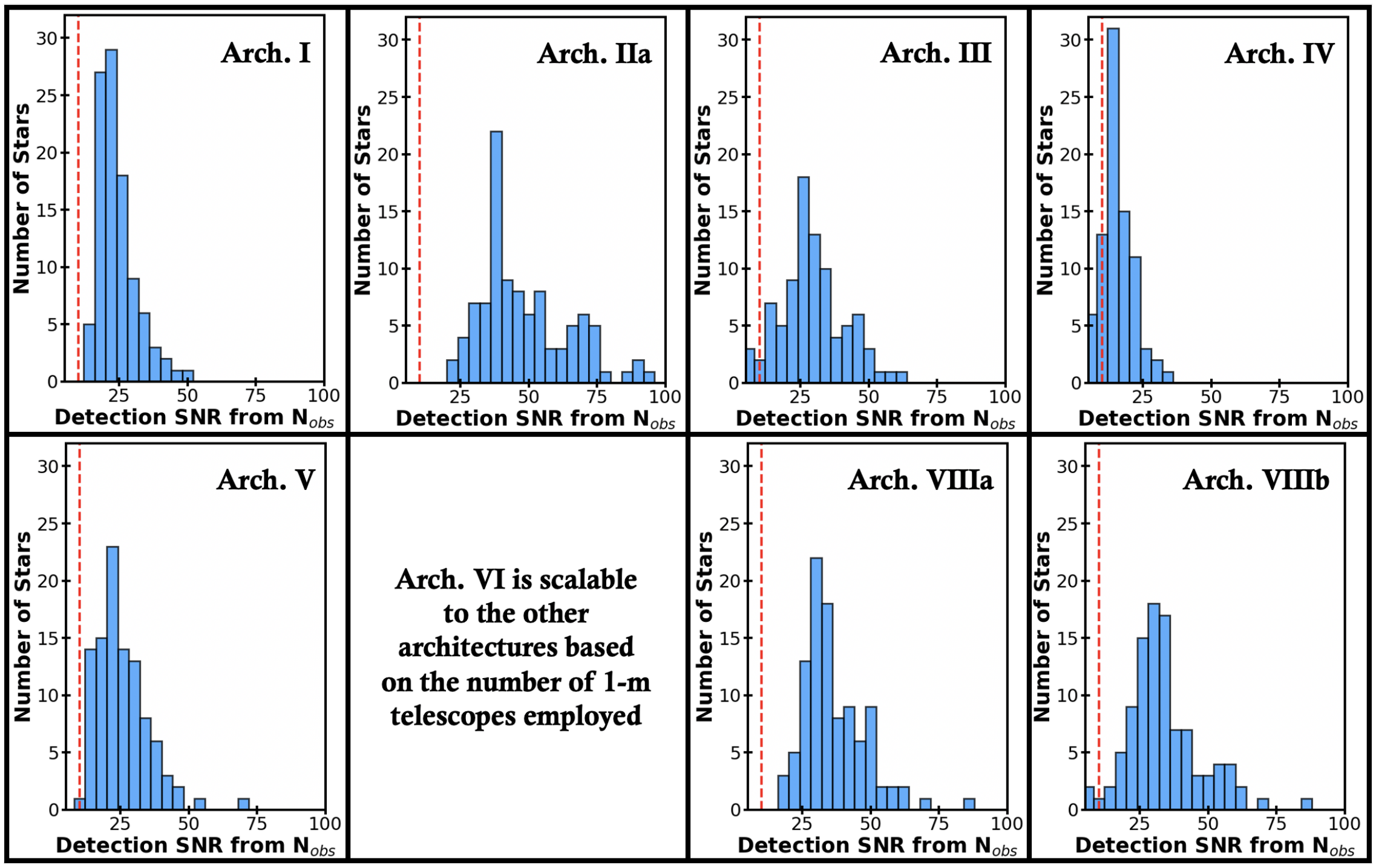

**FIGURE 4-1.** Histograms for the detection SNR achieved when surveying the 101 green star targets, where each target is assumed to harbor one Earth analog. Each panel corresponds to a different survey architecture. The minimum required SNR of 10 is shown by the red dashed line. We emphasize that these histograms present the best possible case estimates given the optimistic assumptions adopted in Section 4.2 (in particular the assumption that stellar variability can be adequately characterized and mitigated. Figure by Jennifer Burt.

All simulated architectures are able to detect Earth analogs around the majority of host stars with several achieving full recovery. It is clear that multiple telescopes in both hemispheres are required to survey all stars, minimize the impact of weather losses, and offer improved cadence, which may be needed to mitigate stellar variability given its various timeframes. Many architectures offer similar performance: a network of 6×2.4 m telescopes accomplishes nearly the same result as 4×3.5 m telescopes in Architectures VIIIa and VIIIb. Additionally, for the architectures that utilize larger (10 m) telescope apertures, the higher precision requirement for these facilities is approximately offset by the collecting area advantage and so they do not observe stars any faster. This, combined with the limited time allocation on these facilities means that they are only able to observe a small fraction of the number of stars.

While the presented results show the aperture/facility aspect of an EPRV survey is likely solvable, these should be considered 'preliminary results' that are likely the best-case scenario. Further research is needed in several areas to validate the assumptions adopted including stellar variability (required instrument design parameters, observing cadence and mitigation strategies) and hardware development. Improved constraints will allow simulations with increased realism to be developed before the optimal architecture for an EPRV survey is selected. It will be important for the EPRV community to reevaluate proposed architectures and potential adaptations that leverage the knowledge gained through the research recommended in other sections of this report.

# 5 Architecture Assessment

In combination with the survey simulation results, each program architecture was assessed against the program-defined criteria (Section 3.1) using a systematic approach for decision-making (Section 5.1). Scientific and logistical considerations of each architecture were assessed and potential risks identified.

## 5.1 The Kepner-Tregoe Method

The EPRV Working Group used a systematic approach for decision-making adapted from Kepner-Tregoe methods (Kepner & Trego 1965). The process required the Working Group to:

- Formulate and agree on a decision statement that is used to determine the evaluation and success criteria (Section 2),
- Agree on evaluation criteria and the relative importance of these criteria (Section 3.1),
- Document the trade space options and their descriptions (Section 3.2),
- Evaluate these options against the criteria,
- Reach consensus on the evaluation,
- Document the risks and opportunities of the options, and
- Make recommendations accounting for the risks and opportunities.

The definition and evaluation of criteria was done by consensus of the entire Working Group.

Figure 5-1 illustrates the format of the decision-making tool the Working Group used. The purpose of employing this technique was not to recommend one specific architecture, but rather to glean the relative strengths and deficiencies of each approach so as to advise the path forward.

| **Decision Statement** | | | | | | | | | |
|---|---|---|---|---|---|---|---|---|---|
| **Description** | | | | *Option 1* | | *Option 2* | | *Option 3* | |
| | *Feature 1* | | | | | | | | |
| | *Feature 2* | | | | | | | | |
| | *Feature 3* | | | | | | | | |
| **Evaluation** | **Musts** | | | | | | | | |
| | *M1* | | | ✔ | | ✔ | | ✔ | |
| | *M2* | | | ✔ | | ? | | ? | |
| | *M3* | | | ✔ | | ✔ | | ✗ | |
| | **Wants** | *Weights* | | | | | | | |
| | *W1* | *w1%* | | *Rel score* | | *Rel score* | | *Rel score* | |
| | *W2* | *w2%* | | *Rel score* | | *Rel score* | | *Rel score* | |
| | *W3* | *w3%* | | *Rel score* | | *Rel score* | | *Rel score* | |
| | | *100%* | Wt sum => | *Score 1* | | *Score 2* | | *Score 3* | |
| | **Risks** | | | C | L | C | L | C | L |
| | *Risk 1* | | | M | L | M | L | | |
| | *Risk 2* | | | H | H | M | M | | |
| **Final Decision, Accounting for Risks** | | | | | | | | | |
| C = Consequence, L = Likelihood | | | | | | | | | |

**FIGURE 5-1.** Kepner-Tregoe decision making tool chart format used by the EPRV Working Group.

The Working Group evaluated the relative ranking of each architecture by considering the ability of each concept to meet the adopted requirements as listed in Table 3-1 (Section 5.2). Desired attributes listed in Table 3-2 were also assessed accounting for the anticipated cost to implement and the concept risk (Section 5.3).

## 5.2 Likelihood of Meeting Program Requirements

Each architecture concept was qualitatively evaluated based on its prospects for meeting the technical criteria, survey criteria, and programmatic requirements listed in the program requirements. Evaluation was given in terms of "*yes- can meet requirements,*" "*likely to meet,*" "*possible,*" "*unlikely,*" or "*cannot meet requirement*" with initial assessment being made by individual Analysis Groups where appropriate and then agreed to by the full Working Group. The results of the evaluation are listed in Table 5-1.

Many developed architectures showed promise to achieve the goal of measuring the masses of temperate terrestrial planets orbiting Sun-like stars. Several architectures however were deemed to fail or be unlikely to meet these needs: 0a (existing plans), 0b (existing plans with new funds), III (1×10 m + 1×8 m), and VII (Novel Technologies). For certain criteria, there was insufficient data at the present time to accurately assess the architecture and this was noted as "unknown." A general conclusion from the evaluation of the architecture concepts is that, *in order to enable studies of terrestrial planets orbiting Sun-like stars, new capabilities beyond current technologies and instruments will be required.*

For the architectures deemed feasible of meeting the program requirements, assessment of the desired attributes including cost and risk was completed (Section 5.3). Architecture 0b, the nominal baseline from which these architectures all build upon, was also assessed.

## 5.3 Likelihood of Meeting Desired Program Attributes

The desired attributes (Table 3-2) were weighted by importance with each architecture being evaluated relative to each other for their perceived likelihood of being most to least able to achieve the desired attributes. The weight assigned to six of the attributes meant that they were noted to be key in the overall assessment of an individual architecture. The ranking and evaluation table for the architecture concepts is shown in Table 5-2.

A score for each criteria was determined by multiplying the weight of the criteria and the relative evaluation between the architectures. Evaluation was assessed against the architecture deemed the best for each criteria using four terms "best (10 points),","small difference (8–9 points)," "significant difference (5–6 points)," and "very large difference (0–3)." Many criteria were deemed to be similar between all of the architectures, which are noted as a 'wash' while those that diverged between architectures were noted to be 'driving' overall ranking assessment.

To assess certain criteria, cost models were developed for each architecture (Section 5.3.1). Risk was also assessed as part of developing a final architecture ranking (Section 5.3.2).

### *5.3.1 Cost Model*

To assess the 'least estimated cost' desired attribute and to assess overall program realism, rough order of magnitude (ROM) cost estimates were established for each architecture undertaking an ERPV development program and subsequent survey. Preliminary estimates were also used to identify cost drivers and to modify the draft architectures accordingly. The cost estimate resolution and uncertainty due to the conceptual natures of the architectures may be inadequate for true cost planning purposes, and provides only relative cost information between architectures.

**TABLE 5-1.** Assessment of architectures against requirements criteria (Table 3-1).

| Architecture | | 0a | 0b | I | IIb | III | IV | V | VI | VII | VIIIa/b |
|---|---|---|---|---|---|---|---|---|---|---|---|
| Architecture Description | | Existing Plans | Existing Plans with Funds | 6×2.4 m | 6×4 m | 1×10 m +1×8 m | VIII + 25 m | 6×3 m (AO) | 6×1 m Arrays | Novel Tech-nologies | Hybrid |
| | **Technical Criteria** | | | | | | | | | | |
| **M0a** | **Determine the feasibility** by 2025 to detect (with a well-characterized and sufficiently small false discovery rate) and measure the mass (msin(*i*) with <=10% fractional precision) of <=1 Earth mass planets that orbit a 1 $M_{Sun}$ main sequence star and receive insolation within 10% $Insolation_{Earth}$. | unlikely | likely | likely | likely | likely | likely | likely | likely | unknown | likely |
| **M0b** | **Demonstrate the feasibility** to detect (with a well-characterized and sufficiently small false discovery rate) and measure the mass (msin(*i*) with <=10% fractional precision) of <=1 Earth mass planets that orbit a 1 $M_{Sun}$ main sequence star and receive insolation within 10% $Insolation_{Earth}$ prior to 2030 Decadal Survey. | no | unlikely | possible | possible | possible | possible | possible | possible | unlikely | possible |
| | **Survey Criteria** | | | | | | | | | | |
| **M1a** | Design and execute a set of **precursor surveys and analysis activities** on the 'green' and 'yellow' stars on the evolving target star list and on the Sun. | no | yes | yes | yes | yes | yes | yes | yes | yes | yes |
| **M1b** | **Demonstrate the feasibility** to survey each of the 'green' stars on the evolving target list at the level of M0b. | no | no | unknown | unknown | unlikely | likely | unknown | unknown | unknown | likely |
| | **Programmatic (Current Surveys Meet L1 Reqt)** | | | | | | | | | | |
| **M2** | Meet Intermediate Milestone: **By 2025, demonstrate on-sky feasibility** with capabilities in-hand to detect *K* down to 30 cm/s for periods out to few hundred days using a statistical method that has been validated using simulated and/or observed spectra time-series | unlikely | likely | likely | likely | likely | likely | likely | likely | likely | likely |
| **M4** | **Capture Knowledge** from current and near-future generation of instruments, surveys, analysis, and coordination activities to help inform development of future EPRV instruments. | no | yes | yes | yes | yes | yes | yes | yes | yes | yes |

**TABLE 5-2.** Assessment of desired attributes (Table 3-2) for architectures which were deemed feasible . Architecture 0a (Existing Plans), III (1×10 m +1×8 m) and VII (Novel Technologies) were not evaluated as they were deemed unable or unlikely to meet the requirement criteria. Specific criteria were 'key' to overall rankings due to their weight while several had significant differences between the architectures and these were noted as 'driving.' Overall scoring of the architectures is noted with an initial ranking being based on desired attributes and a second ranking accounting for risk.

| Architecture | | | | | 0b | | I | | IIb | | IV | | V | | VI | | VIIIa/b | |
|---|---|---|---|---|---|---|---|---|---|---|---|---|---|---|---|---|---|---|
| Architecture Description | | | | | Existing Plans with Funds | | 6×2.4 m | | 6×4 m | | VIII + 25 m | | 6×3 m (AO) | | 6×1 m Arrays | | Hybrid | |
| | | Key | Driving | Weights | Score | | Score | | Score | | Score | | Score | | Score | | Score | |
| | **Relative Science** | | | **37** | **168** | | **254** | | **294** | | **370** | | **310** | | **245** | | **370** | |
| **W1** | Survey as many 'yellow' stars as possible on the evolving target list | K | D | 9 | 0 | VL DIFF | 6 | SIG DIFF | 6 | SIG DIFF | 10 | BEST | 6 | SIG DIFF | 5 | SIG DIFF | 10 | BEST |
| **W2** | Measure masses of temperate terrestrial planets orbiting M stars, not on yellow list | | D | 4 | 6 | SIG DIFF | 6 | SIG DIFF | 8 | SMALL DIFF | 10 | BEST | 10 | BEST | 6 | SIG DIFF | 10 | BEST |
| **W3** | Use follow-up of transiting temperate terrestrial planets to inform the mass-radius relation from key transit discoveries | K | D | 8 | 6 | SIG DIFF | 6 | SIG DIFF | 8 | SMALL DIFF | 10 | BEST | 9 | SMALL DIFF | 6 | SIG DIFF | 10 | BEST |
| **W4** | Validate methods of stellar variability mitigation, telluric mitigation, and statistical validation, key for the EPRV method, including using follow-up of transiting planets | K | D | 16 | 6 | SIG DIFF | 8 | SMALL DIFF | 9 | SMALL DIFF | 10 | BEST | 9 | SMALL DIFF | 8 | SMALL DIFF | 10 | BEST |
| | **Relative Schedule** | | | **17** | **120** | | **170** | | **160** | | **150** | | **165** | | **170** | | **165** | |
| **W5** | **Schedule**: Start the precursor M1a surveys **as soon as possible**, so as to maximize impact at PDR on design of direct imaging missions (e.g., HabEx, LUVOIR) | K | | 12 | 10 | *WASH* | 10 | *WASH* | 10 | *WASH* | 10 | *WASH* | 10 | *WASH* | 10 | *WASH* | 10 | *WASH* |
| **W6** | **Schedule:** Start the survey as soon as possible, so as to maximize impact at PDR on design of direct imaging missions (e.g., HabEx, LUVOIR) | | D | 5 | 0 | FAILS | 10 | BEST | 8 | SMALL DIFF | 6 | SIG DIFF | 9 | SMALL DIFF | 10 | BEST | 9 | SMALL DIFF |
| | **Relative Difficulty** | | | **20** | **65** | | **150** | | **80** | | **80** | | **160** | | **160** | | **112** | |
| **W7** | Prefer the architecture with the greatest relative probability of success to meet stellar variability requirement | K | D | 10 | 6 | SIG DIFF | 6 | SIG DIFF | 10 | BEST | 10 | BEST | 8 | SMALL DIFF | 6 | SIG DIFF | 10 | BEST |
| **W8** | Relative difficulty to secure required telescopes/instruments, fraction of time, and observing cadence and coordination between telescopes | | D | 5 | 0 | N/A | 10 | BEST | 8 | SIG DIFF | 2 | VL DIFF | 6 | SIG DIFF | 8 | SMALL DIFF | 9 | SIG DIFF |
| **W9** | Prefer the architecture with the greatest probability of success of achieving the survey referenced in M1b | | D | 5 | 1 | VL DIFF | 8 | SMALL DIFF | 10 | BEST | 10 | BEST | 8 | SMALL DIFF | 5 | SIG DIFF | 10 | BEST |

| Architecture | | | | | 0b | | I | | IIb | | IV | | V | | VI | | VIIIa/b | |
|---|---|---|---|---|---|---|---|---|---|---|---|---|---|---|---|---|---|---|
| Architecture Description | | | | | Existing Plans with Funds | | 6×2.4 m | | 6×4 m | | VIII + 25 m | | 6×3 m (AO) | | 6×1 m Arrays | | Hybrid | |
| | **Relative Cost** | | | **16** | **N/A** | | **160** | | **80** | | **80** | | **160** | | **160** | | **112** | |
| **W10** | Prefer least estimated cost | K | D | 16 | 0 | N/A | 10 | **BEST $364M** | 5 | SIG DIFF $554M | 5 | SIG DIFF $576M | 10 | **BEST $337M** | 10 | **BEST $365M** | 7 | SIG DIFF/ SM DIFF $496M |
| | **Other Factors** | | | **10** | **80** | | **97** | | **92** | | **80** | | **94** | | **85** | | **90** | |
| **W11** | Take advantage of opportunities for international collaboration and draw from as broad of a pool of relevant expertise and observing facilities as possible | | | 2 | 10 | *WASH* | 10 | *WASH* | 10 | *WASH* | 10 | *WASH* | 10 | *WASH* | 10 | *WASH* | 10 | *WASH* |
| **W12** | Maximize use of, and knowledge and understanding of, existing facilities (observatories), infrastructure, and hardware (including detectors) | | | 3 | 10 | **BEST** | 9 | SMALL DIFF | 8 | SMALL DIFF | 5 | SIG DIFF | 8 | SMALL DIFF | 5 | SIG DIFF | 8 | SMALL DIFF |
| **W13** | Maximize broader impacts in society | | | 1 | 0 | N/A | 10 | *WASH* | 10 | *WASH* | 10 | *WASH* | 10 | *WASH* | 10 | *WASH* | 10 | *WASH* |
| **W14** | Encourage free exchange of ideas, including data and source codes | | | 2 | 10 | *WASH* | 10 | *WASH* | 10 | *WASH* | 10 | *WASH* | 10 | *WASH* | 10 | *WASH* | 10 | *WASH* |
| **W15** | Implement as a coordinated and distributed program | | D | 1 | 0 | N/A | 10 | **BEST** | 8 | SMALL DIFF | 5 | SIG DIFF | 10 | **BEST** | 10 | **BEST** | 6 | SIG DIFF |
| **W16** | Encourage collaboration between the subdisciplines in stellar astrophysics, heliophysics, instrumentation, statistics, and earth sciences (mitigating tellurics) | | | 1 | 10 | *WASH* | 10 | *WASH* | 10 | *WASH* | 10 | *WASH* | 10 | *WASH* | 10 | *WASH* | 10 | *WASH* |
| | ***Subtotal*** | | | **100** | | | | | | | | | | | | | | |
| ***Total Score*** | | | | | ***433*** | | ***831*** | | ***816*** | | ***840*** | | ***879*** | | ***785*** | | ***932*** | |
| ***Ranking by Score*** | | | | | | | ***3*** | | ***3*** | | ***3*** | | ***2*** | | ***4*** | | ***1*** | |
| **Ranking by Score Accounting for Risk** | | | | | | | **2** | | **3** | | **5** | | **2** | | **2** | | **1** | |

The ROM cost estimates were based on a simplified bottom-up model, in which each architecture was broken down into a product tree of individual components such as telescopes, spectrographs, auxiliary equipment, software systems, and personnel. For each component, the resources needed were further broken down into research and development (R&D) effort, development, construction, operations, and science. These were assigned an estimated cost based on experience with existing facilities. Where an architecture calls for multiple identical units, estimated savings in engineering and volume purchases were taken into account. The cost of preparatory and auxiliary science programs was similarly estimated from a breakdown into individual work packages, and simplified assumptions for the typical cost of staff (engineers, researchers) including overheads, and of computing time.

Different architectures selected whether to refurbish existing telescopes or build new ones, with choices driven by cost and data homogeneity. Building uniform hardware will provide the highest reliability and data homogeneity, but may be cost-prohibitive. Modifying existing facilities will incur more cost in spectrograph design (due to differing port adapters, etc.) and reduction pipelines, and incurs some additional risk in the case that the data is not sufficiently uniform to be combined effectively.

This cost model was consolidated into a single spreadsheet (Appendix C) in order to ensure internal consistency and comparability of the architectures under consideration. All amounts given are in present-year (i.e., 2020/21) dollars with total cost estimates shown in Table 5-3.

**TABLE 5-3.** Cost estimates for evaluated architectures. A range of operating costs of facilities were assumed to provide boundaries on cost.

| Architecture | | I | IIb | IV | V | VI | VIIIb |
|---|---|---|---|---|---|---|---|
| Architecture Description | | 6×2.4 m | 6×4 m | VIIIb + 2x25 m | 6×3 m (AO) | 6×1 m Arrays | Hybrid |
| Cost estimate ($M) | Lower | 364 | 554 | 576 | 337 | 365 | 496 |
| | Higher | 407 | 745 | 837 | 380 | 396 | 637 |

The limitations of the present model include the following important points:

1. The cost of constructing a telescope may vary widely depending on the accessibility of the site and the availability of preexisting infrastructure. This was not captured in the cost model; a "typical" cost for construction at a developed site was used.
2. Telescope cost estimates were based on aperture only, ignoring potential savings if single-purpose designs (only one focus with very small field-of-view) are used.
3. Where the use of existing telescopes is foreseen, only the actual cost of operation, but not the depreciation of the facilities was included in the architecture cost estimate.
4. It was assumed that the cost of spectrograph hardware does not depend sensitively on the RV precision requirements. The validity of this assumption in the sub-m/s regime is not currently known.
5. Cost estimates do not include any provisions for contingency or risk mitigation.

**The nominal cost of a comprehensive EPRV program to detect and characterize Earth analogs was found to be comparable to that of a MIDEX class flight mission.**

**RECOMMENDATION:** As more detailed planning progresses in the future, more sophisticated cost models should be developed and employed.

### 5.3.2 *Risk Assessment*

The Working Group compiled a list of the perceived risks to an EPRV program and evaluated these for each architecture. The risk assessment was performed using the Goddard Space Flight Center (GSFC) risk matrix standard scale (Goddard Procedural Requirements: Risk Management GPR 7120.4D, 2012) with the risk likelihood and consequence each being given a value between 1 and 5 and the risk score being the product of the two. The complete risk table is provided in Appendix D with a summary provided in Table 5-4.

Many risks identified applied uniformly across all architectures. While some had minimal impact, several were deemed to be of significant consequence or probability and drove the overall program. These include:

- Stellar variability of the Sun may not be representative of target stars in list/stellar variability cannot be adequately mitigated
- Insufficient personnel to execute program
- Difficulty in supporting non-U.S. participants
- Knowledge retention in the field

Certain key risks differentiated between architectures and this was considered in developing a final architecture ranking. Examples of differentiating risks include:

- Inability to obtain sufficient observing time/cadence
- Insufficient aperture/observing time to acquire the SNR necessary to achieve science the requirements
- Unlikely to obtain high enough SNR or high enough resolution spectra for science goals
- Maturity of required technology
- Facility construction timeline

**TABLE 5-4.** Summary of the perceived risk and evaluation of individual architectures. The total risk score was calculated by summing the value for each identified risk for each architecture. The individual risk value was based on the product of the perceived likelihood (1–5) and the consequence (1–5) of that risk. Full details can be found in Appendix D.

| Architecture | I | IIb | IV | V | VI | VIIIb |
|---|---|---|---|---|---|---|
| Architecture Description | 6×2.4 m | 6×4 m | VIIIb + 2×25 m | 6×3 m (AO) | 6×1 m Arrays | Hybrid |
| Risk score (lower is better) | 216 | 227 | 301 | 282 | 230 | 243 |

Overall, the lowest risk scores were for Architectures I, IIb, VI, and VIIIb, in that order. Architecture V earned the second highest risk assessment due to the relative technical immaturity and concern over adequate performance blueward of 500 nm. Architecture IV was deemed to have the highest overall risk due to schedule uncertainty and limited observing cadence.

## 5.4 Architecture Assessment Conclusions

The systematic assessment of the science capabilities, technical requirements, and programmatic constraints of nominal EPRV instrument architectures demonstrates the existence of multiple plausible pathways to enable a survey with the statistical precision required to detect Earth-analog exoplanets. Key assumptions have been made in arriving at this conclusion: the adequate mitigation of stellar variability, improvements in instrument systematics, correction of telluric contamination, and advances in RV data extraction techniques. While proposed architectures have been ranked to provide a notional sense of the current best path forward, significant gaps remain in the overall realism of the studies completed. While appropriate for this level of early study, further research and data will be required to ensure enhanced realism in future studies. The iterative approach adopted during this study to define final architectures highlights the extensive parameter space that needs to be studied in detail to choose the final optimal architecture for an Earth-analog EPRV survey.

For several architectures, there is currently insufficient data to accurately assess all of the defined program requirements (Table 5-1) and desirable attributes (Table 5-2). **Addressing these knowledge gaps through near-term investment will be key to the overall success of any EPRV program**. Coordinated data collection from newly commissioned instruments and instruments currently under development will contribute significantly towards this goal. Developing coordinated observing programs between facilities and enabling routine solar EPRV observations will allow for improved assessment of instrument systematics while providing data to allow Sun-as-a-star EPRV studies. This investment will be a significant step towards validating the assumptions used in this study and is consistent with all proposed architectures building upon Architecture 0b (augmentation and coordination of existing facilities).

Several technical and programmatic risks have been highlighted in specific program architectures as well as more broadly. These range from technology immaturity (e.g., single-mode fiber technologies), hardware availability (e.g., detector availability), and telescope availability (e.g., building new telescopes or refurbishing existing facilities). **Near-term investment in these areas is recommended in order to reduce key technical risks and to allow a more accurate assessment of the highlighted risks to further improve future architecture assessments.**

**The success of all architectures requires the effective mitigation of stellar variability, which poses the overall highest risk to a future EPRV program. Significant investment in resources, knowledge retention, and international collaboration will be required to deliver a successful program.** An overall cost estimate for the proposed architectures are comparable to that of a MIDEX class flight mission.

# 6 An EPRV Roadmap

The architecture assessment process highlighted several immediate and longer-term key needs that need to be addressed in enabling the capability to detect Earth-analog systems. Given these findings, the EPRV Working Group developed a notional 15-year implementation roadmap and schedule for a funded development program. The overall program timeline is shown in Figure 6-1. An extensive discussion of program needs is presented in individual reports by the Analysis Groups (Appendix A).

The EPRV Roadmap consists broadly of three stages of activities. Stage 1 comprises activities over approximately 5 years that are needed to determine the feasibility of mitigating major sources of systematic errors (i.e., stellar variability, telluric contamination, and instrumental error) at the required level to detect Earth analogs (see Table 6-1). If it is determined that these systematic error sources can be reasonably mitigated, a second stage would be warranted and would have a high chance of success.

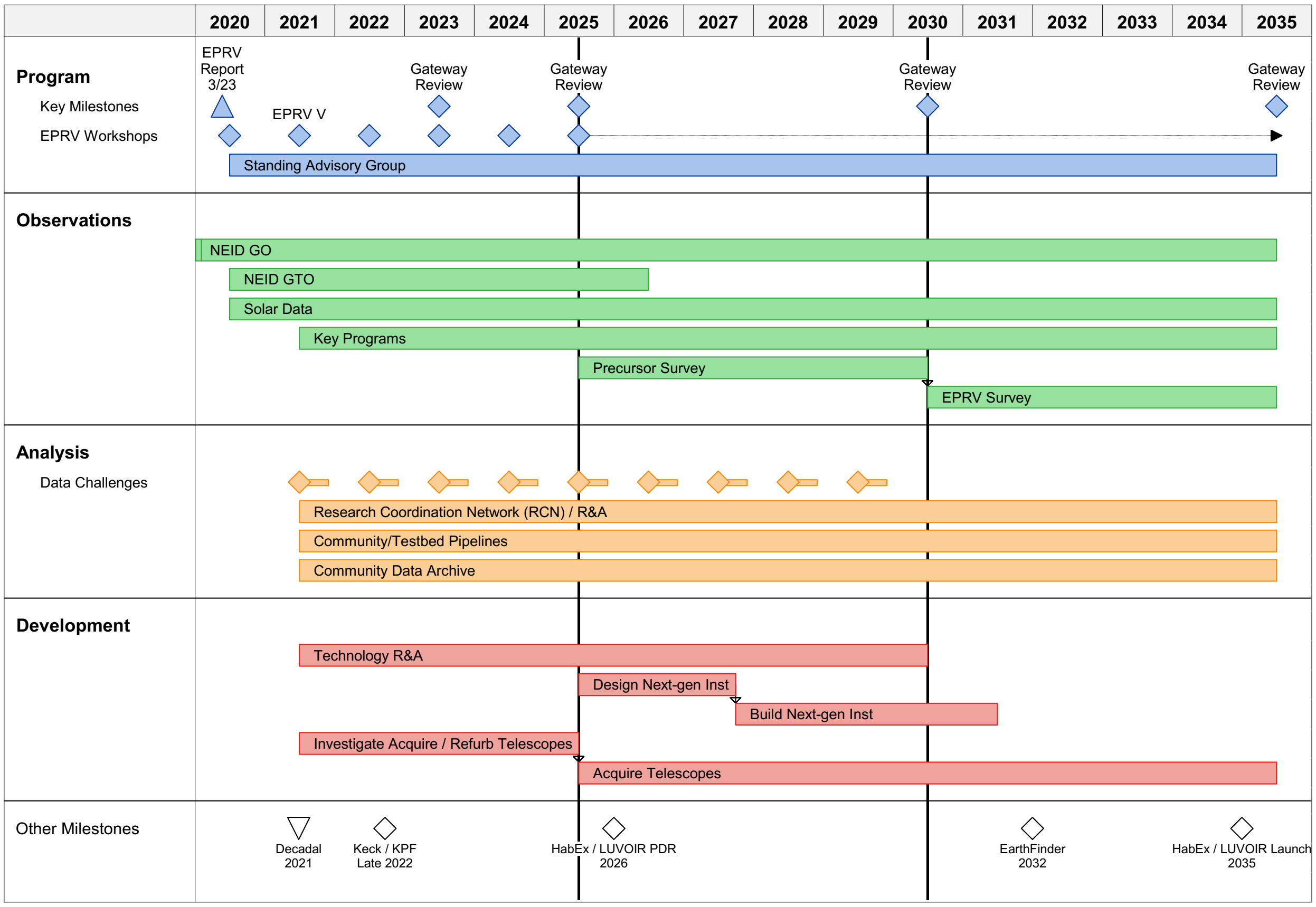

**FIGURE 6-1.** Overall EPRV Roadmap timeline.

Stage 2 is a 5–10-year precursor survey of specific target stars using current generation EPRV instruments that meet the resolution, bandwidth, and precision requirements defined during Stage 1 of the program. It is suggested that the target list will be managed by the NASA ExEP Science Office and nominally follows the target stars adopted for this study (Section 3.1.1). The precursor survey allows for detailed characterization of host star variability, verification of stellar variability mitigation strategies, and provides prioritization criteria for future program elements. The nature of the precursor survey dictates that it is likely to discover new exoplanet systems, however, this is not its primary goal. The precursor survey should be started as soon as feasible to maximize the baseline of observations.

**TABLE 6-1.** Near-term Roadmap (1–5 years) to determine the feasibility and likelihood of a successful EPRV survey.

| |
|---|
| **Stellar Variability:** |
| Support the adoption of solar feeds for spectrographs that demonstrate >100k spectral resolution, SNR > 300 and <50 cm/s stability over a day. |
| Determine how well variability mitigation strategies built from Sun-as-a-star knowledge translate to other spectral types present in the proposed list of target stars. |
| Determine the level of RV precision and accuracy enabled by stellar variability mitigation strategies and modeling and the corresponding implications for the planet mass determinations. |
| **Instrumentation:** |
| Maximize knowledge gained from current instruments still under development, construction, commissioning, and that have recently begun operations. |
| Establish or use existing testbeds for full instrument system development and component testing and characterization, and verification of software-based instrument simulators and select pipeline modules. |
| Determine whether it will be possible to secure a sufficient number of CCDs for a full EPRV survey in a timely manner. |
| Determine whether CMOS (complementary metal-oxide semiconductor) detectors are an acceptable replacement for CCD detectors. |
| Investigate the continued availability of gratings required for high-resolution spectrographs. |
| Determine if there are alternative, cheaper, and more robust methods of fabricating gratings. |
| Investigate the feasibility of securing robust, long-lived, high-stability calibration sources. |
| **Architecture:** |
| Determine whether extreme AO in the visible combined with fiber injection of single-mode fiber-fed diffraction limited spectrographs presents a viable and more desirable option than traditional seeing-limited RV instruments. |
| Determine whether it would be practical to retrofit existing telescopes for dedicated, robotic operation for EPRV observations, or whether new telescopes are required. Determine which of these options presents the least risk and requires the fewest resources. |
| Identify suppliers for multiple, large-aperture telescope systems and conduct site selection surveys, should the building of new telescopes be required for the full EPRV survey. |
| **Telluric Line Contamination:** |
| Determine whether telluric contamination lines in spectra can be adequately mitigated, over a sufficiently broad wavelength range. |
| **Extensive and Detailed Theoretical Analysis:** |
| Determine what combination of spectral bandwidth, resolution, SNR and cadence is sufficient for the detection of Earth-analog systems. |
| Explore the advantages of spectropolarimetry in EPRV observations (e.g., as implemented in SPIRou). |
| Specify, based on lessons learned from observational data, the required quality of the spectra (including resolution and SNR) to detect Earth analogs, which then constrains the required effective apertures, observing time, and cadence. |
| **Software:** |
| Support the development of a well-designed, well-engineered, and actively maintained open-source pipeline with demonstrated ability to retrieve state of the art results on EPRV data from multiple instruments. |
| **Programmatics:** |
| Ensure the needed staffing of personnel with expertise in PRV, heliophysics, and stellar variability to conduct the necessary analysis as well as conduct an EPRV survey. This necessarily involves establishing formal collaborations with non-U.S. entities, and creating attractive employment paths for early career (graduate students, postdoctoral fellows, and non-tenure track researchers) experts in PRV science and technology. |
| Establish a Research Coordination Network and Standing Advisory Committee. |

In parallel with the precursor survey, Stage 2 includes an ongoing research and development effort to deliver the improved observational capabilities and techniques that are needed to enable an EPRV survey of Earth-analog systems.

Stage 3 of the Roadmap undertakes the EPRV habitable-zone Earth survey. This survey utilizes newly constructed EPRV instruments with improved performance over the current generation, which meet the design and performance requirements needed to address stellar variability and to detect and measure the masses of Earth analogs around the set of target stars. This final program stage will span a further 5–10-year period. The precise architecture and format of the program would be informed by Stages 1 and Stage 2 of the EPRV Roadmap. Some technical and construction program elements are required to commence during the latter phases of Stage 2 to meet the overall Roadmap timeline.

The Roadmap schedule includes gateway reviews (notionally every 2 to 5 years) that will evaluate the progress towards the primary goals of each stage of the EPRV initiative. Gateway Reviews will enable the refocusing of investment in key areas if required to advance progress, allow the adjustment of program goals, and would also include potential “off-ramps,” which could be invoked should it be deemed that a pathway towards a relevant stage goal is unlikely to meet its goals. The Roadmap also includes periodic meetings and workshops to engage the broad scientific community.

## 6.1 Addressing Technology and Programmatic Gaps: Required Investments

The EPRV Working Group has collectively, and as individual Analysis Groups, identified technology and knowledge gaps as well as key strategic elements that require investment to enable a future ground-based Earth-analog EPRV survey. Detailed discussion from each of the appropriate Analysis Groups is provided in Appendix A with key points being summarized below. **The EPRV Working Group recommends the establishment of a dedicated research and analysis (R&A) program for sustained EPRV research to support these efforts.**

1. **The most significant obstacle to achieving EPRV capabilities is the intrinsic variability of planet host stars.** To address this, we must radically advance our understanding of the underlying stellar physics and its impact on RV measurements. **We recommend immediately implementing a *long-term, large-scale, interdisciplinary* research and analysis program in this area**. Such a program will include observational (solar and stellar), numerical, and theoretical efforts, and address key science questions by order of priority and importance. This research program is essential to determine the optimal pathway for stellar variability mitigation in a future ground-based EPRV survey.
2. **The extent of research required exceeds the capacity of the present EPRV community.** The scale of effort needed to address the stellar variability problem requires not only increasing the size of the EPRV community, but also increased cooperation within the community. Growing the number of experts prepared to advance the state-of-the-art will require training new graduate students, and ensuring that there are attractive postdoctoral opportunities for them to build on their EPRV expertise. However, there is a more immediate issue, in that **there are not presently enough personnel in the U.S. alone with expertise in high-precision RV, and solar and stellar physics to conduct the necessary analyses in the next few years.** Given both the scale of facilities and the breadth of expertise required, detecting Earth analogs will necessarily be an international endeavor, and the establishment of both domestic and international collaborations is crucial to increase the likelihood of success.
3. **Given the timescale required to develop and carry out a strategic EPRV initiative, knowledge retention in the field of EPRV science and technology is a key issue.** To date, such retention has been challenging, as a great deal of the work thus far has been conducted by graduate students and postdoctoral research fellows, many of whom have gone on to other fields following their academic programs. **The current cycle of short-term appointments is not conducive to the success of this extended program. A deliberate and urgent effort must be made to retain early career experts in high-precision RV science (including solar and stellar physics) and technology. We need to provide viable and attractive** **career opportunities, including (but not limited to) long-term, EPRV-dedicated postdoctoral positions of 4–5 years.**

4. **Establishment of an EPRV Research Coordination Network and Standing Advisory Committee is advised to help with collaboration and coordination of efforts** (e.g., appropriate overlap of stellar target observations). While each existing and upcoming EPRV instrument will have its own science program, by coordinating, the community will improve its ability to make meaningful comparisons across different instruments. Long-term oversight by a Research Coordination Network of current and future EPRV instruments globally would enable EPRV teams to establish and maintain strong, long-lasting synergies across institutions/countries and disciplines (solar, stellar, exoplanetary). Such guidance will enable the community to have shared, coordinated target lists and observing strategies/cadences amongst the instruments. This level of coordination will maximize the quality of the solar and stellar science that is needed to break the stellar variability barrier, and improve methodology and algorithms for analysis of spectroscopic time-series to detect and characterize low-mass planets.
5. **An EPRV program should curate and leverage datasets and knowledge gained from existing state-of-the-art instruments, as well as the next generation instruments currently under development, construction, and commissioning**:
   a. Where feasible, leverage knowledge and experience gained from current EPRV instruments including HARPS/HARPS-N/HARPS3 (Mayor et al. 2003; Cosentino et al. 2012; Thompson et al. 2016), ESPRESSO (Pepe et al. 2010), EXPRES (Jurgenson et al. 2016), MAROON-X (Seifahrt et al. 2018), and NEID (Schwab et al. 2016).
   b. Perform high-cadence observations with solar feeds mounted on several spectrographs that demonstrate <50 cm/s stability over a day. The minimum specifications should be daily cadence, spectral resolution of greater than 100,000 and SNR per resolution element of greater than 300. Continuous solar monitoring (achieved using multiple spectrographs at overlapping longitudes) is crucial to track subtle instrumental systematics and validate stellar variability mitigation techniques over at least one solar cycle.
   c. To the extent possible, coordinate observations with major instruments on a small set of bright standard stars on a variety of timescales in order to create a comprehensive dataset that can be used to disentangle instrumental systematics from other sources of systematic uncertainties (with the same minimum specifications listed in 5b).
   d. Understand the potential performance and extensibility to the visible band of single-mode fiber-fed, diffraction-limited instruments with adaptive optics systems like the PAlomar Radial Velocity Instrument (Gibson et al. 2020; Vasisht et al., in prep) and iLocater (Crepp et al. 2016) as a path to lower cost, essentially telescope-aperture independent spectrograph architectures for EPRV.
   e. Coordinate the data availability, data analysis techniques, statistical tools, and results from these projects.
6. Establishing confidence in detections of low-mass planets will be increasingly challenging. Therefore, **we view it as critical for the EPRV community to carry out a set of EPRV data challenges designed to evaluate the effectiveness and reliability of the advanced analysis methods that are being developed.** In order to reach this point, we recommend that the EPRV community design a comprehensive roadmap for a *series* of data challenges that would lead to an improved ability to extract science from EPRV observations and a better understanding of the capabilities and limitations of EPRV data analysis as a function of key properties of instruments, target stars, and survey strategy. The cadence of these data challenges should be sufficient to address the relevant open questions within the desired timeline. We anticipate one or two data challenges per year will be required. The topics of these individual challenges should be staggered such that they each engage different subsets of the EPRV community, in order to avoid undue burdens on its members.

7. Gaps in RV data pipelines and analysis tools should be addressed:
   a. **There should be a designated centralized repository of RV datasets** in a curated form (standardized formats) where researchers can apply their tools to publicly released datasets from multiple instruments. Examples of current archives that could be adapted for this purpose are the NExScI Exoplanet Archive and the Data Analysis Centre for Exoplanets hosted by University of Geneva.
   b. **There should be a publicly available software pipeline package of data reduction and analysis algorithms.** This package would include a collection of pipelines, analytic tools, and modules that researchers could mix and match with their own algorithms. These should be seamlessly coordinated with the central RV data repository discussed in (a). Ideally, this package would contain a uniform data analysis pipeline that could be used to reduce and process data across instruments, and would be modular, customizable, and open-source for continued community-led development.
8. **EPRV technology testbeds, including full end-to-end systems coupled to solar feeds, should be established.** These testbeds would enable technology maturation ranging from key components to a full instrumental system architecture, analogous to those developed for exoplanet direct imaging programs. Such testbeds will provide a critical component of a technology maturation program that is needed to achieve improved instrumental precision over the current generation of instruments.

To address telluric line contamination in an EPRV survey, **extensive laboratory spectroscopic study is needed in the near-infrared and visible spectral regions using very long gas absorption cells, high intensity light sources, and cavity ring-down spectroscopy (CRDS) systems** to improve line lists and depths. Further, input line lists from the target stars in the EPRV survey will be needed for spectrum fitting.

# 7 Conclusions

We have presented the findings of the Extreme Precision Radial Velocity Working Group, commissioned by NASA and the NSF in response to the National Academy of Science's ESS report. The ESS recommended that **"NASA and NSF should establish a strategic initiative in extremely precise radial velocities to develop methods and facilities for measuring the masses of temperate terrestrial planets orbiting Sun-like stars,"** and the EPRV Working Group was tasked with recommending a ground-based program architecture and implementation plan to achieve this goal.

Based on the Working Group activities that spanned nearly 10 months, we have presented our recommendations for an overall program that has the potential to achieve the goal of measuring the masses of temperate terrestrial planets orbiting Sun-like stars. These recommendations were presented virtually to representatives from NASA and NSF on March 23, 2020.

A summary of the primary findings and recommendations of the study are as follows:

1. There exist multiple plausible system architectures (telescope size, location, and instrument) that could successfully acquire a set of measurements with the statistical precision required to detect Earth analogs, if stellar variability mitigation, telluric mitigation, and instrumental accuracy goals are met.
2. The primary technical risk to the EPRV method for exoplanet characterization at the present time is the inability to adequately correct for stellar variability well below an RV precision of ~10 cm/s on a range of timescales.
3. Of equal importance to addressing the stellar variability question is the urgent need to sustain and grow the expertise and personnel needed to execute an EPRV program. To this end, viable and attractive career opportunities, including (but not limited to) long-term, EPRV-dedicated postdoctoral positions of 4–5 years, are required. Furthermore, engagement with broad expertise, both domestically and internationally, will also be required.
4. A dedicated R&A program for sustained EPRV research should be established including a Research Coordination Network and Standing Advisory Committee.
5. An EPRV program must maximize knowledge gained from current instruments still under development, construction, and commissioning.
6. Gaps in RV data pipelines and analysis tools must be addressed.
7. Key technology gaps must be addressed, including securing availability of critical hardware and establishing EPRV technology testbeds.
8. The potential cost of a comprehensive EPRV program to detect and characterize Earth analogs is comparable to that of a MIDEX class flight mission.

The Working Group has developed a notional 15-year Roadmap to address the current technical and knowledge gaps, enabling a pathway to providing the capability of detecting habitable-zone Earths orbiting nearby F, G, and K spectral-type stars. The program consists of three stages: assessment and mitigation of current major sources of systematic errors, a precursor survey, and a full EPRV survey with the capability to detect and measure the masses of Earth analogs around a defined set of target stars. The first stage of the program, spanning approximately 5 years, will demonstrate that the sources of systematic error can reasonably be mitigated allowing future program stages to proceed.

As part of the first program stage, there is an urgent near-term need for investment and coordination to address key risks, maximize knowledge gained from existing EPRV instruments, and to retain and grow personnel in the field. This investment will serve as an important stepping stone for supporting the future program elements, which will in turn enable the extreme precision radial velocity observational capabilities needed to address several key strategic goals of exoplanet research.

The ability to obtain precise exoplanet masses is critical in determining planet bulk properties, formation history and potential habitability. Further, the detection and mass measurements of Earth-mass planets in the habitable zones of nearby Sun-like stars using precursor observations can be used to improve the efficiency and/or yields of future direct imaging missions. It is widely acknowledged that the EPRV technique will, for the foreseeable future, be the dominant method in providing these measurements. Therefore, advancing the capabilities of EPRV is of critical strategic importance to addressing the science goals of the exoplanet field in general. This scientific need was not only highlighted in the 2018 National Academy of Science's ESS report, but was also endorsed by the Astronomy and Astrophysics 2010 Decadal Survey "New Worlds, New Horizons." The findings, recommendations, and notational 15-year Roadmap developed in this report provide a framework for achieving this goal.

# A Subgroup-Specific Findings and Recommendations

## A.1 EPRV Instrumentation and Facilities

The Instrumentation Analysis Group was tasked to identify key technical needs, development strategies, possible architectures, and hardware risks in delivering the overall EPRV Roadmap. The findings and recommendations of the Instrumentation Analysis Group are presented in this appendix including a discussion of key technical needs, technology gaps and potential instrument architectures. Items relating to facilities as assessed by the Resource Evaluation Group are also included.

The EPRV Working Group adopted the following requirements for EPRV instruments in order to achieve a sub 10 cm/sec precision when measuring the line-of-sight radial velocity of a star other than the Sun. Several parameters are driven by inputs from the Stellar Variability Analysis Group (Section A.2).

- High resolution fiber-fed spectrographs (R = 100k minimum, >130k sought)
- Signal-to-noise ratio > 300 (goal: 800–1000 at 550 nm)
- Spectral coverage from 380 nm to 930 nm as a minimum, and potentially further into the NIR to help further distinguish achromatic Doppler RV signal due to planets from chromatic stellar variability signal (Plavchan et al. 2020)
- Uniform longitude coverage in both hemispheres
- Solar feed on all instruments to aid the characterization of instrument systematics
- Improved NIR precision for removal of telluric contamination
- High-stability visible band calibration sources with <1 cm/s RV precision

Table A-1 shows how architecture-specific wavelength regime and aperture size affects EPRV instrumentation considerations.

The architectures proposed as part of the EPRV Working Group study (Table 3-3), included both seeing-limited designs leveraging improved technologies (Architectures I–IV, VI) and new, diffraction-limited architectures (Architecture V). There exists a significant heritage for seeing-limited EPRV instruments whereas diffraction-limited instruments are only beginning to come online. Section A.1.1 discusses technology maturation paths for both types of systems.

Regardless of the spectrograph architecture, better understanding of and improved performance of critical hardware is required. Large-format detectors are essential for EPRV, as are reliable means of robust, efficient, cost effective, high precision calibration, and low wavefront error, high-efficiency gratings. These and other technology needs for an EPRV program are discussed in Sections A.1.2–A.1.9 and summarized in Table A-2. Other programmatic considerations, including maximizing the knowledge gained from existing instruments, is discussed in Sections A.1.10–A.1.12.

**TABLE A-1.** EPRV technical considerations by wavelength and telescope aperture.

| Wavelength | Aperture |
|---|---|
| **Near UV (320 nm < λ < 390 nm)** | **Small Apertures (*D* ~ 1 m)** |
| • Scientific potential may be high but undemonstrated<br>• Optics increasingly complicated and expensive, coatings problematic<br>• Technology Readiness Level (TRL) ~6 or 7, but potentially high cost and technical risk | • Cheap, fast, easy to build "farms" like MINERVAs<br>• Very high resolution comes for free<br>• Several operational system exist today (pathfinder CORALIE)<br>• Schedule, operational risk – low, but technical risk of achieving 1 cm/sec – high |
| **Blue-Visible (390 nm < λ < 520 nm)** | **Medium Aperture (*D* ~ 2–4m)** |
| • Fiber feed at larger apertures (AO, slicing) still undemonstrated (TRLs 3–6)<br>• Mode scrambling technology immature<br>  ▪ Impact on RV precision weakly demonstrated (TRLs 4–6)<br>• Calibration technology not mature:<br>  ▪ Thorium–argon (ThAr) hollow cathode lamp (HCL) (TRL 9) no longer an option, but commercial LFCs (TRL 7) dubious, and several promising new technologies (TRLs 3– 4 ). Etalons are possible, but require calibration themselves | • Long heritage ELODIE -> SOPHIE, HARPS, HARPS-N<br>• Newest generation of PRV instruments: EXPRES, NEID |
| **Red-Visible (520 nm < λ < 950 nm)** | **Large Aperture (*D* ~ 6.5–10m)** |
| • Single-mode fibers, AO more achievable<br>• LFC technology easier (broadened from NIR)<br>• More telluric line contamination | • Difficult design with high resolution and simple fiber feed<br>• High resolution hard to achieve without pupil/image slicing/AO – still undemonstrated<br>• Hard to get uniform sampling, highly time-specific observations, and access (very few national assets in this aperture range (Gemini/1/2 Keck) |
| **Near NIR (950 nm < λ < 2.5 µm)** | **Extremely Large Aperture (*D* ~ 25–40 m)** |
| • Detectors, optics, mode scrambling are lower TRL, less demonstrated<br>• Performance better than ~1.5 m/s undemonstrated<br>• May not be necessary to characterize stellar variability for solar analogs<br>• Appropriate for M-dwarf stellar population | • Timeline for PRV capability at TMT, GMT, or ELT probably starts after 2030, so not relevant to the current Working Group charter |
| | **Hybrid Options (mixed *D* ~ 2–4 m and *D* ~ 6.5–10 m class)** |
| | • Could help achieve desired observing cadences, hemispheric coverage |

## *A.1.1 Spectrographs*

### A.1.1.1 Seeing Limited EPRV Spectrographs

The standard paradigm for the past decade in PRV instruments in the visible band has been the use of seeing-limited, cross-dispersed echelle spectrographs. These instruments are fed by specialized multimode optical fibers, which deliver light from the telescope focal plane into the spectrometer. Instruments such as HARPS (Mayor et al. 2003) were the first to combine the illumination stability of multimode optical fibers with the thermomechanical stability offered by placing the spectrometer inside a thermally controlled vacuum chamber to reliably reach ~1 m/s measurement accuracy on nearby stars. The latest generation of state-of-the-art instruments, such as NEID (Schwab et al. 2016), EXPRES (Jurgenson et al. 2016), and ESPRESSO (Pepe et al. 2014), have leveraged a suite of new technologies, including specialized multimode fibers to further stabilize the spectrometer illumination, novel wavelength calibration sources to provide improved precision and accuracy in removing instrumental variability, and improved environmental control systems to reach ~50 cm/s on nearby stars over moderate (several month) timescales (Brewer et al. 2020; Pepe et al. 2021). Given the fact that they are already built and funded, these instruments offer the clearest, most timely, most expedient, and likely most fruitful pathway to reaching the instrument precisions that will be required for a future EPRV survey. We strongly endorse strategies and investment for leveraging the capabilities of these instruments to further the EPRV goals recommended in this study.

The primary drawback for seeing-limited designs is one of scalability. For a seeing-limited instrument, the resolution, R, of a spectrometer is given by

$$R = 2D_{grating} \tan\alpha / (D_{tel}\theta_{slit}),$$

or alternatively, the required size of the grating scales as

$$D_{grating} = R(D_{tel}\theta_{slit}) / 2\tan\alpha.$$

Thus, for a given resolution, R, the instrument size scales roughly as

$$D_{grating}^2 \propto (D_{tel}\theta_{slit})^2 \sim A\Omega,$$

where $A$ is the telescope area and $\Omega$ is the solid angle of the entrance beam. The $A\Omega$ product or “étendue” is a conserved quantity in any optical system. In the seeing-limited case where the minimum slit size, $\theta_{slit}$, is perhaps 0.5" in an unsliced system, the size scales as $D_{tel}$ resulting in a physically large instrument that can be challenging to stabilize thermally and mechanically in an evacuated environment. As such, the majority of seeing-limited instruments on large (>8 m) telescopes utilize image or pupil slicers to maintain high efficiency and high spectral resolution, and minimize instrumental footprint.

**Recommendation:** Enable a program of characterization for new cutting-edge seeing-limited instruments to quantify current sources of error and identify future investment needs.

### A.1.1.2 Diffraction Limited EPRV Spectrographs

The need for high-resolution demanded by stellar variability mitigation strategies means it is potentially advantageous to break the telescope aperture-to-spectrograph beam scaling relation in order to limit the size of spectrographs used with increasing telescope apertures. The use of adaptive optics to illuminate an instrument in the diffraction-limited regime has the potential to provide these benefits. Diffraction-limited spectrometers can be made very compact relative to a seeing-limited instrument for a given spectral resolution. In a diffraction-limited system $\theta_{slit}=\lambda/D_{tel}$ and $A\Omega \sim \lambda^2$ and thus, the size of diffraction-limited spectrometers is independent of telescope size (an identical spectrograph design could be employed on either a 3 m telescope or an ELT). This makes them smaller with potentially reduced cost. In a typical diffraction-limited design such as iLocater for the Large Binocular Telescope (LBT) (Crepp et al. 2016), PARVI for the Palomar 5 m (Gibson et al. 2020; Vasisht et al., in prep), or High-resolution Infrared Spectrograph for Exoplanet Characterization (HISPEC) for the Keck telescope (Mawet et al. 2019), AO is used to illuminate single-mode optic fibers which are a fraction of the size of the multimode fibers used in seeing-limited instruments (~5 µm vs. 50–100 µm). The pupil size for these R~100,000–150,000 systems is ~25 mm compared with the seeing-limited R~75,000 SPIRou instrument on Canada France Hawaii Telescope (CFHT) with its 150 mm pupil (Thibault et al. 2012).

These new or planned diffraction-limited instruments operate in the near-infrared where current AO system performance supports efficient injection. Improvements in AO system performance and technologies are required to extend performance into the optical where the stellar information content for Sun-like stars is highest.

There are three primary advantages of a diffraction limited design. First, higher resolution may be possible in diffraction-limited instruments than for seeing limited spectrographs for comparable instrument sizes and beam diameters. This feature may allow resolution of the detailed shapes of individual spectral lines to help identify and mitigate sources of stellar variability (Fischer et al. 2016) and deconvolution of telluric line shape contributions. A second advantage is the prospect of illumination stability via the use of single-mode fibers, where the properties of the output beam are decoupled from the input. Finally, for a multi-telescope program, a diffraction-limited spectrograph design is telescope independent and may be advantageous for cost and the data reduction process.

The primary drawbacks of a diffraction limited spectrometer compared to a seeing limited spectrograph are:

1. An adaptive optics system is required and therefore impacts cost and complexity, and induces potential risk as high performance is yet to be routinely demonstrated in the visible.
2. There is a lack of on-sky demonstrated performance with diffraction limited PRV spectrographs to date. Currently, diffraction limited RV systems are just becoming established in the near-infrared. The first of these, PARVI, was commissioned in 2019. The next, iLocater, has commissioned its fiber injection system (Crass et al. 2021) with the full instrument currently in construction.
3. The potentially reduced efficiency of coupling starlight into single-mode fibers compared to seeing-limited instruments, particularly in the optical and red-optical where stellar information content is highest for Sun-like stars, can negatively impact performance. This could be somewhat offset by other throughput savings (e.g., from not needing optics to mitigate modal noise or imperfect fiber scrambling), but will require further study to quantify (Plavchan et al. 2015).

Because of its relative technical immaturity, it is important to explore the feasibility of a diffraction-limited approach early in an EPRV program given their potential advantages. Ultimately, a determination of whether to pursue seeing-limited systems, diffraction-limited systems, or a combination of both in a hybrid architecture must be made and this requires data to make a technically informed decision.

**RECOMMENDATION:** Explore the feasibility of an adaptive optics-assisted, diffraction-limited approach to EPRV early in a technology development program. Viability of this approach may allow for smaller, less expensive, telescope size-independent RV spectrographs. Improving the performance of the NIR arm of such spectrographs to the 10 cm/s regime may be useful in mitigating stellar variability.

### A.1.2 Detectors

Current PRV instruments utilize CCDs in the optical and hybrid CMOS detectors (HxRG) in the near-infrared. The RV precision achievable with these detectors is hindered by numerous effects including saturation, limited full well capacity, fringing, pixel size variations, “tree rings,” cross-talk, cosmic rays, “bigger/fatter” effect, stitching errors, persistence, imperfect charge transfer efficiency (CTE), intrapixel structure, readout time lost, readout noise, pixel-to-pixel variations, flat-fielding, long- and short-term thermal stabilization, and deformation during readout. The corresponding RV error is estimated to be up to ~40 cm/s on Habitable Zone Planet Finder (HPF; Ninan et al. 2019; Bechter et al. 2019) for its H2RG detectors, and 8.1 cm/s for NEID (Halverson et al. 2016) from its CCD detectors. While some dominating error sources of HxRG detectors have been mitigated as part of the detector development program of the Roman Space Telescope (Mosby et al. 2020), these have yet to be fully assessed in the context of both general high resolution spectroscopy and EPRV science.

The use of CCDs in existing PRV instruments means they serve as the nominal baseline for future visible wavelength EPRV instruments. However, with changing commercial demands and a decrease in the manufacturing base for these products, there is concern that in the timeline of the proposed program, commercial availability of this detector type may become challenging. It is prudent to therefore consider new visible detector formats (e.g., CMOS) or ensure a future supply of large-format CCDs for EPRV instruments as well as broader astronomy research programs.

**RECOMMENDATION:** Enable a program of thorough detector characterization for any EPRV instrument to inform the data pipeline so that detector effects can be adequately removed from the RV data. Identify a reliable supply chain for future EPRV detectors. The suitability of CMOS detectors for EPRV instruments should be assessed as part of this need.

### *A.1.3 Calibration Systems*

Given the minimum spectral coverage adopted in this study, the ideal EPRV spectrograph calibration source must have spectral coverage from 380 nm through 930 nm, with line spacing in the 10–30 GHz range, and uniform intensity across the full bandpass that can be matched to the intensity of the stellar target. Wavelength coverage further into the infrared may be required if it is deemed necessary for stellar variability mitigation. Fractional frequency stability should be better than $\sim 3\times 10^{-11}$ over ~100 s integration times or longer, corresponding to an RV accuracy of 1 cm/s. Maintaining intensity stability across the calibration source spectrum is also crucial, as any variations can add systematic noise to the instrumental wavelength solution. In addition to meeting these requirements, a calibration source should be robust, long-lived, and fiber-coupled for instrument interface. The spectral range and RV precision capability of PRV calibration sources deployed on various spectrographs in the field are shown in Figure A-1.

Classically, RV spectrograph calibration has relied on atomic hollow cathode lamps (HCL) or molecular absorption cells (e.g., $I_2$) for precise wavelength determination. However, these methods have a variety of shortcomings when pushing towards the highest precisions, and have usually been limited at the ~1 m/s RV precision level on-sky.

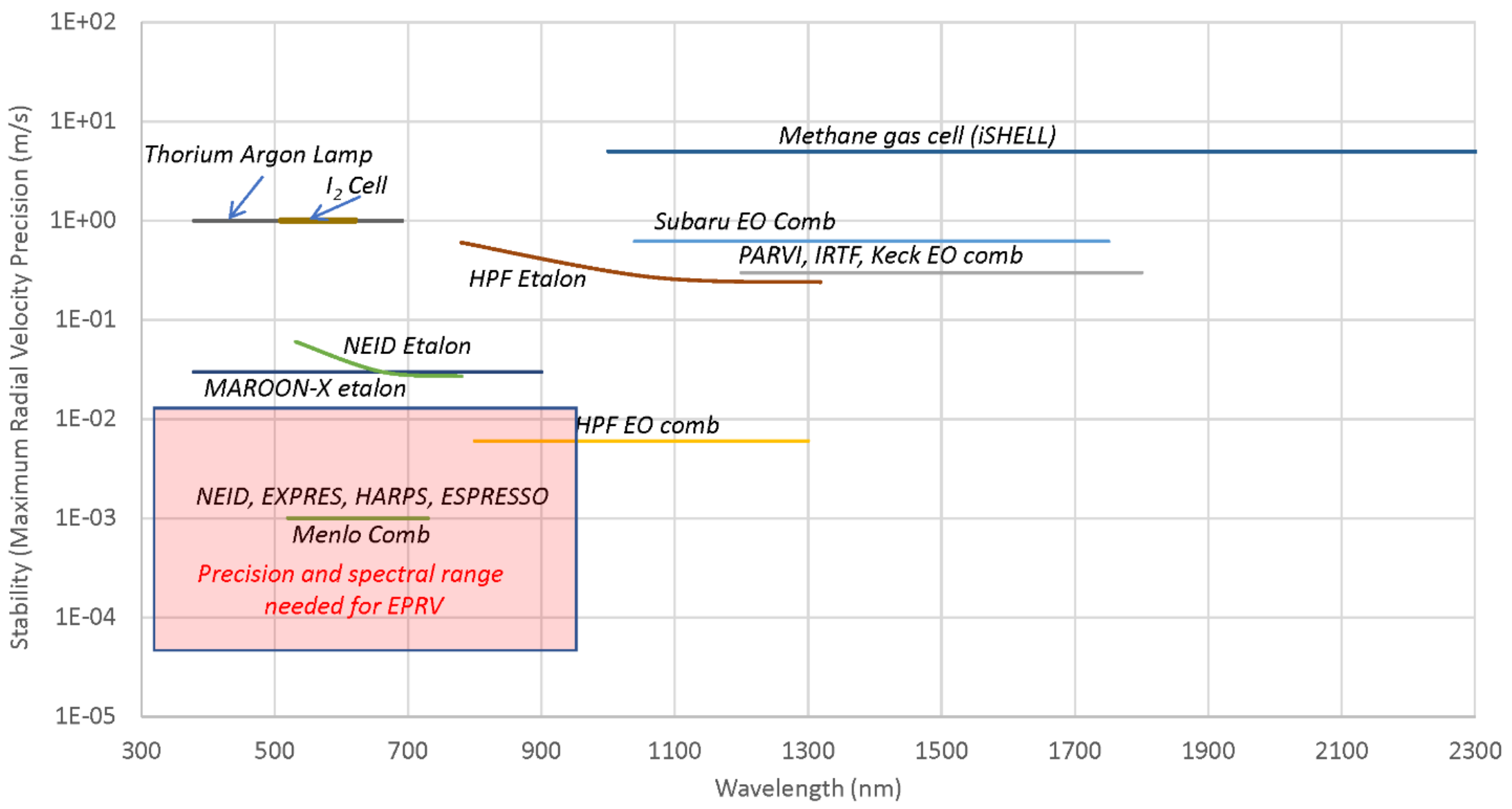

**Figure A-1.** PRV calibration sources – RV precision and spectral range for a range of current technologies and instruments. Data from Butler et al. (1996), Tokovinin et al. (2013), Schwab et al. (2015), Fischer et al. (2016), Kashiwagi et al. (2016), Yi et al. (2016), Cale et al. (2019), Metcalf et al. (2019), Jennings et al. (2020) and Probst et al. (2020). Figure by Stephanie Leifer.

More recently, broadband optical LFCs have been developed for the highest-precision RV applications. LFCs intrinsically produce a uniformly spaced, dense grid of laser lines, each with a frequency known to better than $10^{-12}$ fractional accuracy. LFCs represent the pinnacle of RV calibration systems, providing wide bandwidth calibration at levels of precision far better than those set by other instrument systematics. Menlo Systems GmbH, has developed broadband 'astrocombs' specifically for optical RV applications that have been implemented on the NEID, EXPRES, ESPRESSO, and Veloce (Gilbert et al. 2018) spectrographs. The design employs mode filtering of amplified, low repetition rate (~100–200 MHz) fiber combs through a series of 3 Fabry-Perot filter cavities, thus eliminating ~99% of the comb lines to achieve the sparse line spacing (10–30 GHz) needed to match typical EPRV spectrograph resolutions (R > 100,000). Astrocombs are highly complex devices that require significant engineering efforts to make them 'turn-key' and suffer from several drawbacks. These devices are relatively expensive (~$1M), and have yet to demonstrate both long-term operability at the observatory and reasonable performance at wavelengths blueward of 500 nm. Furthermore, these systems require periodic maintenance to replace consumable components, such as the photonic crystal fiber (PCF) that enables spectral broadening of the combs, compounding the high costs.

Other methods of achieving reliable visible band, 10–30 GHz repetition rate LFCs for EPRV applications are being investigated by multiple groups; most of these approaches involve nonlinear spectral broadening and second harmonic generation of NIR frequency combs generated through either electro-optic modulation (EOM) of a CW laser, or in high-Q disk or ring microresonators through nonlinear optical processes – so-called Kerr microcombs (Del'Haye et al. 2007; Kippenberg, Holzwarth & Diddams 2011), or a combination of both, i.e., pulse-pumped microcombs. NIR astrocombs have been implemented with great success as exemplified by the 800 nm to 1300 nm EOM frequency comb the Habitable Planet Finder instrument at the Hobby Eberly Telescope (Metcalf et al. 2019) that, at the time of this writing, has been operating nearly continuously for 3 years (Frederick et al. 2020). The Subaru and Palomar Observatories each also host a NIR EOM frequency comb for their respective PRV instruments. However, broadening NIR combs into the visible range with 10–30 GHz line spacing is challenging because at these high pulse repetition rates, it is difficult to achieve the threshold pulse energies needed to realize the non-linear optical effects without substantial pulse amplification. Broadening well into the blue-visible has been demonstrated with low repetition rate combs. Thus, exploring methods for reducing the line density of such combs using, for example, pulse rate multiplication (Haboucha et al. 2011) is an interesting avenue.

It should also be noted that for any comb architecture, the power per mode varies across the comb spectrum and must be spectrally "flattened." Flattening is typically achieved using spatial light modulator (SLM) technology. However, development efforts to engineer the power profile in arrayed photonic waveguide devices are also being explored to reduce volume.

Although not as stable as frequency combs, Fabry-Perot etalons are being used as spectrograph calibration sources for on-sky observations in multiple state-of-the art PRV planet-hunting and characterization facilities (ESPRESSO, CARMENES (Calar Alto high-Resolution search for M dwarfs with Exoearths with Near-infrared and optical Échelle Spectrographs), HARPS, NEID, MAROON-X). The ideal etalon leverages fully single-mode operation, has broad wavelength coverage that extends into the blue (<500 nm), is contained in a compact design that is easily thermally stabilized, provides high line brightness and good uniformity, and is referenced to a proven frequency standard. Previous PRV etalon calibration sources, such as those developed for the NEID, HPF (Terrien et al. 2021), ESPRESSO (Schmidt et al. 2021), HARPS (Wildi, Chazelas & Pepe 2012), CARMENES (Bauer, Zechmeister & Reiners 2015), and MAROON-X (Stürmer et al. 2016) instruments have satisfied some subset of these requirements. The line positions of these etalons are set by physical parameters, like etalon spacer thickness and mirror coating properties, and they inherently rely on an external reference for absolute accuracy. These systems are passively stabilized through precise environmental control (temperature, vacuum) to produce a stable output spectrum, and require comparison to an absolute calibration standard (e.g., an atomic transition or a LFC), to verify their long-term stability. Others are actively tracked against an atomic standard, for example, MAROON-X's etalon is referenced to a hyperfine transition of rubidium at 780 nm (Stürmer et al. 2016). However, it is only referenced to one wavelength, leaving the stability of etalons lines far from this point in question due to material dispersion; this is a particular concern at bluer wavelengths. Further, frequency drifts can occur for many reasons in passively stabilized Fabry–Pérot (FP) etalons, including aging of materials like the Zerodur spacers in vacuum-gap etalons (Wildi et al. 2009), long-term degradation of the dielectric mirror coatings that form the resonator, changing environmental conditions (temperature and pressure), as well as changes in the illumination (i.e., thermo-mechanical stability of the light injection). Some of these changes can result in a wavelength-dependent drift of the calibration spectrum (Jennings et al. 2020). Monitoring and, if desired, active control of a FP etalon tailored for astronomical research at sub-m/s accuracy thus requires precise metrology (Schwab et al. 2015). The long-term stability of existing systems has yet to be fully explored, though this is an active area of research for instruments that use both a broadband etalon and a frequency-referenced LFC (HPF, NEID). Beyond 'classical' etalon systems, NASA is currently investing in the development of an advanced hybrid comb-etalon concept based on crystalline $CaF_2$ and $MgF_2$ whispering gallery mode (WGM) resonators that may overcome some of the challenges of traditional FP etalons. The WGM etalons will be referenced to comb-stabilized lasers.

**RECOMMENDATION:** NASA and the NSF should continue to invest in robust, long-lived, high-stability wavelength references for calibration of EPRV spectrographs in the visible and NIR portion of the spectrum. In particular, frequency combs with robust performance blueward of 500 nm and monitoring of etalon drifts across the full instrument bandpass and from the reference/locking wavelength are important to an EPRV instrumentation program.

### A.1.4 Gratings

Traditional seeing-limited and diffraction-limited spectrographs would benefit from improvements in grating technology. To minimize sensitivity to illumination variations, and maximize throughput and spectral resolution, PRV spectrographs require echelle gratings with low wavefront error and high efficiency, both of which are very challenging to achieve in practice. Echelle spectrographs are designed to operate at high angle of incidence and very high diffraction order, and hence the grating must have very accurate groove placement (for low wavefront error) and very flat groove facets (for high efficiency). For decades, echelle gratings have been fabricated by diamond ruling, but it is difficult to achieve all aspects of the performance required for PRV instruments with this technique. Newer grating fabrication techniques using lithographic methods to form the grooves may be a promising approach. Immersion gratings, while limited to IR applications (J–M photometric bands), could boost spectral resolutions to >200,000. In seeing limited PRV spectrographs, the grating dimensions scale with the size of the telescope aperture for a fixed spectral resolution. Thus, in practice, it has been necessary to stitch smaller grating together in order to achieve the total required diffraction aperture for R~100,000, even for moderately-sized telescopes (D > 2.5 m). The registration of these stitched gratings often introduces significant wavefront error, and complicates the fabrication process.

In summary, future EPRV studies rely on high efficiency, steep blaze angle echelle gratings to achieve high spectral resolutions of R > 100,000 for both seeing-limited and diffraction-limited systems. Desires for future gratings used in EPRV systems include:

- Size (seeing-limited): ~200×1200 mm (width × length for the clear aperture at the blaze angle)
- Size (diffraction-limited): ~50×200 mm (width × length)
- Steep blaze angles (>76 deg or >R4) for achieving higher spectral resolutions >100,000 (e.g., R6 for a 150,000 resolution).
- Higher efficiency, both by reducing diffraction effects and improving coatings. The state of the art is ~50–60%, while >70% is sought.
- Better wavefront error (<lambda/8) across the aperture. This is important for diffraction-limited, adaptive optics-fed systems where maintaining the point spread function (PSF) profile is required to achieve high resolutions (Bechter et al. 2021), as well as seeing-limited instruments that strive to achieve sharp image quality to mitigate the effects of variable pupil illumination.
- Lower line density echelles < 13 lines per mm to be more compatible with detector array widths in diffraction limited spectrographs.

**RECOMMENDATION:** Develop grating fabrication techniques as alternatives to the traditional diamond-ruled process to achieve large-format, high-efficiency, steep blaze angle, low wavefront error echelle gratings for both seeing-limited and diffraction-limited spectrographs. These are required to achieve high spectral resolutions of R > 100,000 with high throughput and high image quality.

### A.1.5 Adaptive Optics

Without the Earth's atmosphere, light propagating from astronomical objects would have virtually planar wavefronts. However, nonuniformities in the refractive index of the atmosphere caused by temperature and pressure fluctuations (turbulence) distort the wavefront and reduce the quality of the images. AO systems measure this distortion and in real-time correct the wavefront. The result is vastly improved image quality on ground-based telescopes compared to seeing-limited operation.

Typical AO systems are composed of three main components: a wavefront sensor, wavefront corrector (such as a deformable mirror (DM)), and a controller/feedback system to actively adapt the wavefront corrector on millisecond timescales. For EPRV applications that use single-mode fibers (Section A.1.1.2), the challenge is development of an AO system that provides the required performance over small fields of view in the spectral bandpass of interest (towards blue wavelengths) and suitable for the selected telescope aperture. Increasing the level of correction will increase the overall coupling efficiency of starlight into single-mode optical fibers.

When coupling light from AO-equipped telescopes into single-mode fibers, overall fiber coupling efficiency can be expressed as the product of the incident Strehl ratio and the theoretical fiber coupling efficiency, where the latter value is typically 60–80% due to losses from diffraction from telescope apertures. This value may be improved through the following avenues:

- Developing optics (i.e., PIAA lenses) for reshaping the beam into a Gaussian shape with minimal loss of light, allowing for theoretical coupling efficiencies close to 100% (see, e.g., Jovanovic et al. 2017a);
- Removing non-common path aberrations; and
- Identifying the optimal method for maintaining fiber coupling and positioning (e.g., with tip/tilt, nodding, or other methods).

Strehl ratio improvements can be accomplished by:

- Reducing the fitting error of the wavefront with a high number of actuators and more stroke in deformable mirrors;
- Wavefront sensor choice/development for reduced noise characteristics;
- Latency reduction through judicious choice of real-time controllers (both firmware and processors). Up to 2 kHz is a reasonable present-day assumption; and
- Automated parameter correction for adaptive/predictive control. This may not be necessary for small telescopes, and differentiates requirements between larger apertures e.g., 3-meter class telescopes vs. 10-meter class telescopes.

Figure A-2 presents predictions of a model of Strehl ratio which might be achievable with an advanced AO system across the visible and NIR bands on a 3 m telescope with 1" seeing.

AO system design is a function of telescope size, and therefore must be planned for the specific telescope and PRV instrument. Thus, the recurring cost of duplicate AO systems to equip a global network of identical PRV telescopes facilities could be lower than designing unique AO systems for various existing facilities. Even in the case of differing AO system designs for heterogeneous telescopes, by using a diffraction-limited input for spectrograph illumination, identical spectrographs can be employed at all locations, thereby offering a potential engineering and software cost benefit compared to (necessarily) heterogeneous seeing-limited designs.

The critical capability affecting the performance of a diffraction-limited EPRV system is the ability of an AO system to provide adequate optical coupling into the single-mode fiber (SMF) *at visible wavelengths*. This is important because expected EPRV target stars (typically FGK spectral types) have the majority of their Doppler information content shortward of 600 nm. Figure A-2 shows how PRV precision (in m/s) degrades as the shortest operable wavelength moves to longer wavelengths. High Strehl AO in the NIR has been demonstrated at numerous telescopes including the Keck 10 m telescope and the Palomar 5 m telescope with peak K-band Strehl values of 0.65 and 0.85. Strehl values of 0.60 have been demonstrated in the Y-band at the LBT 8.4 m telescope. These values were obtained for bright targets like the stars explored in the EPRV simulation study (Section 4). Attaining comparable values at visible light has yet to be demonstrated routinely, but significant efforts are underway at a number of observatories, including at the Magellan 6 m telescope (Close 2016; Close et al. 2018) where the MagAO-X system (new extreme AO system for the Magellan Clay 6.5 m telescope) has achieved Strehl ratios of ~50% at 900 nm in initial commissioning with an eventual expected performance of 70% at Hα (656 nm) in median seeing conditions (Males et al. 2020). Additional 'extreme-AO' systems include SCExAO (Subaru Coronagraphic Extreme Adaptive Optics), which is on-sky at the Subaru Telescope in Hawaii (Lozi et al. 2020), and SOUL (Single conjugated adaptive Optics Upgrade), which is undergoing commissioning at the LBT in Arizona (Pinna et al. 2016).

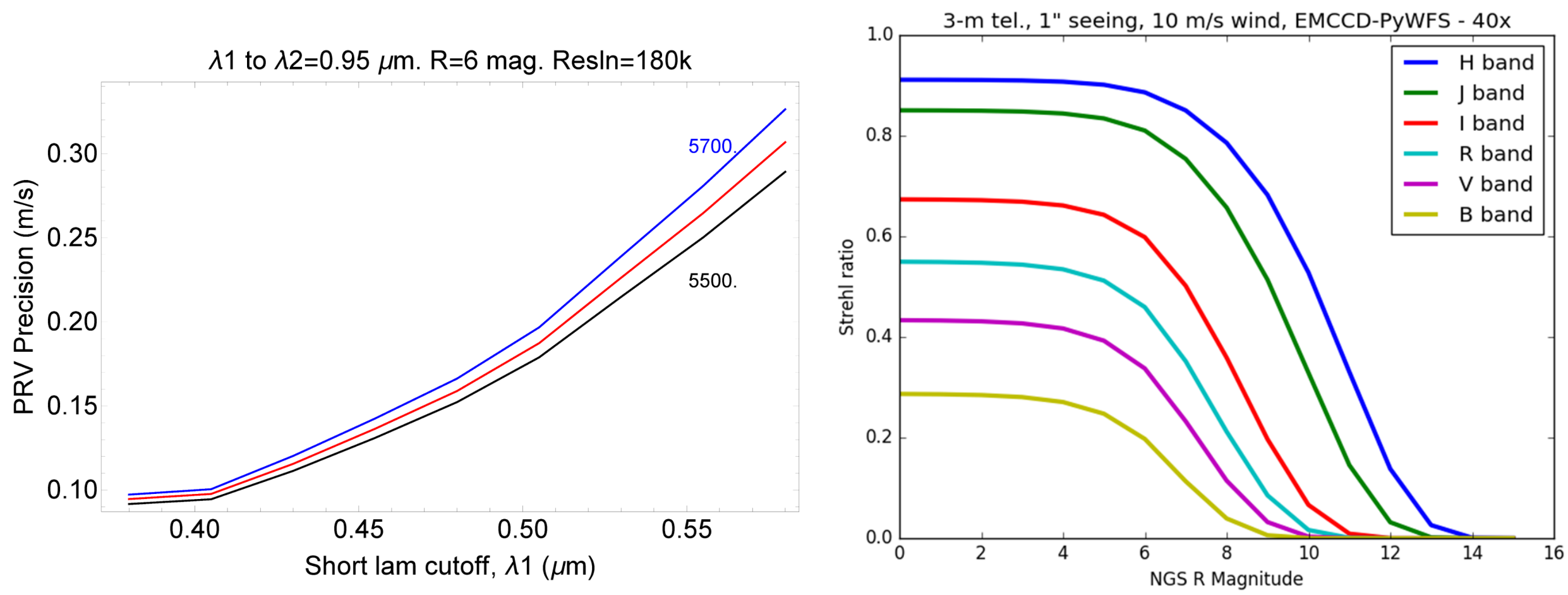

**FIGURE A-2.** *Right:* The Strehl ratio performance of a 3 m telescope with a Natural Guide Star Adaptive Optics (NGS-AO) system operating under good seeing conditions, 1” with a 10 m/s wind speed. The assumed system employs a 1600 element deformable mirror and Pyramid wavefront sensor using an EMCCD (electron multiplying charge-coupled device) that can operate at high speed with no read noise. Predictions of the Strehl ratio are given as a function of target R-band magnitude at various wavelengths. *Left:* The wavelength grasp of the EPRV instrument affects the achievable precision for a given integration time. As the performance of an AO system degrades at shorter wavelengths, we can expect the RV performance to degrade as well. The figure shows the effect of cutting off the short wavelength performance at progressively longer wavelengths, from 0.4 to 0.55 µm for stars of three different effective temperatures.

While AO systems are typically considered for their role in enabling diffraction-limited capabilities, they also play a role in supporting seeing-limited observations. AO systems can be used to improve image concentration and to enhance near and far field stability for increased wavelength-scale stability in high dispersion spectrographs (Mello et al. 2018). This also provides a potential method to reduce the size of a seeing-limited spot, which allows for reduced fiber diameters in seeing-limited instruments on large-aperture telescopes.

**RECOMMENDATION:** A parametric study of adaptive optics system design as a function of aperture size, observing site (average seeing), spectral band, and target star magnitude should be conducted to advise the design of EPRV architecture-specific AO systems. A visible band AO system should then be developed in conjunction with a testbed spectrometer to achieve the necessary coupling into optical fiber and taking into account atmospheric dispersion correction (ADC) and PIAA pupil mapping.

### A.1.6 Fibers

The intrinsic illumination stability of the spectrometer sets a fundamental limit on its measurement floor. If the image of the star varies at the entrance to the spectrometer due to atmospheric effects, telescope guiding errors, or intrinsic fiber properties, so too does the recorded stellar spectrum, leading to spurious RV offsets. Current seeing-limited PRV instruments use multimode optical fibers, which provide some degree of azimuthal image ‘scrambling’, to efficiently deliver stellar light from the telescope focal plane to the spectrometer input. Novel-core-geometry fibers, in concert with dedicated optical double-scramblers, are often used to further homogenize and stabilize the telescope illumination pattern in both the image and pupil planes. However, these systems still demonstrate measurable sensitivity to incident illumination variations from the telescope and atmosphere.

As spectral resolution requirements increase, the commensurate increase in seeing-limited instrument size can become impractical for large (>8 m) telescopes. As such, the community has turned to implementing image and pupil 'slicers' to reformat the near or far fields of light entering the spectrometer. These slicers preferentially redistribute starlight exiting the fiber to maintain high spectral resolution, high efficiency, and compact spectrometer size.

An advantage of using single-mode fibers for illumination is their spatial stability in illumination, which overcomes the 'modal-noise' that must be suppressed in the multimode fibers. Two polarization states in SMF do remain however, and these must also be controlled in order to avoid polarization noise in an EPRV system (Halverson et al. 2015; Bechter et al. 2020).

**RECOMMENDATION:** An extended study of existing multimode fiber system performance, fiber scrambling systems, and image and pupil slicer should be undertaken to identify required areas of investment. Studies should include an assessment of chromatic scrambling effects. For single-mode fiber systems, strategies to mitigate effects of polarization should be demonstrated and performance quantified.

### *A.1.7 Optical Bench Materials*

As we move towards reducing instrument error contributions in RV measurements, the material used for instrument construction may become increasingly important. Adopting an intrinsically stable, optimized material for spectrograph optomechanical systems will help reduce the environmental stability requirements for instruments with increased volumes while also improving performance for current instrument scales. The next generation of instruments already under development are beginning to utilize this approach with the Keck Planet Finder being fabricated using aged Zerodur (Gibson et al. 2018), G-CLEF using carbon-fiber composite (Szentgyorgyi et al. 2018), and iLocater having optomechanics fabricated with Invar (Crepp et al. 2016). Silicon carbide is another potential material that may allow for increased stability. When considering a material selection, it is important to consider the instantaneous material characteristics at the instrument operating temperature in addition to how this will interact with the system optical components to environmental changes.

**RECOMMENDATION:** To maximize intrinsic spectral stability in EPRV instruments, explore the use of known and exotic materials with low CTE and/or high conductivity for instrument optical benches and optomechanics.

### *A.1.8 Advanced Concepts/Photonics*

Development of advanced spectrograph designs may provide important new options for future EPRV surveys. An example is the development of photonic spectrographs based on lithographically formed arrayed waveguide gratings (Jovanovic et al. 2017b). Such monolithic devices would occupy a small fraction of the volume of existing PRV spectrographs and potentially offer lower cost. Externally dispersed spectrograph designs (e.g., VERVE (Vacuum Extreme Radial Velocity Experiment), Van Zandt et al. 2019) may offer advantages as well.

Other photonic technologies may also be applicable to EPRV applications. The use of photonic lanterns can be used to convert a multimode fiber input into separate SMF outputs by sampling multiple positions centered on the near-diffraction limited input. By using the SMF outputs from such a system, these can illuminate separate traces of a diffraction-limited spectrometer (Schwab et al. 2014; Mawet et al. 2019). Photonic lanterns are an attractive technology that can provide an interface between single-mode and multimode optical fibers. This allows a potential pathway to increase diffraction-limited EPRV capabilities into the blue visible.

**RECOMMENDATION:** Continued investment in advanced EPRV technology, including externally dispersed spectrographs, photonic lanterns, and arrayed waveguide grating (AWG) spectrographs may circumvent limitations of current technology paths.

### *A.1.9 Testbed Development*

State-of-the-art PRV instruments have typically been designed as individual builds for specific applications/telescopes. These tools are built, sent to remote astronomical telescope sites, and after a relatively brief commissioning and verification phase, are scheduled for scientific observation. Thus, there is limited opportunity for technologists to measure the influence of different components and configurations on overall instrument performance.

In contrast to the radial velocity exoplanet community, the direct imaging community has invested in at least two powerful testbeds for adaptive optics and coronagraphy. This has facilitated rapid technological progress toward high contrast imaging capabilities with ever-smaller inner working angles. Testbeds are essential to astronomical instrumentation R&D for a number of reasons, specifically:

- Laboratories are good places to do experiments, while observatories are suboptimal for a variety of reasons:
  - Travel to remote observatories is expensive.
  - Daytime access to facilities is circumscribed by telescope maintenance operations, and nighttime access is extremely difficult.
  - Technical infrastructure is very limited at a working observatory.
- Instrumentation built for R&D can be designed to be flexibly modified and configured to attain operating conditions rapidly to allow for rapid turnaround. In contrast, instrumentation that is designed to work on a telescope is tightly circumscribed by technical and environmental consideration.
- Testbeds allow engineers to test technology internal to the spectrograph at the component level, for example, detectors and gratings.

End-to-end EPRV testbeds, including simulated RV sources and complete software suites for operation and the data pipeline, would enable an R&D program that permits studies of different technologies – e.g., an intercomparison of multiple calibration lights sources, possibly simultaneously. It would permit parametric determination of influence functions. Optimally, two such testbeds for each architecture type (i.e., seeing-limited and diffraction-limited) would be built, so as to provide independent cross checks of performance. A separate detector and/or grating testbed may be needed.

**RECOMMENDATION:** Commit to dedicated testbeds for EPRV instrumentation and data pipeline software development in accessible laboratory locations, and wherever possible, make use of existing facilities for EPRV instrument component level testing such as for detector characterization.

### *A.1.10 Existing PRV Instruments*

The EPRV community has recently begun observations with a new generation of EPRV spectrographs (e.g., EXPRES, ESPRESSO, HPF, NEID). Several additional EPRV instruments are in development and construction. It is imperative to compare the performance of different instrument designs and analyze the impact of various hardware components and design choices on system performance. Initially, comparisons should focus on instrumental stability and RV performance. In the longer term, it will be just as important to evaluate how well each of these instruments is able to measure and mitigate stellar variability. Since different instruments have taken different design approaches (e.g., prioritizing spectral resolution or wavelength coverage), they are complementary and can inform future instrument design choices.

Comparing the ability of each instrument to mitigate stellar variability will require many observations of each star in question. The science programs which justified their construction inevitably will use different survey strategies (e.g., choice of targets, SNR per exposure, observing cadence). In order to make meaningful comparisons, it is critical that each instrument obtain a substantial number of high-quality observations for at least a few RV benchmark stars, including the Sun (see Section 6 and Section A.2), and would be particularly valuable for comparing which instrument designs perform best as a function of target star properties (temperature, metallicity, and brightness). Such comparisons will provide valuable information when making design choices for the next generation of EPRV instruments.

**RECOMMENDATION:** Carry out comprehensive cross-comparisons of the performance of current and near-term EPRV instruments (e.g., ESPRESSO, EXPRES, KPF, MAROON-X, NEID). Maintain engineering expertise needed for the development and construction of EPRV instruments beyond instrument delivery, which can contribute to understanding the instrument performance characteristics, in order to identify opportunities for improvements in future instrument design. Thoroughly characterize the final instrument performance using both the instrument teams, which can perform the most detailed and well-informed analyses of their own instruments, and external groups, which can ensure independent, uniform analysis methodologies that can be meaningfully applied to multiple instruments. Identify both successes and lessons learned, and socialize these with the community. Identify and document characteristics of each instrument that are a consequence of its specific design, including, but not limited to: temporal instrument characteristics (absolute and relative fiber-to-fiber drift), detector characteristics, wavelength calibration strategy, etc. Adopt and document performance metrics beyond RV precision, such as wavelength fidelity, PSF stability, and flux conserving extraction.

### A.1.11 Telescope Acquisition/Repurposing

The EPRV Working Group developed and analyzed a representative set of telescope aperture networks that met the objectives of the initiative (Section 3.2). These architectures ranged in size from dense arrays on 1-meter telescopes to collections of several multi-meter apertures. Each architecture covered both northern and southern hemispheres with multiple telescopes in each hemisphere (longitudinal diversity) in order to be resilient against weather losses and provide independent verification of candidate detections. In some cases, a candidate architecture had identical apertures (homogeneous architecture), in others there were multiple aperture sizes (heterogeneous architecture). All had as a baseline to be fully robotized where possible. These candidate aperture networks illustrate some options, but are not exhaustive of all possible architectures that will meet the EPRV goals.

For any telescope facility that will not be 100% utilized by an EPRV survey program, careful consideration should be given to the optimal strategy for its use. Currently, many telescopes cannot switch instruments during the course of a night and, for those that do, it can be inefficient, and thus instrument changes are generally minimized. The prospects for requiring increased EPRV observing cadence may be improved at many observatories without impacting other instruments if an interface is engineered to allow a PRV fiber feed from the telescope beam. As PRV observations can be done for short intervals spread over a night, this would allow for large time allocations for other science observations while still maintaining a high EPRV observing cadence.

The optimal architecture for a future Earth-analog EPRV survey is not known at this time and will depend on multiple factors, including but not limited to: availability of existing apertures, ability to repurpose, refurbish and robotize those telescopes, the time available on those telescopes, the timing of their availability, and their geographic location. The developed architecture must be flexible to permit a variety of aperture and location options.

A portfolio of candidate EPRV telescopes is needed. The NSF, with the largest portfolio of US telescopes, will likely be the optimal partner in such an effort, which could be modeled on the NASA-NSF Exoplanet Observational Research (NN-EXPLORE) partnership for joint agency collaboration.

**RECOMMENDATION:** Develop a candidate list of apertures through discussions with observatory custodial agencies, consortia, and institutions, including international entities. Convene a set of workshops to bring together telescope operators, builders, and custodians to learn from past experiences on repurposing, refurbishing and robotizing telescope systems with emphasis on capabilities, costs, timeline and lessons learned. Conduct site visits to inspect telescope facilities and meet with potential partners.

### A.1.12 References

## A.2 Stellar Variability

### A.2.1 Overview

Our incomplete understanding of the physics of the outer layers (photosphere, chromosphere, corona) of Sun-like stars poses the most significant obstacle to confirming and characterizing Earth-like worlds via EPRV. ***The spectra of Sun-like stars are variable on a broad range of timescales due to temperature and velocity fluctuations in their outer layers. These fluctuations are primarily driven by convection and magnetic fields and their interplay. This variability induces correlated RV signals with 0.1–100 m/s amplitudes spanning timescales from seconds to decades*** (e.g., see reviews by Fischer et al. 2016; Dumusque et al. 2017; Collier Cameron 2018; Cegla 2019; Hatzes 2019; and references therein).

When searching for planets with a stabilized spectrograph, traditionally, the stellar spectra are cross-correlated with a template and the RV is determined from the centroid of the cross-correlation function (e.g., Baranne et al. 1996; Pepe et al. 2002). Consequently, changes in the stellar line shapes can be mistaken as Doppler shifts. Hence, time-variable inhomogeneities on stellar surfaces induce spurious RVs and mask or mimic planetary signals (e.g., CoRoT-7 "d": Hatzes et al. 2010; Lanza et al. 2010; Pont et al. 2011; Haywood et al. 2014).

Furthermore, bulk velocity flows in the photospheres of Sun-like stars can induce coherent Doppler shifts, with magnitudes that depend on the depth in the photosphere and thus manifest differently in different lines. In particular, Sun-like stars have a radiative core and a convective outer envelope. In the outer envelope, hot bubbles of plasma (granules) rise to the surface, cool and sink down into intergranular lanes; these motions induce blueshifts and redshifts accordingly, resulting in a net blueshift and RV variability up to the m/s level. This variability alone is an order magnitude larger than the signal from a potential Earth twin (see Cegla 2019 for further details).

Granulation and supergranulation are manifestations of convection. The granules excite acoustic modes (e.g., p-modes, r-modes) that give rise to surface oscillations that introduce net shifts on the same order (see reviews by Stein 2012; Nordlund, Stein & Asplund 2009; Rincon & Rieutord 2018). The convection can also be suppressed by magnetic fields, and the net blueshift altered, resulting in faculae and spots (Figure A-3; e.g., Schrijver & Zwaan 2000, Chap. 1 and 12; Foukal 2004, Chap. 5 and 8). Faculae are bright regions where significant magnetic fields have altered the opacity so that the deeper, hotter regions of the star are visible; spots are the same, but the even greater magnetic field strengths suppress the convection so significantly that these regions are cooler and darker than the surrounding photosphere. Magnetic active regions can induce spurious RVs of tens of m/s (Saar & Donahue 1997; see reviews by Collier Cameron 2018; Hatzes 2019). In comparison to sun/starspots, faculae have been studied relatively little, but they dominate the surfaces of slowly rotating, Sun-like stars (e.g., Radick et al. 1983; Lockwood et al. 2007) and drive RV variations of the Sun (Meunier et al. 2010b; Haywood et al. 2016; Milbourne et al. 2019). Even magnetically quiet stars experience stellar "noise" (e.g., RV observations presented in Isaacson & Fischer 2010; Motalebi et al. 2015; Luhn, Wright & Isaacson 2020). ***Consequently, intrinsic stellar variability currently precludes the confirmation and characterization of Earth-analogs*** (NAS, 2018, pp. 92–93).

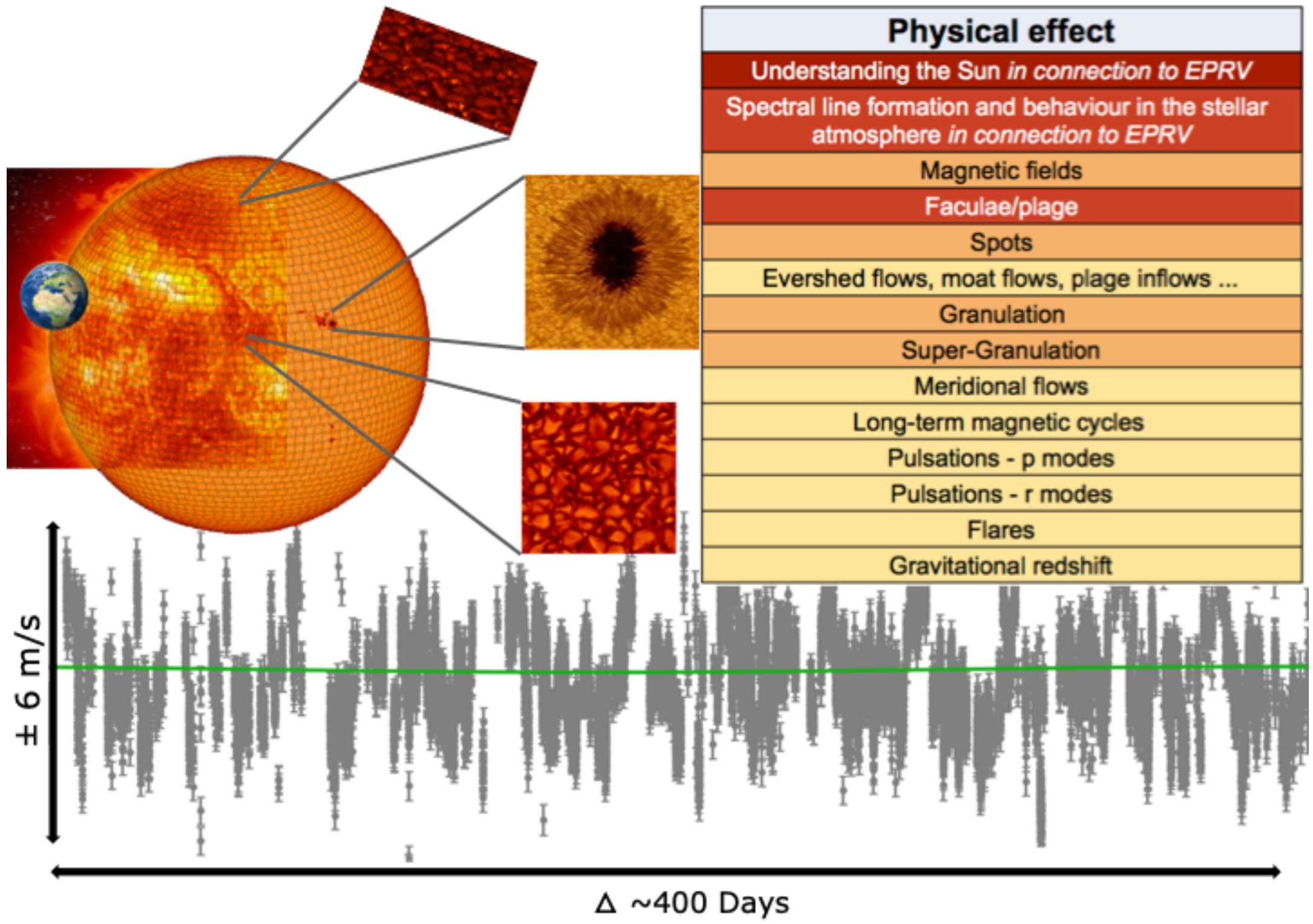

**FIGURE A-3.** *Top left:* Artist impression of the stellar variability challenge to EPRV. *Top right:* physical phenomena known to induce RV variability, color-coded by importance to EPRV (ranging from red: urgent & critical; to yellow: long-term & significant). *Bottom:* HARPS-N RV observations of the Sun seen as a distant, point-like star (gray points; Collier Cameron et al. 2019), and signal of an Earth twin (green line). *Credits:* NASA, ESA, SDO/HMI, MURaM, Big Bear Solar Observatory, solar RV observations from HARPS-N, Cegla/Haywood/Watson.

A Roadmap for overcoming stellar variability would pursue a robust research and analysis program to address key physics (listed in a priority matrix shown in Table A-3) that are critical to reaching the EPRV level:

1. **Acquire fundamental solar/stellar knowledge:** At heart, this is understanding convection and its interplay with magnetic fields, as it underpins each of the 12 physical phenomena listed in Figure A-3.
2. **Develop and apply stellar models and mitigation frameworks:**
   a. Sun-as-star/stellar models;
   b. Variability mitigation frameworks; and
   c. Inform precursor survey.
3. **Develop RV observation and analysis toolkits/strategies** for use by the exoplanet community. Disentangle stellar and planetary signals to push detection/characterization limits towards Earth analogs, including precise and accurate determination of their masses.

**TABLE A-3.** Areas of solar and stellar physics that must be addressed to meet the EPRV challenge, ordered by priority, timeline, and importance.

| | High Importance | Medium Importance | Low Importance | Critical Empirical Datasets (including, but not limited to) |
|---|---|---|---|---|
| **Near term**:<br>70% solar<br>30% stellar | How does convection interact with magnetic fields? How do surface phenomena (ranked by importance: granulation/faculae/plage, supergranulation, spots/active-region velocity flows, meridional flows, r-modes) drive Sun-as-a-star RV variations? How are magnetic fields generated? How does the solar/stellar photosphere connect to the chromosphere? Understand line formation and behavior to a level of detail necessary to create the next generation of physically motivated solar/stellar models and instrumentation. | How do solar/stellar surface phenomena and their RV impact change over the magnetic cycle? Identify new, robust observable solar/stellar variability indicators for RV variations to inform future instrumentation, observational surveys/strategies. Explore data-driven techniques for solar and stellar variability mitigation in EPRV. | How do flares and gravitational redshift impact solar/stellar RV variations? Can we improve existing p-mode mitigation techniques? | **Spectroscopic daily observations of the Sun as a star** (including HARPS-N/solar, HELIOS, NEID, EXPRES).<br>Stellar spectroscopic time series (including archival HARPS data).<br>**Require <50 cm/s RV precision and daily/nightly cadence for at least one full season.** |
| **Medium term:**<br>60% solar<br>40% stellar | How does solar knowledge (observations/theory/simulations) connect to stellar knowledge? What instrumentation/ simulations/precursor surveys are needed to answer the unknowns from above? Continue "high importance" efforts from near term. | How do stellar surface phenomena (including granulation/faculae/ plage, supergranulation, spots/active-region velocity flows, meridional flows, r-modes) change as a function of surface gravity and surface temperature? Continue "medium importance" efforts from near term. | Design physically motivated RV models for M-dwarfs. Develop and apply RV observation and analysis toolkits/strategies to M-dwarfs hosts and key transiting systems. | Spectroscopic daily observations of the Sun as a star (including all above + ESPRESSO).<br>Archival HARPS, NEID, ESPRESSO data (+published HARPS-N & EXPRES). **Precursor survey requirement: <20–30 cm/s RV precision daily/nightly cadence for at least one full season.** |
| **Long term**:<br>40% solar<br>60% stellar | Develop and apply stellar models and mitigation frameworks (RV and others such as photometry, spectropolarimetry, etc.) as a function of surface gravity and surface temperature. Incorporate models and frameworks into RV observation and analysis toolkits/strategies for use by the exoplanet community. | Improve and optimize RV observation and analysis toolkits/strategies for use by the exoplanet community. | How does stellar variability impact observations of exoplanet atmospheres and exoplanetary habitability? | **Require <2 cm/s instrumental RV precision with daily/nightly cadence over a season** (e.g., HIRES/ELT; G-CLEF/GMT). |

### *A.2.2 Findings on Intrinsic Stellar Variability*

#### A.2.2.1 Stellar Variability Error Budget

The Stellar Variability subgroup determined an error budget, shown in Table A-4, that lists the physical processes known to induce RV variability, and clearly displays a meter-per-second "wall" that currently precludes the confirmation and characterization of Earth-analogs orbiting old, slowly rotating, relatively inactive Sun-like stars. The main findings relevant to EPRV from this subgroup are:

- Currently, the biggest obstacle is understanding the inhibition of convection from magnetically active regions.
- Existing variability indicators do not trace stellar variability down to sub m/s precision, including the log $R'_{HK}$ and line profile shape measurements (e.g., bisector or BIS, full-width at half-maximum or FWHM).
- The unsigned magnetic flux is an excellent tracer of inhibition of convection in Sun-like stars, but cannot currently be measured in slowly rotating, relatively inactive stars like the Sun.
- Granulation and super granulation are not easily binned out (time averaged).
- P-modes can likely be binned out to sub 10 cm/s.

The RV behavior of several phenomena remains poorly characterized on the Sun and other stars. Additionally, there may well be additional physical processes at play that contribute to RV variability, which are yet to be identified and characterized.

#### A.2.2.2 Current State of the Art for Mitigating Stellar Variability to Determine Planet Masses

Current RV analyses typically treat stellar variability as (un)correlated *noise* (e.g., see the *RV Community Challenge* – Dumusque 2016; Dumusque et al. 2017). Planet masses are typically determined using statistical tools that allow us to incorporate minimum prior knowledge (such as the star's rotation period) into a robust and flexible function (e.g., Gaussian process regression: Haywood et al. 2014; Rajpaul et al. 2015; Faria et al. 2018; Fulton et al. 2018; Gilbertson et al. 2020; moving average regression: Feng et al. 2017; Anglada-Escudé et al. 2016; to name a few techniques and studies). Correlated noise frameworks allow us to incorporate the uncertainty that arises from the star's magnetic variability in our determination of planet masses, while making minimal assumptions on the exact form of stellar-induced RV signals. Correlated noise techniques provide the most statistically robust mass determinations to date, and are mainstream among the exoplanet community (e.g., Dumusque et al. 2017; Díaz et al. 2016; Damasso et al 2019). Recently, promising frameworks have been developed that harness the power of data-driven machine learning, including analyses in the wavelength domain (Jones et al. 2017; Collier Cameron et al. 2021; Rajpaul et al. 2020; Holzer et al. 2021) that build on Principal Component Analysis spectral analysis (Davis et al. 2017), and neural networks (de Beurs et al. 2020). The use of purely statistical (non-physically motivated) tools on their own cannot currently mitigate stellar variability down to the precision necessary to determine precise masses of small planets on long orbits (e.g., Dumusque et al. 2017, Fig. 13). To break through the variability barrier at the EPRV level, we must, in tandem with applying robust, sophisticated statistical tools, improve our knowledge of stellar surface phenomena and develop novel, physically motivated models to account for stellar RV variability.

**TABLE A-4.** Physical processes known to induce RV variations. The values given are approximate and average estimates. Interpretation of the column headers: "Timescale": typical periodicity of the effect based on solar-type stars; "Lifetime": coherence of the effect based on solar-type stars; "Ideal verification test": observation strategy needed to determine whether affected by this phenomenon in stellar observations; "Current mitigation method": technique currently in use to mitigate this effect; "Uncorrected typical RV peak-to-peak variation" peak-to-peak variation of the RVs based on solar-type stars; "Current best residual error": if a current mitigation method exists, the residual peak-to-peak variation that is left after mitigating; "Expected RV improvement" improvement (actual or anticipated) to be expected if a research and analysis program is pursued successfully.

| Physical Phenomenon | Timescale | Lifetime | Ideal Verification Test | Current Mitigation Method | Uncorrected Typical RV Variation | Current Best Residual Error | Expected RV Improvement | Selection of Relevant Literature (the lists below highlight notable relevant works; they are not intended to be comprehensive.) |
|---|---|---|---|---|---|---|---|---|
| **Granulation** | Over all timescales; mainly seconds to hours | 5–8 minutes | Dense EPRV observations within each night | Simultaneous modelling or other methods, e.g., shape of spectral lines, multi-epoch binning | Few m/s RMS 0.4–0.8 m/s | 0.5 m/s | <10 cm/s | Elsworth et al. 1994; Palle et al. 1995; Dumusque et al. 2011; Meunier et al. 2015, 2019; Cegla et al. 2013, 2018, 2019; Cegla 2019; Sulis et al. 2016, 2017a,b |
| **Supergranulation** | Hours to days | Few days | Dense EPRV observations spanning multiple days using observatories at complementary longitudes | Simultaneous modelling or other methods, e.g., shape of spectral lines, multi-epoch binning | Few m/s RMS 0.3–1.1 m/s | 0.8 m/s | Unknown | Palle et al. 1995; Dumusque et al. 2011; Rieutord & Rincon 2010; Meunier et al. 2015, 2019; Rincon & Rieutord 2018 |
| **Rotationally modulated magnetic variability – faculae/plage** | Rotation period (20–30 days) | 3-4 months | Nightly EPRV observations spanning at least 3–4 rotation periods | Treat as (un-)correlated noise, e.g., Gaussian process-type regression: decorrelate against contemporaneous proxies (spectroscopic, photometric) | 1–10 m/s | 1 m/s | Unknown | Meunier et al. 2010a,b; Aigrain et al. 2012; Haywood et al. 2014; Borgniet et al. 2015; Rajpaul et al. 2015; Davis et al. 2017; Thompson et al. 2017; Wise et al. 2018; Dumusque 2018; Milbourne et al. 2019; Collier Cameron et al. 2019, 2021; Cretignier et al. 2020; ISSI International Team #453 (2018) (https://www.issibern.ch/teams/earthlike world/) |
| **Rotationally modulated magnetic variability – spots** | Rotation period (20–30 days) | 2 weeks – 2 months | Nightly EPRV observations spanning at least 3–4 rotation periods | Treat as (un-)correlated noise, e.g., Gaussian process-type regression: decorrelate against contemporaneous proxies (spectroscopic, chromatic, photometric) | 1–10 m/s | 1 m/s | Unknown | Saar & Donahue 1997; Lagrange et al. 2010; Makarov et al. 2009; Aigrain et al. 2012 |
| **Evershed flows, moat flows, plage inflows** | Rotation and magnetic cycle | Spot lifetime (days) | Observe a planet transiting over a spot or observe a spot while coming on and going off the disc. Doppler imaging at extreme high resolution (~500k) | None | Unknown | Unknown | Unknown | Solanki 2003; Löhner-Böttcher & Schlichenmaier 2013; Braun 2019 |

| Physical Phenomenon | Timescale | Lifetime | Ideal Verification Test | Current Mitigation Method | Uncorrected Typical RV Variation | Current Best Residual Error | Expected RV Improvement | Selection of Relevant Literature (the lists below highlight notable relevant works; they are not intended to be comprehensive.) |
|---|---|---|---|---|---|---|---|---|
| **Meridional flows** | Magnetic cycle (years, decades) | Incoherence over years | Long-term EPRV observations, at least few each month | None | 1–2 m/s | Unknown | Unknown | Komm et al. 1993; Ulrich 2010; Meunier & Lagrange 2020 |
| **Long-term magnetic cycles** | Years, decades | Years, decades | Long-term EPRV observations, at least few each month | Decorrelate with contemporaneous proxies (spectroscopic, chromatic, photometric) | 10 m/s | 1–2 m/s | Unknown | Radick et al. 1983; Lockwood et al. 1984, 2007; Lovis et al. 2011; Meunier et al. 2010a,b; Meunier & Lagrange 2019; Meunier, Lagrange & Cuzacq 2019 |
| **Pulsations – p-modes** | 5 minutes | Few days | Dense spectra for spectroscopic asteroseismology or dense, high-precision photometry | Exposing for ~1–2 pulsation timescales | Few m/s | <10 cm/s | <10 cm/s | Chaplin et al. 2019; further references in review by Cegla 2019 |
| **Pulsations – r-modes** | 10–20 days | Months | Dense spectra for spectroscopic asteroseismology | None | Few m/s | Unknown | Unknown | Lanza et al. 2019 |
| **Flares** | Unpredictable (currently) | Minutes to hours | H alpha, Ca H&K, Ca infrared triplet | Discard data | Few tens cm/s | 0 cm/s | Unknown | Reiners 2009 |
| **Gravitational redshift** | Double effect: magnetic cycle and rotation period | Rotation period and variability cycle | Stellar radius variation measurements to 0.01% accuracy; transit duration variations | None | Few tens cm/s | Unknown | Unknown | Cegla et al. 2012; Loeb et al. 2014 |

### A.2.2.3 The Importance of Solar Observations and Theory

The Sun is the only star whose surface we can image directly at high resolution. It is a unique test case to examine the physical phenomena responsible for RV variability beyond the m/s barrier. For example, since the solar disk is well resolved, it is possible to compare observations of select regions displaying particular variability features with whole-disk integrations (e.g., Dravins, Lindegren & Nordlund 1981; Dravins, Larsson & Nordlund 1986; Cavallini, Ceppatelli & Righini 1985; Cegla et al. 2018; Löhner-Böttcher et al. 2019). Such analyses enable us to test 3D magneto-hydrodynamic (MHD) simulations of solar and stellar surfaces (e.g., Chiavassa et al. 2017; Norris et al. 2017; Bjørgen et al. 2019; Cegla et al. 2019; see further references in Section 4 of the review by Stein 2012). MHD simulations of stellar surfaces will play an important part in developing the next generation of stellar variability models, so it is essential that we test them rigorously against observations. In the case of the Sun, current physically motivated models account for rotationally-modulated RV variations down to ~0.8 m/s (e.g., Milbourne et al. 2019; Meunier, Lagrange & Desort 2019; Dumusque 2018; Herrero et al. 2016). These residuals are an order of magnitude larger than the RV amplitude of an Earth-mass planet in the habitable zone of a Sun-like star. There are still many physical processes, known and unknown (e.g., see discussion in Haywood et al. 2020) that we cannot yet model adequately, both in the Sun and Sun-like stars. We must examine and understand these in detail and incorporate our understanding into exoplanet analysis techniques.

Extensive solar observations have been taken by the HARPS-N spectrograph. The first three years of HARPS-N solar data has recently been released, providing valuable information on solar variability over multi-year timescales (Dumusque et al. 2021) and enabling the development and testing of new approaches to mitigating solar variability (Collier Cameron et al. 2021; de Beurs et al. 2020). Further HARPS-N solar data will be released on a rolling 1-year basis. Other high RV precision, high spectral resolution spectrographs have collected solar spectra over various lengths of time, including: the Göttingen Solar Radial Velocity Project feeding a Fourier transform spectrometer (FTS; Lemke & Reiners 2016); HELIOS feeding HARPS (Dumusque 2019); Potsdam Echelle Polarimetric and Spectroscopic Instrument (PEPSI) at LBT (Strassmeier et al. 2018); the Lowell Solar Telescope feeding EXPRES; and soon a solar feed for NEID. We encourage the development of a solar feed on ESPRESSO as it would provide spatially resolved solar observations with another EPRV instrument. Beyond their use to understand solar variability, solar feeds on EPRV spectrographs are advantageous as they can help identify and characterize instrumental systematics. At present, instrumental precision still precludes the study of solar/stellar variability below the sub-m/s level; hence, it is crucial for the next generation of ultra-stabilized instruments (e.g., EXPRES, NEID, ESPRESSO) to employ solar feeds to diagnose this variability to the 10 cm/s level necessary for an Earth-analog detection.

Complementary spatially resolved solar observations by high-resolution spectrographs could be contributed by other non-EPRV facilities (e.g., the Daniel K. Inouye Solar Telescope (DKIST), Institut für Astrophysik, Göttingen (IAG) Solar Observatory; the Laser Absolute Reference Spectrograph (LARS), Leibniz-Institut für Sonnenphysik (KIS)). While these are clearly quite different from astronomical observations of other stars, they could prove valuable data for refining and validating astrophysical models for solar spectra (e.g., granulation, supergranulation, plage/faculae and active-region flows).

**RECOMMENDATION:** Observing the Sun as a star continuously through multiple EPRV and PRV instruments (e.g., HARPS-N, NEID, EXPRES) should be an immediate high priority. Once public, such data should be made available via the NExScI community archive. The construction and operation of solar feeds for additional high-priority EPRV instruments is highly recommended. NExScI should explore obtaining permission to archive additional solar data from instruments not federally funded (e.g., EXPRES, HARPS, ESPRESSO). The collection, reduction, and analysis of daily solar data for at least one solar cycle from multiple high-priority instruments is critical for a viable EPRV program.

### A.2.2.4 Optimal Instrument and Survey Specifications for Stellar Variability

To extrapolate the lessons learned from solar RV observations to other stars, we must collect high-cadence, high-spectral resolution, and high-SNR observations. Such stellar datasets are crucial to validate and apply our mitigation strategies on stars with different spectral types, metallicities and magnetic variability behaviours. It is important to observe several benchmark stars intensively with multiple instruments to distinguish between astrophysical and instrumental effects.

Our recommendations for instrument and survey specifications are listed in Table A-5. Nightly cadence is required to sample the stellar rotation period and evolution of stellar surface features. We highlight the need for a resolving power >100k and SNR > 300. We recommend high cadence observations of the Ca II H and K lines, as emission in these lines can be a useful variability indicator; measurements of other lines that trace stellar variability may be necessary. Moreover, all EPRV instruments should have a solar feed and share standardized data products (including RVs, cross-correlation functions (CCFs), variability indicators, 1D/2D echellograms, extracted spectra, wavelength solutions, sky spectra, telluric models/masks, CCF masks, and telemetry). We emphasize the need for EPRV teams to engage in global coordination (cf. "Wants" 11, 12, and 14 (Table 3-1); Risks 24 and 25) and for the community to carry out a coordinated precursor survey (cf. "Wants" 1 and 5 (Table 3-2)); see Risk Assessment Table in Appendix D.

**TABLE A-5.** Group D recommendations for instrument and survey specifications.

| | Minimum Requirement | Optimal |
|---|---|---|
| **Cadence** | Nightly | 3x a night |
| **Spectral Resolution R** | 100k | 130–180k |
| **SNR** | 300 at 550 nm | 800–1000 at 550 nm |
| **Activity Indicator** | Ca HK (390 nm) | Ca HK + more |
| **Supplementary Obs.** | Solar telescope | |
| **Call to Action:** | Precursor survey | |
| | Standardized data products | |
| | Plan for global coordination | |
| | Increase R&A Effort | |

**RECOMMENDATION:** In the near term, pursue a program to obtain high-cadence observations of a common set of 4–10 RV-benchmark stars in a coordinated program (including simultaneous RV observations) at multiple PRV facilities around the world. The minimum specification for RV observations is nightly cadence, spectral resolution >100k and SNR > 300. All data products should be made publicly accessible as they will serve to test ground for stellar variability mitigation strategies.

### A.2.2.5 Required Research and Analysis Investment

We quantified the R&A effort required to mitigate stellar variability from a typical level of 200 cm/s down to 5–10 cm/s at the period of the planetary orbit, i.e., the same level of improvement that was assumed across all architectures as outlined in Section 3.2. Our estimates of the necessary person power and computing power are shown in Table A-6. As a community, we must significantly improve our understanding of stellar surface phenomena and their subsequent impact on EPRV observations to determine sufficient stellar mitigation strategies. The computational power estimates are extrapolated from the current literature, and partially scaled from central processing unit (CPU) usage at the Rossland Centre for Solar Physics in Oslo. At present, we do not know how much of the solar/stellar atmosphere must be fully modelled or if common simplifying approximations (e.g., 1D vs. 3D radiative transport, assumptions of local thermal equilibrium) are sufficient, nor do we know how many stellar lines must be analyzed in a full 3D magnetohydrodynamic environment or how well data-driven algorithms will perform at the EPRV level, etc. Hence, the estimates of the required computational power are currently accurate to two orders of magnitude and will evolve over time. The WG identifies R23: *'Not enough staffing to execute program'* as high perceived consequence and high likelihood. The corresponding values used for the budget estimates discussed in the Cost Model Table (Appendix C) are on the lower end of the possible range of estimates.

**RECOMMENDATION:** Immediately launch a long-term, large-scale, multidisciplinary R&A program to understand and model RV solar/stellar variability. Such a program will span observational, numerical, and theoretical fronts, and address the science questions outlined in Table A-3 by order of priority and importance.

The Stellar Variability Analysis Group outlines the following possible pathways to recruiting and retaining an EPRV workforce. We recommend investment in R&A for early career researchers who may work on Guaranteed Time Observations of current and future EPRV instruments, that will be protected for a limited proprietary period, rather than insisting on immediate public data releases. We strongly recommend that funding agencies issue long-term postdoctoral positions of 4–5 years rather than the typical 2–3 years. The long-term duration is necessary not only to sample the longer-term astrophysical processes, but also to provide enough data to develop stellar variability mitigation methods to sufficient fidelity and in sufficient time for in-depth, thorough research and analysis. Long-term funding will improve career and personal stability, thereby increasing retention rates in key knowledge areas of the nascent EPRV field; the WG notes that Risk 25: *'Knowledge retention in the field'* has a high perceived consequence and likelihood (see Risk Assessment Table in Appendix D). Long-term funding will also contribute to creating a more inclusive and diverse work environment (NAS 2018, pp.115–118: Reducing Barriers to Scientific Excellence).

We further recommend that funding agencies create a few long-term management positions dedicated to the oversight of current and future EPRV instruments globally. This will enable EPRV teams to establish and maintain strong, long-lasting synergies across institutions/countries and disciplines (solar, stellar, exoplanetary). Such management will enable the community to have shared, coordinated target lists and observing strategies/cadences amongst the instruments. The WG identified Risk 24: '*Difficulty in funding non-US participants*' as high perceived consequence and likelihood (see Risk Assessment Table in Appendix D). Our suggested level of coordination will maximize the quality of the solar and stellar science that must be undertaken to break the stellar variability barrier in EPRV. These recommendations are in line with Want 11: '*Take* *advantage of opportunities for international collaboration and draw from as broad a pool of relevant expertise as possible,*' and Want 14: '*Encourage free exchange of ideas, including data and source codes.*'

**RECOMMENDATION:** Funding agencies and the community should establish a dedicated, funded Research Coordination Network to address stellar variability, and establish long-term, EPRV-dedicated postdoctoral positions of 4–5 years.

**TABLE A-6.** Estimates of the person power and computing power necessary for a robust stellar variability research and analysis effort.

| Physical Effect/Problem | Current Best Residual RV (m/s) | Typical Timescale (full lifetime is often longer) | Personnel (FTE) | Computational Load (kilo-core-hours) |
|---|---|---|---|---|
| Understanding the Sun better (observation and theory) in connection to EPRV | | | 58 | 100000 |
| Line formation and behavior in the stellar atmosphere in connection to EPRV | | | 20 | 50000 |
| Magnetic fields | several m/s | Weeks to decades | 28 | 1000 |
| Faculae/plage | ~1 m/s | 1 month (stellar rotation period) | 121 | 2000 |
| Spots | ~1 m/s | 1 month (stellar rotation period) | 18.6 | 200000 |
| Evershed flows, moat flows, plage inflows | TBD | 1 month (stellar rotation period) | 7.2 | 50000 |
| Granulation | 0.4–0.8 m/s | Minutes to hours | 15.2 | 200000 |
| Super-granulation | 0.4–0.8 m/s | Hours to 1–2 days | 15.2 | 200000 |
| Meridional flows | TBD | Years to decades | 10 | 100 |
| Long-term magnetic cycles | ~1–2 m/s | Years to decades | 6.4 | 1 |
| Pulsations – p-modes | 0.1 m/s | Minutes | 0.8 | 10 |
| Pulsations – r-modes | <0.5 m/s | 10–20 days | 6.7 | 10000 |
| Flares | n/a | Unpredictable, minutes | 2 | 1 |
| Gravitational redshift | TBD | Months to years | 1.6 | 100 |
| **Total** | | | **310.7** | **813212** |

## A.2.3 References

## A.3 Observing Strategies

The Observing Strategies Analysis Group was tasked with assessing the viability of different survey strategies and proposed architectures in achieving the science goals of the proposed future EPRV program. While all proposed observing architectures proved capable of detecting an Earth analog in the completed simulations (Section 4), these simulations involved many assumptions and idealized conditions that may impact their overall realism. In order to make informed decisions about future Earth analog survey efforts, the EPRV community will need to perform more extensive simulations to determine the relative impact of factors such as facility location, telescope allocation, and scheduling strategy, among others.

Specific areas that require improved treatment in future survey simulation efforts include:

- Development of realistic models of weather losses, including partial night losses and localized weather systems.
- Development of a simulated observing network, where each observing node is aware of what other nodes have observed and can fold that information into their target selection.
- Investigation of the tradeoffs between different survey strategies, considering variations in cadence, number of targets, SNR, etc. Simulations should be performed to inform design choices for EPRV instruments and address questions including:
  - What combination of spectral resolution, SNR, and wavelength coverage maximizes survey efficiency?
  - What level of instrumental stability will be required?
  - How will these choices affect the ability of an EPRV survey to distinguish planets from false positives while controlling the false discovery rate to an acceptable level?
  - How will these choices affect the precision, accuracy, bias, and coverage of planet mass estimates?
- Development of a modular stellar variability simulation tool that allows for the injection of multiple types of stellar variability spanning a number of timescales and amplitudes as identified by the user.
- Future RV measurement simulations should include:
  - A realistic treatment of wavelength regions that are lost to tellurics.
  - Wavelength-dependent instrument throughput profiles.
  - Robust checks of simulated RV precision versus true RV precision from facilities such as HARPS, NEID, EXPRES, and ESPRESSO to ensure that the empirical calculations produce realistic results.

General survey-related recommendations that emerged from the Working Group:

- EPRV facilities should be distributed evenly across the globe to allow access to all EPRV target stars, combat weather losses, and ensure that stars can be observed at least once per night. If candidate signals are detected at one facility, RV observations from the other facilities can then be used to determine whether the signal is Keplerian or due to instrument systematics or observational aliases.
- A list of EPRV standard stars should be identified by the community and these stars should be targeted at least once per night by EPRV facilities in order to develop best practices for combining data from multiple instruments. Intensive (e.g., across an entire night) observing campaigns at different facilities of these standard stars may also be needed (see below).

- Future EPRV facilities will need to either be dedicated or operated in a queue that can switch between facility instruments to meet the high cadence requirements.
- A site survey should be carried out to identify what combinations of existing telescope apertures and locations could feasibly be brought into future EPRV efforts.
- Comparisons should be carried out to identify the relative expense and potential political challenges of building new facilities as compared to updating/upgrading existing facilities. This should include considerations of the cost required to automate and network existing facilities.
- RV observations collected at high cadence may help disentangle purely periodic Keplerian signals from quasi-periodic variability signals in the final RV time-series. Extremely high SNR (~1000 per resolution element) and extremely high resolution (R > 150k) provides a path to disentangling line-shape distortions from achromatic doppler shifts and allow the RVs to be corrected for stellar variability before the RV time-series is generated. Effort should be focused on identifying the turnoff points (if any) in each of these parameter spaces where returns begin to diminish.
- Longer-term funding options (5–10 years) should be encouraged to facilitate more nuanced studies and to keep early career scientists engaged in the field.

## A.4 Data Pipeline and Analysis: From Pixels to Planets

The detection of exoplanets and measurement of their masses from a time series of spectra requires sophisticated data processing and complex statistical models. Most previous research and analysis of RV data has focused on analyzing a time-series of true velocities contaminated by "noise." With EPRV instruments, the dominant source of "noise" will be stellar variability. Since many complex physical processes are occurring simultaneously, residual instrumental systematics, bulk velocity, and stellar variability will need to be modeled simultaneously. Formal statistical models are essential in order to verify that algorithms perform as desired. For example, it will be critical to understand false discovery rates when algorithms are applied to simulated data generated from forward models or to synthetic datasets based on real data.

### A.4.1 Summary Recommendations

The EPRV Working Group found that there are two fundamental challenges in the radial velocity data analysis landscape that must be met to enable detection of low-mass planets. These challenges can be summarized as the following questions:

1. Can we leverage uniform and sophisticated analysis of data from current and upcoming facilities to establish instrumental, algorithmic, and Earth-based limitations in approaching <10 cm/s single measurement precision over multi-year baselines?
2. Can stellar variability be understood well enough to "correct" for its contribution to the spectroscopic measurements of the radial velocity signal?

Each of the recommendations listed in Table A-7 aims to: (1) address these challenges directly, (2) quantify how well the methods being developed work, or (3) explore the implications of progress in mitigating instrumental and stellar variability for the design of future surveys, instruments, and/or facilities.

### A.4.2 Data Processing Gap List Description

A substantial R&A program will be necessary to develop, refine, and evaluate multiple strategies for mitigating various forms of stellar variability. Developing, evaluating, and comparing such strategies will require:

1. Data for training, testing and validating models for stellar variability (see Sections A.4.2.1 and A4.2.3);
2. Statistical models appropriate for spectroscopic time-series with stellar variability (see Section A.4.2.4); and
3. A standardized framework for evaluating the effectiveness of each strategy (see Section A.4.2.5). The impact of various hardware components and design choices of PRV instruments should be determined (see Section A.4.2.2).

Once these elements are assembled, it will be possible to make quantitative comparisons between different stellar variability mitigation strategies and assess whether the EPRV community is ready to proceed with a full survey (see Section A.4.2.6). At that point, it would be important to reevaluate potential EPRV architectures, including observatories, instruments, and observing strategies, in light of new information learned through the above R&A. Finally, it will be necessary for the EPRV community to demonstrate the robustness of detections that rely on a chain of sophisticated analysis processes (see Section A.4.2.7).

**TABLE A-7.** Summary of recommendations for EPRV data pipeline and analyses.

| | |
|---|---|
| Near-Term | **1. EPRV Observations of Sun & Benchmark Stars**<br>Collect, reduce and analyze high-quality observational data for the Sun and a modest number of benchmark stars (4–10) on at least 2–3 high priority* instruments to precisely characterize the instruments and validate algorithms developed to correct instrument systematics and mitigate stellar variability. Data should be publicly archived in a standardized format.<br>• Solar data collection nearly daily for one solar cycle<br>• Support the construction and operation of solar feeds for additional high-priority EPRV instruments<br>* Current "high-priority" instruments include EXPRES, NEID, ESPRESSO, and MAROON-X. This list can be expanded to include other high-resolution (R > 100k) instruments that demonstrate <50 cm/s long-term stability. |
| | **2. Cross-Comparison of Near-Term Instrument Performance**<br>Perform a thorough comparison of existing/upcoming instrument designs and performance:<br>• Maintain engineering expertise ***beyond instrument delivery*** to understand instrument performance characteristics and identify opportunities for improvements in future instrument design.<br>• Perform detailed analyses (by both instrument teams and external groups to ensure independent analyses) of the final instrument performance so that the community can identify successes and lessons learned, and thereby make informed design choices for future EPRV instruments. |
| | **3. Stellar Variability Mitigation**<br>Develop a suite of strategies to allow for accurate and precise estimates of the stellar velocity in the presence of stellar variability. |
| | **4. Statistical Methodology**<br>Develop statistical models appropriate for modeling EPRV data where the dominant source of "noise" will be stellar variability; velocity and stellar variability need to be modeled simultaneously. Understand false discovery rates when applied to simulated data generated from forward models or to synthetic datasets based on real data. |
| | **5. Develop Modular, Open-Source Pipeline for EPRV Science**<br>Establish a community-standard framework that allows researchers to experiment with new RV measurement and interpretation techniques and make meaningful comparisons between current practices and next-generation analysis tools.<br>• Data publicly available at multiple levels (i.e., raw, 1/2D-extracted, dimensionally reduced spectra, CCFs), and as standardized data products in standard formats<br>• Instrument/testbed data on sub-pixel detector properties, calibration stability, etc. for pipeline ingestion<br>• Community pipeline based on heritage of best existing codes and featuring a modular design that includes multiple algorithm choices for key modules |
| Mid-Term | **6. Evaluate Effectiveness of Stellar Variability Mitigation Strategies**<br>Conduct blind tests of the ability to detect planets using different algorithms while controlling the false discovery rate to an acceptable and standardized rate using synthetic datasets. Tests should simultaneously evaluate the precision, accuracy, bias, and coverage of planet mass estimates. |
| | **7. Inform Future Spectrograph and Survey Designs**<br>Assess the effectiveness of: (1) hardware solutions to emergent issues from advanced data analysis; (2) increased number of EPRV observations, spectral resolution, SNR in each spectra, and wavelength coverage; (3) more exacting instrument specifications for distinguishing planets from false positives and measuring their masses. |
| Ongoing /Extended | **8. Series of EPRV Data Challenges**<br>Establish confidence in detections of low-mass planets by conducting data challenges designed to evaluate the effectiveness and reliability of the advanced methods developed.<br>• Formative data challenges for improving ability to extract science from EPRV observations<br>• Summative data challenges for demonstrating the credibility of planet detections and mass measurements<br>Examples include using both simulated and/or real data to compare effectiveness of strategies for mitigation of rotationally modulated signals for sun, granulation, super-granulation, and pulsations for sun, and combined stellar variability for other Sun-like stars. |
| | **9. EPRV Virtual Center or Research Coordination Network**<br>Establish a virtual distributed EPRV center or research network to help the community take a comprehensive and coordinated approach for improving both methodologies and algorithms for analysis of spectroscopic time-series, and the capabilities of current and next-generation EPRV surveys to detect and characterize Earth-mass planets in the habitable zone of Sun-like stars. |

#### A.4.2.1 Validating Models and Methods Developed to Overcome Stellar Variability

As new ideas for mitigating stellar variability are developed, it will be essential to test each method, validate that approaches work on real astrophysical data, and quantify the properties of the residual noise. While data-driven models for modeling spectroscopic variability appear promising, they require large datasets for training, testing, and validation. Given the volume of data required, it is essential that the EPRV community work together to collect and share data for both training and testing purposes.

a. **Sun-as-a-Star PRV & EPRV observations**

Extensive solar observations have been undertaken to date by HARPS-N with this data recently being made publicly available. It provides valuable information about variability over multi-year timescales. Going forward, solar data from other instruments (e.g., NEID, EXPRES) will be particularly valuable due to their increased instrumental stability and complementary strategies for mitigating stellar variability.

b. **Spatially resolved solar EPRV observations**

Complementary spatially resolved solar observations by high-resolution spectrographs could be contributed by other non-EPRV facilities (e.g., Inouye Solar Telescope, Institut für Astrophysik, Göttingen Solar Observatory). While these are clearly quite different from astronomical observations of other stars, they could prove valuable data for refining and validating astrophysical models for solar spectra (e.g., granulation, supergranulation, and pulsations). There was a recent proposal for an ESPRESSO solar feed that would provide spatially resolved solar observations with an EPRV instrument, but it is unclear if or when this project will be completed.

c. **Evaluating algorithms for addressing telluric line contamination**

The algorithms developed to recognize minute changes in the spectral of stars due to stellar variability must work effectively even in the presence of telluric variability. The intensive observations needed to fill gaps Sections A.4.1.1–A.4.1.3 will also provide extensive data about the temporal variability of telluric contamination that can eventually be integrated into validation processes and data challenges.

#### A.4.2.2 Analysis of Data from New Instruments

Several EPRV instruments are, or will soon be, collecting data that will allow for analysis of the impact of various hardware components and design choices. It will be essential to support detailed analyses of the component and system performance of these PRV instruments so as to identify which designs are capable of delivering the required performance and which will need further investment to control/correct systematics, mature hardware, and/or explore alternative approaches.

#### A.4.2.3 Generalize Methods Developed and/or Validated for the Sun to Other Stars

While solar observations are very valuable, they are insufficient for validating methods for detecting and characterizing planets around other stars that differ from the Sun. Given the limited observational resources, it will be necessary to take a two-pronged approach: (1) collect intensive observations of a modest number of RV benchmark stars as soon as possible, and (2) build the capability to simulate the spectroscopic variability of Sun-like stars. These two approaches are complementary, as improved observations will motivate continued improvements in the astrophysical models.

a. **Intensive PRV & EPRV observations of RV benchmark Stars**

While solar observations can provide unparalleled cadence and SNR, there are significant differences between EPRV observations of the Sun and other stars. First, the Sun appears as an extended source resulting in atmospheric effects that are not present in observations of (unresolved) stars. Second, the Earth's eccentric orbit causes the rotation broadening to vary over the course of a year. Additionally, the relative motion of telluric and stellar lines is much smaller for the Sun than for other stars. Given these inevitable differences between EPRV observations of the Sun and other stars, it will be essential to collect very high-cadence, high-resolution, and high-SNR observations of at least a few RV benchmark stars. It is important to make such intensive observations with multiple instruments to help distinguish between astrophysical and instrumental effects. Ideally, the set of benchmark stars would be large enough that we could use these observations to test how well data-driven models can be applied to target stars that differ from the Sun (e.g., temperature, metallicity).

b. **Developing forward modeling for spectroscopic variability of Sun-like stars**

Astrophysical modeling of stellar spectroscopic variability can provide a valuable complement to EPRV observations of the Sun and other stars. The considerable telescope time required to obtain high-cadence, high-resolution, and high-SNR observations of stars other than the Sun motivates the generation of synthetic datasets for training, testing, and validating stellar variability mitigation strategies. While significant work has already begun for starspots and p-mode pulsations, more research is needed for each mechanism. There is even more to be learned about how facula, network, granulation, and long-term magnetic cycles affect stellar spectra. Ideally, forward models would be physically based, but it may also be possible to complement physical models with faster emulators (either physically motivated or data-driven models building on recent progress in machine learning for scientific applications) to construct alternative algorithms for stellar variability.

#### A.4.2.4 Providing Robust EPRV Planet Detections and Accurate Mass Measurements

The gravitational effect of planets on a star's spectroscopic time series can be distinguished from stellar variability in two ways. First, planetary motion induces a pure Doppler shift, while stellar variability affects the spectrum in more complex ways (e.g., individual line shapes). Second, planetary motion (neglecting planet-planet interactions and long-term effects such as orbital decay) is strictly periodic, in contrast to stellar variability which may be stochastic (e.g., pulsations, granulation) or quasi-periodic (e.g., spots and facula rotating across the disk). This naturally leads to two qualitatively different and complementary methods for separating signals due to planets from signals due to stellar variability: wavelength-domain and time-domain. A systematic approach and principled statistical framework(s) for incorporating both is needed.

a. **Wavelength-domain: designing improved stellar variability indicators**

At present, most RV surveys have used variability indicators that were developed for studying stars or were chosen for ease of measurement, rather than for disentangling stellar variability from Doppler shifts. Early research has identified a number of promising stellar variability indicators for RVs. However, much more research is needed to evaluate their utility. Additionally, there is considerable potential that data-driven methods can provide even more powerful indicators for predicting the apparent RV signal due to stellar variability. Again, more research is needed to apply such methods to real data and to understand how they perform as a function of the quality of the data available (e.g., resolution, SNR, number of observations).

b. **Time-domain: jointly modeling planets and variability of stellar spectra**

The simplest path to jointly modeling planetary motion and stellar variability is likely by employing dimensional reduction to convert each spectra into an apparent radial velocity and a set of stellar variability indicators. It then becomes possible to analyze a multivariate time-series for each star, in order to detect planets, distinguish planets from stellar variability, and measure the masses and orbits of such planets.

### A.4.2.5 Assessing Evidence for an *N*th Planet

Given the size of EPRV datasets, one can construct extremely complex statistical models. However, for a model to be practical, the necessary computations must be tractable. While Bayesian parameter estimation (e.g., measuring planet masses and position along its orbit) is likely to be challenging, we anticipate it will not be prohibitive. In contrast, Bayesian model comparison is much more computationally expensive. There will need to be significant research in developing accurate and computationally practical algorithms for evaluating the significance of evidence for an *N*th planet. Based on observations of inner planetary systems by Kepler, we should be prepared to evaluate evidence for up to ~10 planets in a given system.

Until such advances are available, early algorithmic research, survey simulations, and data challenges should focus on the characterization of planet masses, rather than planet discovery. While concerns about aliasing should be addressed prior to commencing a full EPRV survey, we anticipate that the large number of observations required to achieve the desired mass precision will likely result in aliasing being less problematic than in traditional RV surveys.

### A.4.2.6 Meaningful Comparisons of Strategies Require a Community-Standard Framework

Inferring the presence and properties of low-mass planets via EPRV observations will require:

1. Developing data reduction procedures that correct for instrumental effects (e.g., wavelength calibration, line spread function variability, detector properties);
2. Developing stellar spectroscopic indicators that enable the mitigation of stellar variability;
3. Developing a statistical model for spectroscopic time-series that simultaneously models the Doppler shift of planets, stellar variability and telluric variability and
4. Performing inference with high-dimensional models.

As there are multiple pieces of the puzzle, the EPRV community needs to develop a state-of-the-art, community-standard framework for reducing EPRV observations, measuring stellar variability indicators, mitigating stellar variability, and detecting and characterizing planets. Such a pipeline must be modular, well-engineered, open-source, and well-documented. Given the large datasets and complex statistical models that are anticipated, it should also be computationally efficient.

### A.4.2.7 Selecting Strategies to Mitigate Stellar Variability

The EPRV community needs to establish common metrics to assess the performance of algorithms to identify planets in the presence of stellar variability under both idealized and realistic conditions. Only by establishing common routines for low-level data reduction, common statistical models (including priors, likelihoods, and sampling algorithms), and common datasets for training and testing algorithms, will it be practical for different research groups to make apples-to-apples comparisons of their approaches. Researchers can perform their own tests on public datasets that are explicitly intended for making such comparisons (e.g., large simulated EPRV datasets or solar data with planet injections). Periodic evaluation of the state-of-the-art is needed to assess which approaches are promising and merit further effort, and which are ready to be incorporated into the community EPRV pipeline.

#### A.4.2.8 Demonstrating the Credibility of Planet Detections and Mass Measurements Using Data Challenges

It is anticipated that successful algorithms for mitigating stellar variability will be complex. The detection of Earth analogs will require many observations and advanced statistical algorithms. Therefore, it will be challenging for an EPRV survey team to single-handedly build confidence in the most sought-after low-mass planet discoveries, which should ideally undergo many levels of vetting before announcement to the general public. Other large and important scientific projects (e.g., Large Hadron Collider (LHC), Laser Interferometer Gravitational-Wave Observatory (LIGO), Gaia) use strategies such as:

1. Multiple teams independently analyzing the same data,
2. Injecting synthetic signals into data provided to analysis teams, and
3. Blind tests of analysis pipelines.

The EPRV community needs to establish a standard for detecting Earth analogs and demonstrate their readiness via blind data challenges before beginning a full EPRV survey.

#### A.4.2.9 Roadmap for Data Challenges

Early data challenges should focus on addressing specific questions such as:

1. How accurately can wavelength-domain algorithms estimate the injected radial velocities for simulated spectra that include a single type of stellar variability?
2. How does the mass precision depend on the cadence of observations (e.g., one observation each night versus bursts of nights with three observations each night), when the SNR and number of observations are held fixed?
3. How does the claimed false discovery rate of an algorithm compare to the empirical false discovery rate when analyzing an ensemble of simulated datasets generated by injected planets into solar observations?
4. What is the relative effectiveness of strategies for mitigation of: (a) rotationally modulated signals for sun; (b) granulation, super-granulation, and pulsations of the sun; (c) combined stellar variability for other Sun-like stars.

Each data challenge should address specific aspects of the problem, using both simulated and/or real data, so as to compare effectiveness of strategies, learn from each exercise, and improve the state-of-the-art.

The goals of the formative data challenge series are to help the EPRV community improve its ability to extract science from EPRV observations and to enhance the understanding of the capabilities and limitations of EPRV data analysis as a function of key properties of instruments, target stars, and survey strategy. After years of formative challenges, the EPRV community may be ready for one or more summative data challenges to assess how various components integrate together and eventually demonstrate the credibility of discoveries from an upcoming EPRV dream survey. Given the large number of outstanding questions and inevitable interactions of different variables, deciding the optimal approach to challenges, and assessing when the EPRV community is ready for increasingly complex challenges is a substantial undertaking.

a. **Formative data challenges**

Each data challenge will need careful attention to the question(s) being probed, the input data to be provided to participating teams, the outputs that teams are to return, and the metrics by which submissions will be evaluated. After teams submit results to each data challenge, a small team will need to review results of data challenges to perform a meta-analysis of the original questions being probed. The results should be published including a discussion of how the challenge contributed to advancing and/or characterizing the state-of-art and whether/how much further investment is needed to address issues investigated by each data challenge.

b. **Summative data challenges**

Once it appears that there is at least one stellar variability mitigation strategy capable of detecting Earth analogs around Sun-like stars, the EPRV community should engage in blind-tests that ensure the strategy or strategies are robust and have not been over-fit to training data. The false discovery rate for a given pair of stellar variability and statistical models will depend on the data characteristics. Since the quality, number, and cadence of observations will inevitably differ from star to star, extensive simulations will be necessary to accurately characterize the false discovery rate for a realistic EPRV survey. The design of such survey simulations will require considerable care. If done well, the results of such blind tests will inform future investments in instruments, facilities, and algorithms. Additionally, a successful outcome would be a key milestone in deciding when to make the large investments in the facilities required for a full EPRV survey.

c. **Mechanisms for data challenges**

There is a need for an efficient mechanism to enable (1) experimental design and data generation for each data challenge; (2) at least 3–5 teams to participate in each of the EPRV data challenges; (3) meta-analysis of the community responses to each data challenge; and (4) organizing publications, conferences, and/or workshops to help disseminate lessons learned from previous data challenges and/or to help prepare teams for upcoming data challenges.

## A.5 Telluric Line Contamination

When starlight passes through the Earth's atmosphere, telluric lines become convolved with the incident stellar spectrum. This effect can lead to the distortion of spectral lines in stellar spectra, thereby inducing erroneous signals in RV data. Currently, correcting for telluric contamination requires high-accuracy empirical line lists, detailed understanding of molecular opacities, and the integrated telluric line shapes that are imprinted on ground-based astronomical spectra.

The impact of telluric contamination is site specific. Its effect must therefore be considered in any site selections for EPRV surveys/instruments. Some observatories at low elevations and/or high $H_2O$ columns may be excluded from consideration for EPRV surveys, as it may be necessary to restrict $H_2O$ columns to a few mm or less.

Achieving the goals of an EPRV survey will require improved telluric mitigation strategies in the visible and potentially in the NIR. The NIR may be required in addressing stellar variability, particularly for more active targets and for late-type stars. NIR RVs, when measured simultaneously with visible RVs, are a means by which chromatic stellar variability signals, such as spots, can be isolated from any planet signals.

For the past 30 years, NASA has invested in establishing remote Earth-sensing capabilities in the mid- to far-infrared, including the recent GOSAT (Greenhouse gases Observing SATellite) and OCO-2 (Orbiting Carbon Observatory-2) missions, which sense $CH_4$ and $CO_2$ in the 1.6 µm region, and $O_2$ in the 0.76 µm band. While this has advanced our understanding of telluric absorption, further work is necessary to study and ultimately develop strategies to mitigate the impact of telluric line contamination on an EPRV survey focused at visible wavelengths, where $H_2O$ is the primary telluric species.

The key questions driving the mitigation of telluric line contamination of ground-based EPRV datasets are:

1. Can the correction of telluric absorption at visible wavelengths be achieved at a level sufficient for EPRV at the requisite precision using software modeling tools from the atmospheric science community?
2. Can improvements to the software tools make them more applicable to the broadband ground-based spectroscopy problem?
3. Are the existing line lists of sufficient quality for the correction EPRV requires, or is more theoretical or laboratory work necessary? Can data-driven methods remove the need for such line lists?
4. How can telluric correction derived from solar datasets inform the corrections for target EPRV stars?
5. Can data-driven models that rely on ensembles of stellar spectra be utilized across target stars of varying temperature and for data collected across different sites and conditions?

In the near-term, substantial progress can likely be made using existing line lists and code (as is, or slightly expanded). However, eventually, substantial laboratory work may be needed to enhance line lists.

Even with a perfect understanding of the telluric line lists, shapes, and opacities, the removal of temporally varying telluric line contamination from stellar spectra may depend on temporally resolved knowledge of the mixing ratio profile of $H_2O$ at the particular observation site, and the quality of the underlying stellar spectrum. To obtain temporal sampling of atmospheric conditions, EPRV facilities may benefit from strategies including radiosonde launches at a ~6-hour cadence at each observing site that measure pressure, temperature, and relative humidity, and which can then be converted to a local $H_2O$ profile to provide the information for spectrum fitting.[7] Other novel techniques for measuring precipitable water vapor should be investigated (e.g., Blake & Shaw 2011).

Solar data offers a potential strategy for improving telluric correction through model validation. It may, however, have reduced applicability for expanding fitting codes and line lists as these are taken at identical resolutions to RV data, and some aspects of the atmosphere appear spectrally different during the day than at night (e.g., Noxon 1968). Significantly higher resolution spectra are needed to resolve the telluric absorption lines and to make substantial improvements to underlying codes. Observations of standard stars will be required, in addition to solar observations, when developing and testing mitigation techniques.

Standard star datasets provide access to a broader parameter space that overlaps with RV targets and addresses the difference in relative motion of telluric lines compared to stellar lines between the Sun and other stars. Coordinated standard star observing programs with different sites may be beneficial in assessing the validity of modelling tools. These studies and observations should commence in the near-term using solar feeds on existing PRV spectrometers with future expansion to coordinated standard star observations.

**RECOMMENDATION:** To address telluric line contamination in an EPRV survey, extensive laboratory spectroscopic studies should be undertaken in the near-infrared (1–2.5 µm) and visible spectral regions using very long gas absorption cells, high intensity light sources, and cavity ring-down spectroscopy (CRDS) systems to improve line lists and depths. Further, investments should be made to improve input line lists that will be needed for spectral fitting of target stars in the EPRV survey. As an EPRV survey will likely be conducted in the visible band, $H_2O$ line lists are of prime importance. In parallel, efforts should be made to research mitigating tellurics via data driven methods that could complement and/or reduce the need for extensive laboratory spectroscopic study. This includes the collection and analysis of solar and standard star datasets from existing PRV instruments.

### A.5.1 References

[7] Sameshima et al. (2018) established the requirements for empirical telluric correction in terms of air mass and simultaneity in the NIR (900 nm to 1350 nm). They found that time variability of water vapor has a large impact on the accuracy of telluric correction and minimizing the difference in time from that of the telluric standard star is important especially in near-infrared high-resolution spectroscopic studies.

# B EPRV TARGET STARS USED IN SURVEY SIMULATION

## B.1 Star List Source

| | | Starshade Rendezvous (Obj. #1) | Starshade Rendezvous (Obj. #3) | HabEx (4 m) | LUVOIR-B (8 m) | LUVOIR-A (15 m) | Combined List (duplicate targets excluded) |
|---|---|---|---|---|---|---|---|
| Number of stars | | 16 | 16 | 149 | 156 | 283 | 309 |
| Median V magnitude | | 3.46 | 5.72 | 5.04 | 5.31 | 5.60 | 5.56 |
| Median Distance (pc) | | 5.90 | 16.98 | 12.81 | 12.18 | 15.32 | 15.09 |
| Spectral Type | A | 4 | 0 | 4 | 0 | 2 | 7 |
| | F | 2 | 2 | 59 | 51 | 100 | 106 |
| | G | 6 | 8 | 47 | 51 | 96 | 103 |
| | K | 4 | 5 | 32 | 35 | 52 | 59 |
| | M | 0 | 1 | 4 | 16 | 30 | 30 |
| Number of "Sun-like" stars (F7–K9): | | **10** | **14** | **111** | **115** | **196** | **213** |

| | Number of Stars in Combined List | Number of Stars | | | | | | | | |
|---|---|---|---|---|---|---|---|---|---|---|
| | | | vsin(*i*)<5 km/s | | | | vsin(*i*) = 5–10 km/s | | | |
| | | vsin(*i*)<3 km/s | Habex Deep List or 50 Star List | 2+ other Surveys | 1 other Survey | Total | Habex Deep List or 50 Star List | 2+ other Surveys | 1 other Survey | Total |
| "Sun-like stars" (F7–K9) | 213 | 116 | 34 | 72 | 63 | 169 | 4 | 8 | 16 | **28** |
| Stars F7–M | 244 | 136 | 37 | 83 | 74 | 194 | 4 | 10 | 17 | **31** |

- Green targets (106 stars)
  - Spectral types F7–K9 and
  - vsin(*i*) < 5 km/s and
  - On the HabEx 'deep survey' or '50 highest priority stars' lists (Gaudi et al. 2020), or on at least 2 other mission concept target lists (including LUVOIR-A, LUVOIR-B, HabEx 'master list,' Starshade Rendezvous)
- Yellow targets (105 stars)
  - Spectral types F7–M
  - vsin(*i*) 5–10 km/s and/or
  - Appears on only one mission concept list
- Red targets
  - Spectral type hotter than F7 and/or
  - vsin(*i*) > 10 km/s

## B.2 Simulation Star List

| | HIP# | HD# | Distance (pc) | Vmag | vsin(*i*) | $M_{Sun}$ | Spectral Type | $T_{eff}$ | Notes | Planets? |
|---|---|---|---|---|---|---|---|---|---|---|
| 1 | HIP_16537_ | HD_22049_ | 3.216 | 3.72 | 1.9 | 0.78 | K2V | 5050 | active | Y |
| 2 | HIP_104217_ | HD_201092_ | 3.495 | 6.05 | 1.8 | 0.63 | K7V | 4045 | slightly active | N |
| 3 | HIP_104214_ | HD_201091_ | 3.497 | 5.20 | 1.8 | 0.68 | K5V | 4339 | slightly active | N |
| 4 | HIP_108870_ | HD_209100_ | 3.639 | 4.69 | 1.4 | 0.72 | K4V(k) | 4649 | active | N |
| 5 | HIP_8102_ | HD_10700_ | 3.650 | 3.49 | 0.9 | 0.94 | G8V | 5331 | quiet | Y |
| 6 | HIP_49908_ | HD_88230_ | 4.869 | 6.60 | 2.7 | 0.63 | K7V | 4131 | active | N |
| 7 | HIP_19849_ | HD_26965_ | 4.985 | 4.43 | 0.9 | 0.86 | K0.5V | 5202 | slightly active | Y |
| 8 | HIP_88601_ | HD_165341_ | 5.123 | 4.03 | 3.7 | 0.88 | K0-V | 5394 | active | N |
| 9 | HIP_96100_ | HD_185144_ | 5.755 | 4.67 | 1.8 | 0.87 | K0V | 5318 | slightly active | N |
| 10 | HIP_73184_ | HD_131977_ | 5.882 | 5.72 | 3.5 | 0.72 | K4V | 4744 | active | N? |
| 11 | HIP_84478_ | HD_156026_ | 5.950 | 6.33 | 3.3 | 0.68 | K5V(k) | 4600 | active | N |
| 12 | HIP_3821_ | HD_4614_ | 5.953 | 3.46 | 3.4 | 1.14 | F9V | 5904 | | N |
| 13 | HIP_15510_ | HD_20794_ | 6.043 | 4.26 | 0.9 | 0.97 | G6V | 5398 | quiet | Y |
| 14 | HIP_99240_ | HD_190248_ | 6.108 | 3.55 | 2.0 | 0.94 | G8IV | 5566 | quiet | N |
| 15 | HIP_114622_ | HD_219134_ | 6.532 | 5.57 | 1.8 | 0.75 | K3V | 4833 | quiet | Y |
| 16 | HIP_72659_ | HD_131156_ | 6.733 | 4.54 | 3.5 | 0.96 | G7V | 5527 | active | N |
| 17 | HIP_12114_ | HD_16160_ | 7.235 | 5.79 | 1.3 | 0.75 | K3V | 4829 | quiet | N |
| 18 | HIP_3765_ | HD_4628_ | 7.435 | 5.74 | 1.8 | 0.78 | K2V | 5015 | slightly active | N |
| 19 | HIP_2021_ | HD_2151_ | 7.459 | 2.82 | 3.4 | 1.08 | G0V | 5873 | quiet | N |
| 20 | HIP_7981_ | HD_10476_ | 7.605 | 5.24 | 0.1 | 0.85 | K1V | 5196 | quiet | N |
| 21 | HIP_113283_ | HD_216803_ | 7.608 | 6.48 | 2.6 | 0.72 | K4Ve | 4555 | active | N |
| 22 | HIP_85295_ | HD_157881_ | 7.715 | 7.54 | 3.5 | 0.63 | K7V | 3941 | active | N |
| 23 | HIP_61317_ | HD_109358_ | 8.440 | 4.24 | 2.8 | 1.08 | G0V | 5887 | quiet | N |
| 24 | HIP_64924_ | HD_115617_ | 8.555 | 4.74 | 1.8 | 0.97 | G6.5V | 5537 | quiet | Y |
| 25 | HIP_1599_ | HD_1581_ | 8.587 | 4.23 | 4.9 | 1.11 | F9.5V | 5977 | | N |
| 26 | HIP_32984_ | HD_50281_ | 8.749 | 6.58 | 2.7 | 0.79 | K3.5V | 4758 | | N |
| 27 | HIP_99825_ | HD_192310_ | 8.799 | 5.73 | 0.6 | 0.77 | K2+V | 5104 | quiet | Y |
| 28 | HIP_23311_ | HD_32147_ | 8.848 | 6.22 | 1.4 | 0.75 | K3+V | 4745 | quiet | N |
| 29 | HIP_17378_ | HD_23249_ | 9.041 | 3.52 | 1.0 | 0.87 | K0+IV | 5144 | quiet | N |
| 30 | HIP_64394_ | HD_114710_ | 9.129 | 4.23 | 4.5 | 1.11 | F9.5V | 6034 | active | N |
| 31 | HIP_15457_ | HD_20630_ | 9.140 | 4.84 | 4.5 | 0.98 | G5V | 5749 | active | N |
| 32 | HIP_105858_ | HD_203608_ | 9.262 | 4.21 | 3.4 | 1.14 | F9V_Fe-1.4_CH-0.7 | 6150 | active | N |
| 33 | HIP_57443_ | HD_102365_ | 9.292 | 4.89 | 2.7 | 1.02 | G2V | 5655 | quiet | Y |
| 34 | HIP_56452_ | HD_100623_ | 9.544 | 5.96 | 0.9 | 0.88 | K0-V | 5241 | quiet | N |
| 35 | HIP_56997_ | HD_101501_ | 9.579 | 5.31 | 2.3 | 0.94 | G8V | 5528 | active | N |
| 36 | HIP_81300_ | HD_149661_ | 9.920 | 5.77 | 1.6 | 0.87 | K0V(k) | 5248 | active | N |
| 37 | HIP_8362_ | HD_10780_ | 10.043 | 5.63 | 0.9 | 0.90 | G9V | 5354 | active | N |
| 38 | HIP_68184_ | HD_122064_ | 10.078 | 6.49 | _ | 0.75 | K3V | 4851 | quiet | N |
| 39 | HIP_29271_ | HD_43834_ | 10.215 | 5.08 | 2.3 | 0.96 | G7V | 5569 | quiet | N |
| 40 | HIP_14632_ | HD_19373_ | 10.541 | 4.05 | 3.6 | 1.08 | G0V | 5968 | quiet | N |
| 41 | HIP_10138_ | HD_13445_ | 10.787 | 6.12 | 1.9 | 0.82 | K1.5V | 5217 | active | Y |
| 42 | HIP_57757_ | HD_102870_ | 10.929 | 3.59 | 3.6 | 1.14 | F9V | 6083 | | N |
| 43 | HIP_64797_ | HD_115404_ | 10.985 | 6.49 | 3.3 | 0.76 | K2.5V(k) | 5081 | active | N |
| 44 | HIP_86400_ | HD_160346_ | 11.000 | 6.53 | 1.5 | 0.75 | K3-V | 4808 | slightly active | N |
| 45 | HIP_10644_ | HD_13974_ | 11.008 | 4.84 | 2.0 | 1.08 | G0.5V_Fe-0.5 | 5786 | active | N |
| 46 | HIP_88972_ | HD_166620_ | 11.096 | 6.38 | 0.6 | 0.78 | K2V | 5048 | quiet | N |
| 47 | HIP_3093_ | HD_3651_ | 11.137 | 5.88 | 1.8 | 0.86 | K0.5V | 5303 | quiet | Y |
| 48 | HIP_42808_ | HD_74576_ | 11.186 | 6.58 | 2.7 | 0.76 | K2.5V(k) | 5005 | active | N |
| 49 | HIP_47080_ | HD_82885_ | 11.203 | 5.40 | 2.3 | 0.94 | G8Va | 5511 | active | N |
| 50 | HIP_48331_ | HD_85512_ | 11.285 | 7.67 | 0.9 | 0.65 | K6V(k) | 4400 | active | Y |
| 51 | HIP_72848_ | HD_131511_ | 11.510 | 6.00 | 3.9 | 0.86 | K0.5V | 5291 | active | N |

| | HIP# | HD# | Distance (pc) | Vmag | vsin(*i*) | $M_{Sun}$ | Spectral Type | $T_{eff}$ | Notes | Planets? |
|---|---|---|---|---|---|---|---|---|---|---|
| 52 | HIP_77257_ | HD_141004_ | 11.819 | 4.42 | 3.3 | 1.09 | G0-V | 5900 | quiet | N |
| 53 | HIP_69972_ | HD_125072_ | 11.841 | 6.66 | 0.9 | 0.83 | K3IV | 4903 | quiet | N |
| 54 | HIP_15330_ | HD_20766_ | 12.039 | 5.53 | 2.7 | 1.01 | G2.5V_Hdel1 | 5712 | active | N |
| 55 | HIP_15371_ | HD_20807_ | 12.046 | 5.24 | 2.7 | 1.07 | G1V | 5852 | slightly active | N |
| 56 | HIP_80686_ | HD_147584_ | 12.177 | 4.90 | 2.4 | 1.14 | F9V | 6030 | active | N |
| 57 | HIP_41926_ | HD_72673_ | 12.182 | 6.38 | 1.8 | 0.85 | K1V | 5243 | quiet | N |
| 58 | HIP_40693_ | HD_69830_ | 12.564 | 5.95 | 2.2 | 0.93 | G8+V | 5442 | quiet | Y |
| 59 | HIP_43587_ | HD_75732_ | 12.590 | 5.96 | 2.2 | 0.95 | K0IV-V | 5270 | quiet | Y |
| 60 | HIP_58576_ | HD_104304_ | 12.697 | 5.54 | 1.8 | 0.94 | G8IV | 5510 | quiet | N |
| 61 | HIP_85235_ | HD_158633_ | 12.793 | 6.44 | 1.3 | 0.87 | K0V | 5327 | quiet | N |
| 62 | HIP_10798_ | HD_14412_ | 12.834 | 6.33 | 2.7 | 0.94 | G8V | 5476 | slightly active | N |
| 63 | HIP_80337_ | HD_147513_ | 12.908 | 5.37 | 2.2 | 1.07 | G1V_CH-0.4 | 5858 | active | Y |
| 64 | HIP_22263_ | HD_30495_ | 13.241 | 5.49 | 2.9 | 1.04 | G1.5V_CH-0.5 | 5840 | active | N |
| 65 | HIP_98036_ | HD_188512_ | 13.699 | 3.71 | 2.7 | 0.94 | G8IV | 5223 | quiet | N |
| 66 | HIP_544_ | HD_166_ | 13.779 | 6.07 | 3.6 | 0.94 | G8V | 5458 | active | N |
| 67 | HIP_53721_ | HD_95128_ | 13.802 | 5.03 | 3.1 | 1.07 | G1-V_Fe-0.5 | 5894 | quiet | Y |
| 68 | HIP_17420_ | HD_23356_ | 13.955 | 7.10 | 3.0 | 0.76 | K2.5V | 4930 | slightly active | N |
| 69 | HIP_16852_ | HD_22484_ | 13.963 | 4.29 | 3.7 | 1.14 | F9IV-V | 5971 | quiet | N |
| 70 | HIP_97944_ | HD_188088_ | 14.107 | 6.22 | 1.8 | 0.90 | K2IV(k) | 4600 | active | N |
| 71 | HIP_79672_ | HD_146233_ | 14.131 | 5.49 | 2.7 | 1.02 | G2Va | 5814 | quiet | N |
| 72 | HIP_102422_ | HD_198149_ | 14.265 | 3.41 | 1.7 | 1.23 | K0IV | 5002 | subgiant | N |
| 73 | HIP_101997_ | HD_196761_ | 14.677 | 6.36 | 1.8 | 0.95 | G7.5IV-V | 5414 | quiet | N |
| 74 | HIP_75181_ | HD_136352_ | 14.688 | 5.65 | 2.4 | 1.02 | G2-V | 5664 | quiet | Y |
| 75 | HIP_49081_ | HD_86728_ | 14.926 | 5.37 | 1.8 | 1.00 | G3Va_Hdel1 | 5753 | quiet | N |
| 76 | HIP_95447_ | HD_182572_ | 14.959 | 5.17 | 1.9 | 0.95 | G7IV_Hdel1 | 5530 | quiet | N |
| 77 | HIP_5862_ | HD_7570_ | 15.177 | 4.97 | 4.7 | 1.14 | F9V_Fe+0.4 | 6111 | quiet | N |
| 78 | HIP_27435_ | HD_38858_ | 15.255 | 5.97 | 2.7 | 1.02 | G2V | 5733 | slightly active | Y? |
| 79 | HIP_77358_ | HD_140901_ | 15.260 | 6.01 | 1.8 | 0.96 | G7IV-V | 5584 | slightly active | N |
| 80 | HIP_107649_ | HD_207129_ | 15.561 | 5.57 | 1.8 | 1.08 | G0V_Fe+0.4 | 5946 | slightly active | N |
| 81 | HIP_86796_ | HD_160691_ | 15.605 | 5.12 | 3.8 | 1.00 | G3IV-V | 5845 | quiet | Y |
| 82 | HIP_88745_ | HD_165908_ | 15.739 | 5.05 | 2.8 | 1.14 | F9V_mw | 6049 | quiet | N |
| 83 | HIP_77760_ | HD_142373_ | 15.832 | 4.60 | 3.4 | 1.08 | G0V_Fe-0.8_CH-0.5 | 5776 | quiet | N |
| 84 | HIP_3909_ | HD_4813_ | 15.880 | 5.17 | 3.9 | 1.21 | F7V | 6203 | hot | N |
| 85 | HIP_98767_ | HD_190360_ | 16.014 | 5.73 | 0.8 | 0.96 | G7IV-V | 5606 | quiet | Y |
| 86 | HIP_32480_ | HD_48682_ | 16.648 | 5.24 | 3.6 | 1.14 | F9V | 6064 | | N |
| 87 | HIP_43726_ | HD_76151_ | 16.850 | 6.01 | 2.4 | 1.02 | G2V | 5781 | active | N |
| 88 | HIP_35136_ | HD_55575_ | 16.867 | 5.54 | 2.9 | 1.14 | F9V | 5849 | quiet | N |
| 89 | HIP_33277_ | HD_50692_ | 17.470 | 5.74 | 2.9 | 1.08 | G0V | 5891 | quiet | N |
| 90 | HIP_62207_ | HD_110897_ | 17.565 | 5.95 | 1.8 | 1.14 | F9V_Fe-0.3 | 5842 | quiet | N |
| 91 | HIP_89042_ | HD_165499_ | 17.753 | 5.47 | 4.2 | 1.08 | G0V | 5950 | | N |
| 92 | HIP_65721_ | HD_117176_ | 17.910 | 4.97 | 3.0 | 0.99 | G4V-IV | 5559 | quiet | Y |
| 93 | HIP_79248_ | HD_145675_ | 17.942 | 6.61 | 2.6 | 0.87 | K0V | 5388 | quiet | Y |
| 94 | HIP_910_ | HD_693_ | 17.990 | 4.89 | 4.8 | 1.18 | F8V_Fe-0.8_CH-0.5 | 6169 | quiet | N |
| 95 | HIP_26394_ | HD_39091_ | 18.280 | 5.65 | 2.7 | 1.08 | G0V | 6003 | quiet | Y |
| 96 | HIP_83389_ | HD_154345_ | 18.294 | 6.76 | 1.8 | 0.94 | G8V | 5442 | slightly active | Y |
| 97 | HIP_48113_ | HD_84737_ | 18.904 | 5.08 | 2.9 | 1.08 | G0.5Va | 5872 | quiet | N |
| 98 | HIP_45333_ | HD_79028_ | 19.659 | 5.18 | 4.8 | 1.08 | G0IV-V | 5973 | quiet | N |
| 99 | HIP_64408_ | HD_114613_ | 20.295 | 4.85 | 2.7 | 0.99 | G4IV | 5670 | quiet | Y |
| 100 | HIP_110649_ | HD_212330_ | 20.454 | 5.31 | 1.8 | 1.02 | G2IV-V | 5739 | quiet | N |
| 101 | HIP_37606_ | HD_62644_ | 22.473 | 5.04 | 4.3 | 0.94 | G8IV-V | 5526 | quiet | N |

# C Cost Model

## C.1 Common Costs Across Architectures

| Phase 1 (2021–2025) | | | |
|---|---|---|---|
| **Item** | **Estimated Cost** | | **Assumptions** |
| Observations with EPRV Instruments | 3375 | k$ | Current generation of instruments |
| Pipeline Development | 1600 | k$ | |
| Data Challenges | 2400 | k$ | |
| Variability R&D | 22737 | k$ | |
| Variability Mitigation | 3900 | k$ | |
| Meetings, etc. | 375 | k$ | |
| Tellurics | 3750 | k$ | |
| Architecture 0b | 11900 | k$ | Baseline for all other architectures |
| Auxiliary Surveys | 10938 | k$ | Photometry, archival, high-resolution imaging, and archival |
| Instrument R&D | 32750 | k$ | From Instrument R&D tab |
| **Total** | **93724** | **k$** | |

| Phase 2 (2026–2030) | | | |
|---|---|---|---|
| **Item** | **Estimated Cost** | | **Assumptions** |
| Observations with EPRV Instruments | 1350 | k$ | |
| Pipeline Development | 2000 | k$ | |
| Data Challenges | 2225 | k$ | |
| Variability R&D | 14731 | k$ | |
| Variability Mitigation | 2175 | k$ | |
| Meetings, etc. | 375 | k$ | |
| Unknowns | 5000 | k$ | Item to address unknowns from phase 1 |
| Architecture 0b | 14250 | k$ | Baseline for all other architectures |
| Auxiliary Surveys | 10375 | k$ | Photometry, archival, high-resolution imaging and archival |
| Total | 52481 | k$ | |

| Phase 3 (2031+) | | | |
|---|---|---|---|
| **Item** | **Estimated Cost** | | **Assumptions** |
| Pipeline Development | 2000 | k$ | |
| Data Challenges | 2225 | k$ | |
| Variability R&D | 9464 | k$ | |
| Meetings etc. | 375 | k$ | |
| Unknowns | 5000 | k$ | Item to address unknowns from phase 1 |
| **Total** | **19064** | **k$** | |

| | | |
|---|---|---|
| **GRAND TOTAL** | **165268.6** | **k$** |

## C.2 Cost of Resources Adopted in Architecture Assessment

| Item | Estimated Cost | | Assumptions | | |
|---|---|---|---|---|---|
| **Staff** | | | | | |
| PhD Student | 75 | k$ / yr | | | |
| Postdoc | 150 | k$ / yr | | | |
| Engineer | 250 | k$ / yr | | | |
| Faculty | 350 | k$ / yr | | | |
| **Spectrograph Operations** | | | | | |
| Seeing-Limited Spectrograph | 1 | k$ / night | Includes maintenance cost, could be higher with LFC | | |
| Single-Mode Spectrograph | 1 | k$ / night | Includes maintenance, could be higher with LFC | | |
| **Spectrograph Development** | | | | | |
| Seeing-Limited Spectrograph | 10000 | k$ | R&D, prototyping, and non-recurrent engineering | | |
| Single-Mode Spectrograph | 5000 | k$ | R&D, prototyping, and non-recurrent engineering | | |
| **Spectrograph Construction** | | | | | |
| Seeing-Limited Spectrograph | 8000 | k$ | Manufacturing, assembly, integration and test (MAIT) only, no non-recurring engineering (NRE) | | |
| Single-mode Spectrograph | 3000 | k$ | MAIT only, no NRE | | |
| **Telescope Operations** | | | | | |
| 0.7 m Telescope | 0.4 | k$ / night | 250 nights/yr, operating cost | 0.1 | M$/yr |
| 1.0 m Telescope | 0.4 | k$ / night | 250 nights/yr, operating cost | 0.1 | M$/yr |
| 2.4 m Telescope | 4 | k$ / night | 250 nights/yr, operating cost | 1 | M$/yr |
| 4 m Telescope | 10 | k$ / night | 250 nights/yr, operating cost | 2.5 | M$/yr |
| 6 m Telescope | 20 | k$ / night | 250 nights/yr, operating cost | 5 | M$/yr |
| 10 m Telescope | 40 | k$ / night | 250 nights/yr, operating cost | 10 | M$/yr |
| 25 m Telescope | 200 | k$ / night | 250 nights/yr, operating cost | 50 | M$/yr |
| **Telescope Development** | | | | | |
| 2.4 m Telescope | 3000 | k$ | R&D, prototyping, and non-recurrent engineering | | |
| 4 m Telescope | 10000 | k$ | R&D, prototyping, and non-recurrent engineering | | |
| **Telescope Construction** | | | | | |
| 0.7 m Telescope | 350 | k$ | Commercial-off-the-shelf (COTS), includes dome and equipment, but not site development | | |
| 1.0 m Telescope | 800 | k$ | COTS, includes dome and equipment, but not site development | | |
| 2.4 m Telescope | 7000 | k$ | MAIT only, no NRE, highly dependent on site | | |
| 4 m Telescope | 40000 | k$ | MAIT only, no NRE, highly dependent on site | | |
| 10 m Telescope | 200000 | k$ | MAIT only, no NRE, highly dependent on site | | |
| Existing Telescope Refurbishment | 2000 | k$ | | | |
| Visible-Light AO for 2.4 m Telescope | 3000 | k$ | | | |
| **Survey Development** | | | | | |
| Data Reduction Pipeline | 3000 | k$ | | | |
| **Survey Operations** | | | | | |
| Single Telescope | 5 | k$ / night | Includes scheduling and data reduction | | |
| 6-Telescope Network | 10 | k$ / night | Includes scheduling and data reduction | | |
| **Computing** | | | | | |
| CPU time | 0.05 | k$ / k-core-hr | Likely value varies from 0.03 to 0.1 | | |

# D Risk Assessment Table

This table shows the risk items and assigned/perceived consequence (“C”) and likelihood (“L”) for each of the EPRV architectures in this study.

| Architecture | | | | I | | | II | | | IV | | | V | | | VI | | | VIII | | |
|---|---|---|---|---|---|---|---|---|---|---|---|---|---|---|---|---|---|---|---|---|---|
| Architecture Description | | | | 6×2.4 m | | | 6×4 m | | | VIIIb + 2×25 m | | | 6×3 m (AO) | | | 6×1 m Arrays | | | Hybrid | | |
| Risk | Risk Description | Key | Driving | C | L | | C | L | | C | L | | C | L | | C | L | | C | L | |
| Key and Driving Risks | | | | | | | | | | | | | | | | | | | | | |
| R1 | Can't get enough/desired observing time/cadence/schedule | K | D | 5 | 1 | 5 | 5 | 1 | 5 | 5 | 5 | 25 | 5 | 1 | 5 | 5 | 1 | 5 | 5 | 1 | 5 |
| R2 | Photon limited | K | D | 5 | 3 | 15 | 3 | 1 | 3 | 3 | 1 | 3 | 5 | 3 | 15 | 5 | 3 | 15 | 3 | 1 | 3 |
| R3 | LUVOIR/HabEx not selected | | | 4 | 2 | 8 | 4 | 2 | 8 | 2 | 2 | 4 | 2 | 2 | 4 | 4 | 2 | 8 | 4 | 2 | 8 |
| R4 | Cannot meet schedule | K | D | 3 | 2 | 6 | 3 | 3 | 9 | 3 | 5 | 15 | 3 | 3 | 9 | 3 | 3 | 9 | 3 | 3 | 9 |
| R5 | Upgrading/repurposing of existing facilities results in more work time, challenges to implementation | | | 3 | 4 | 12 | 3 | 4 | 12 | 3 | 4 | 12 | 3 | 4 | 12 | 1 | 1 | 1 | 3 | 4 | 12 |
| R6 | GMT cost risk and TMT location uncertainty for large aperture options | K | D | 1 | 1 | 1 | 1 | 1 | 1 | 5 | 3 | 15 | 1 | 1 | 1 | 1 | 1 | 1 | 1 | 1 | 1 |
| R7 | Non-robotic operations of telescopes impacts cost, staffing, uniformity | | | 3 | 3 | 9 | 4 | 3 | 12 | 4 | 3 | 12 | 4 | 3 | 12 | 5 | 1 | 5 | 4 | 3 | 12 |
| R8 | AO performance in visible getting below 600 nm, below 500 nm increasingly difficult; need coverage at shorter wavelengths | K | D | 1 | 1 | 1 | 1 | 1 | 1 | 1 | 1 | 1 | 5 | 3 | 15 | 1 | 1 | 1 | 1 | 1 | 1 |
| R9 | Slicing on high resolution, large aperture options, equivalent to many small telescopes (e.g., Minerva but then higher read noise) | K | D | 1 | 1 | 1 | 3 | 2 | 6 | 5 | 2 | 10 | 1 | 1 | 1 | 5 | 3 | 15 | 5 | 2 | 10 |
| R10 | Long integration times and imperfect characterization of system throughput --> barycentric correction challenge | | | 1 | 1 | 1 | 1 | 1 | 1 | 1 | 1 | 1 | 3 | 2 | 6 | 1 | 1 | 1 | 1 | 1 | 1 |
| R11 | Requires new technology not demonstrated in allocated time frame | | | 1 | 1 | 1 | 1 | 1 | 1 | 4 | 2 | 8 | 4 | 3 | 12 | 1 | 1 | 1 | 1 | 1 | 1 |
| R12 | Extrapolation of technologies from Architecture 0 to other architectures may not be valid | K | D | 1 | 1 | 1 | 2 | 2 | 4 | 3 | 3 | 9 | 4 | 4 | 16 | 2 | 2 | 4 | 2 | 2 | 4 |
| R13 | Unlikely to obtain high enough SNR or high enough resolution spectra for science goals | K | D | 5 | 4 | 20 | 5 | 2 | 10 | 5 | 3 | 15 | 5 | 2 | 10 | 5 | 4 | 20 | 5 | 3 | 15 |
| R14 | Unrealistic system efficiency estimation compared to what was submitted | | | 4 | 2 | 8 | 4 | 3 | 12 | 4 | 3 | 12 | 4 | 3 | 12 | 4 | 3 | 12 | 4 | 3 | 12 |
| R15 | Telluric correction in NIR is much worse (> ~900 nm) | | | 1 | 1 | 1 | 1 | 1 | 1 | 2 | 3 | 6 | 3 | 3 | 9 | 1 | 1 | 1 | 1 | 1 | 1 |
| R16 | Lack of broad spectral coverage impacts stellar variability mitigation | | | 3 | 1 | 3 | 4 | 1 | 4 | 3 | 1 | 3 | 4 | 2 | 8 | 3 | 1 | 3 | 4 | 1 | 4 |
| R17 | Lessons learned have to be applied to architecture for success | | | 2 | 1 | 2 | 2 | 1 | 2 | 3 | 2 | 6 | 4 | 3 | 12 | 4 | 3 | 12 | 3 | 3 | 9 |
| R18 | Availability of components from at-risk, sole-source supplier | K | D | 5 | 3 | 15 | 5 | 3 | 15 | 5 | 3 | 15 | 5 | 2 | 10 | 5 | 3 | 15 | 5 | 3 | 15 |

| Architecture | | | | I | | | II | | | IV | | | V | | | VI | | | VIII | | |
|---|---|---|---|---|---|---|---|---|---|---|---|---|---|---|---|---|---|---|---|---|---|
| Architecture Description | | | | 6×2.4 m | | | 6×4 m | | | VIIIb + 2×25 m | | | 6×3 m (AO) | | | 6×1 m Arrays | | | Hybrid | | |
| **Risk** | **Risk Description** | **Key** | **Driving** | **C** | **L** | | **C** | **L** | | **C** | **L** | | **C** | **L** | | **C** | **L** | | **C** | **L** | |
| R19 | Requirement to build new telescopes | K | D | 5 | 3 | 15 | 5 | 4 | 20 | 5 | 4 | 20 | 5 | 3 | 15 | 5 | 2 | 10 | 5 | 4 | 20 |
| R20 | Coordination between different telescope facilities problematic | K | D | 3 | 1 | 3 | 3 | 4 | 12 | 3 | 4 | 12 | 3 | 2 | 6 | 3 | 1 | 3 | 3 | 4 | 12 |
| **Project Risks Common to All Architectures** | | | | | | | | | | | | | | | | | | | | | |
| R21 | Sun's variability is not representative of target stars in list/stellar variability cannot be adequately subtracted | K | | 5 | 3 | 15 | 5 | 3 | 15 | 5 | 3 | 15 | 5 | 3 | 15 | 5 | 3 | 15 | 5 | 3 | 15 |
| R22 | Telluric line contamination cannot be adequately mitigated | | | 4 | 2 | 8 | 4 | 2 | 8 | 4 | 2 | 8 | 4 | 3 | 12 | 4 | 2 | 8 | 4 | 2 | 8 |
| R23 | Not enough staffing to execute program | K | | 5 | 3 | 15 | 5 | 3 | 15 | 5 | 3 | 15 | 5 | 3 | 15 | 5 | 3 | 15 | 5 | 3 | 15 |
| R24 | Difficulty in funding non-U.S. participants | K | | 5 | 5 | 25 | 5 | 5 | 25 | 5 | 5 | 25 | 5 | 5 | 25 | 5 | 5 | 25 | 5 | 5 | 25 |
| R25 | Knowledge retention in the field | K | | 5 | 5 | 25 | 5 | 5 | 25 | 5 | 5 | 25 | 5 | 5 | 25 | 5 | 5 | 25 | 5 | 5 | 25 |
| | | | **SUM** | **216** | | | **227** | | | **292** | | | **282** | | | **230** | | | **243** | | |

# E Acronyms

| | |
|---|---|
| ADC | Atmospheric Dispersion Correction |
| ADU | Analog-to-Digital Unit |
| AO | Adaptive Optics |
| APF | Automated Planet Finder |
| AWG | Arrayed Waveguide Grating |
| BIS | Bisector |
| BW | Beam Width |
| Caltech | California Institute of Technology |
| CARMENES | Calar Alto high-Resolution search for M dwarfs with Exoearths with Near-infrared and optical Échelle Spectrographs |
| CCD | Charge-Coupled Device |
| CCF | Cross-Correlation Function |
| CFHT | Canada France Hawaii Telescope |
| CHEOPS | CHaracterising ExOPlanets Satellite |
| CMOS | Complementary Metal-Oxide Semiconductor |
| CoRoT | Convection, Rotation and planetary Transits |
| COTS | Commercial-off-the-Shelf |
| CPU | Central Processing Unit |
| CRDS | Cavity Ring-Down Spectroscopy |
| CTE | Charge Transfer Efficiency |
| DKIST | Daniel K. Inouye Solar Telescope |
| DM | Deformable Mirror |
| ELT | Extremely Large Telescope |
| EMCCD | Electron Multiplying Charge Coupled Device |
| EO | Electro-Optic |
| EOM | Electro-Optic Modulation |
| EPRV | Extreme Precision Radial Velocity |
| ESA | European Space Agency |
| ESPRESSO | Echelle SPectrograph for Rocky Exoplanets and Stable Spectroscopic Observations |
| ESS | Exoplanet Science Strategy |
| ExEP | Exoplanet Exploration Program |
| ExoPAG | Exoplanet Exploration Program Analysis Group |
| EXPRES | Extreme Precision Spectrometer |
| FP | Fabry–Pérot |
| FTS | Fourier Transform Spectrometer |
| FWHM | Full-Width at Half-Maximum |
| G-CLEF | GMT-Consortium Large Earth Finder |
| GMT | Giant Magellan Telescope |
| GO | Guest Observer |
| GOSAT | Greenhouse gases Observing SATellite |
| GSFC | Goddard Space Flight Center |
| GTO | Guaranteed Time Observer |
| HabEX | Habitable Exoplanet Observatory |
| HARPS | High Accuracy Radial velocity Planet Searcher |
| HARPS3 | High Accuracy Radial velocity Planet Searcher 3 |
| HARPS-N | High Accuracy Radial velocity Planet Searcher for the Northern hemisphere |
| HCL | Hollow Cathode Lamp |
| HD | Henry Draper (Catalogue) |
| HELIOS | HARPS Experiment for Light Integrated Over the Sun |
| HIP | Hipparcos (catalogue) |
| HIRES | High-Resolution Spectrograph |
| HISPEC | High-resolution Infrared Spectrograph for Exoplanet Characterization |
| HMI | Helioseismic and Magnetic Imager |
| HPF | Habitable Zone Planet Finder |
| HZ | Habitable Zone |
| IAG | Institut für Astrophysik, Göttingen |
| ISSI | International Space Science Institute |
| JPL | Jet Propulsion Laboratory |
| KIS | Leibniz-Institut für Sonnenphysik |
| KPF | Keck Planet Finder |
| LARS | Laser Absolute Reference Spectrograph |
| LBT | Large Binocular Telescope |
| LFC | Laser Frequency Comb |
| LHC | Large Hadron Collider |
| LIGO | Laser Interferometer Gravitational-Wave Observatory |
| LRD | Launch Readiness Date |
| LUVOIR | Large Ultraviolet Optical Infrared Surveyor |
| MagAO-X | Extreme AO system for the Magellan Clay 6.5 m telescope |
| MAIT | Manufacturing, Assembly, Integration and Test |
| MAROON-X | M-dwarf Advanced Radial velocity Observer Of Neighboring eXoplanets |
| MHD | Magneto-Hydrodynamic |
| MIDEX | Medium-class Explorer |
| MINERVA | MINiature Exoplanet Radial Velocity Array |
| MODHIS | Multi-Object Diffraction-limited High-Resolution Infrared Spectrograph |
| MuRAM | Max Planck Institute for Solar System Research (MPS) /University of Chicago Radiative Magnetohydrodynamics |
| NAS | National Academy of Sciences |
| NASA | National Aeronautics and Space Administration |
| NCCR | National Centre for Competence in Research |
| NEID | NN-EXPLORE Exoplanet Investigations with Doppler spectroscopy |
| NExScI | NASA Exoplanet Science Institute |
| NIR | Near-Infrared |
| NN-EXPLORE | NASA-NSF Exoplanet Observational Research |
| NOAO | National Optical Astronomy Observatory |
| NRE | Non-Recurring Engineering |
| NSF | National Science Foundation |
| OCO-2 | Orbiting Carbon Observatory-2 |
| PARVI | PAlomar Radial Velocity Instrument |
| PCF | Photonic Crystal Fiber |
| PDR | Preliminary Design Review |
| PEPSI | Potsdam Echelle Polarimetric and Spectroscopic Instrument |
| PIAA | Phase-Induced Amplitude Apodization |
| PRV | Precision Radial Velocity |
| PSF | Point Spread Function |
| R&A | Research and Analysis |
| R&D | Research and Development |
| RCN | Research Coordination Network |
| ROM | Rough Order of Magnitude |
| RV | Radial Velocity |
| SCExAO | Subaru Coronagraphic Extreme Adaptive Optics |
| SDO | Solar Dynamics Observatory |
| SLM | Spatial Light Modulator |
| SMF | Single-Mode Fiber |
| SNR | Signal-to-noise Ratio |
| SNSF | Swiss National Science Foundation |
| SOUL | Single Conjugated Adaptive Optics Upgrade |
| ThAR | Thorium–Argon |
| TMT | Thirty Meter Telescope |
| ToR | Terms of Reference |
| TRL | Technology Readiness Level |
| TTV | Transit Timing Variation |
| UKRI | UK Research and Innovation |
| VERVE | Vacuum Extreme Radial Velocity Experiment |
| WG | Working Group |
| WGM | Whispering Gallery Mode |
| WIYN | Wisconsin-Indiana-Yale-NOAO |